\documentclass[aps,prd,superscriptaddress,preprintnumbers,floatfix]{revtex4}
\usepackage{graphicx,xcolor,amsmath,bm}

\newcommand{\be}{\begin{equation}}
\newcommand{\ee}{\end{equation}}
\newcommand{\bea}{\begin{eqnarray}}
\newcommand{\eea}{\end{eqnarray}}

\def\kt{k_\perp}
\def\pt{p_\perp}

\def\avk{\langle k_\perp ^2\rangle}

\def\avkS{\langle k_\perp^2\rangle_S}

\def\lsim{\mathrel{\rlap{\lower4pt\hbox{\hskip1pt$\sim$}}\raise1pt\hbox{$<$}}}
\def\gsim{\mathrel{\rlap{\lower4pt\hbox{\hskip1pt$\sim$}}\raise1pt\hbox{$>$}}}
\def\nostrocostruttino#1\over#2{\mathrel{\mathop{\kern 0pt \rlap
{\hbox{$#1$}}} \hbox{\kern-.135em $#2$}}}

\begin{document}

\title{Assessing signals of TMD physics in SIDIS
azimuthal asymmetries \\ and in the extraction of the Sivers function}

\author{M.~Boglione}
\email{elena.boglione@to.infn.it}
\affiliation{Dipartimento di Fisica, Universit\`a di Torino,
             Via P.~Giuria 1, I-10125 Torino, Italy}
\affiliation{INFN, Sezione di Torino, Via P.~Giuria 1, I-10125 Torino, Italy}
 \author{U.~D'Alesio}
 \email{umberto.dalesio@ca.infn.it}
 \affiliation{Dipartimento di Fisica, Universit\`a di Cagliari,
              Cittadella Universitaria, I-09042 Monserrato (CA), Italy}
 \affiliation{INFN, Sezione di Cagliari,
              Cittadella Universitaria, I-09042 Monserrato (CA), Italy}
 \author{C.~Flore}
 \email{carlo.flore@ca.infn.it}
 \affiliation{Dipartimento di Fisica, Universit\`a di Cagliari,
              Cittadella Universitaria, I-09042 Monserrato (CA), Italy}
 \affiliation{INFN, Sezione di Cagliari,
              Cittadella Universitaria, I-09042 Monserrato (CA), Italy}
\author{J.O.~Gonzalez-Hernandez}
\email{joseosvaldo.gonzalez@to.infn.it}
\affiliation{Dipartimento di Fisica, Universit\`a di Torino,
             Via P.~Giuria 1, I-10125 Torino, Italy}
\affiliation{INFN, Sezione di Torino, Via P.~Giuria 1, I-10125 Torino, Italy}
%
%

\begin{abstract}
New data on the Sivers azimuthal asymmetry measured in semi-inclusive deep-inelastic scattering 
processes have recently been released by the COMPASS Collaboration at CERN. Their increased precision and their 
particular binning, in terms of $Q^2$ as well as $x$, motivates a new extraction of the Sivers function, 
within the framework of a simple and transparent parametrization. 
Signals of TMD effects visible in the Sivers asymmetries are critically assessed.
A thorough study of the uncertainties affecting the extracted Sivers function is presented, 
including the low- and large-$x$ regions.  

\end{abstract}

\maketitle

\section{\label{Intro} Introduction}

A successful study of the 3D-structure of nucleons depends on our ability to efficiently extract transverse 
momentum dependent (TMD) parton densities from experimental data. These functions encode 
non-perturbative information on the inner composition of hadrons in terms of their elementary constituents, 
and on the dynamical mechanisms which confine partons inside hadronic states. 

The extraction of TMD parton distribution functions (PDFs), however, is a complex task that involves a series 
of steps not often free of pitfalls.
On one side we have a ``theory'' (QCD in our case) which in principle provides a 
full description of the underlying physics relevant to the dynamical processes considered.
Often, however, theory cannot be applied directly because it is not exactly solvable, 
or incomplete, or simply impractical. 

Beyond the unpolarized TMD, the most interesting
and studied polarized TMD-PDF is perhaps the Sivers function~\cite{Sivers:1989cc,Sivers:1990fh}, 
which correlates the motion of unpolarized partons with the 
spin of the parent nucleon, and can be accessed through azimuthal asymmetries in polarized 
Drell-Yan (DY) and Semi-Inclusive Deep-Inelastic Scattering (SIDIS) processes. 
Remarkably, the Sivers function is predicted to have opposite
signs in these two processes~\cite{Brodsky:2002cx,Collins:2002kn}. 
This sign change, in fact, has been the focus of several phenomenological 
analyses, although none has been totally 
conclusive~\cite{Adamczyk:2015gyk,Huang:2015vpy,Anselmino:2016uie,Aghasyan:2017jop}. 

Very recently the COMPASS Collaboration has presented a new re-analysis of their SIDIS 
measurements~\cite{Adolph:2016dvl}, based on a two-dimensional binning: the Sivers asymmetries are 
presented as functions of the kinematic variables, $x$, $P_T$ and $z$, one at a time, 
for four regions of the photon virtuality $Q^2$.
These $Q^2$ ranges correspond to the four regions of the di-muon mass explored
in the ongoing analyses of the COMPASS Drell-Yan measurements~\cite{Aghasyan:2017jop}.

The large number of data, together with a considerably increased precision and a finer binning 
in $Q^2$ as well as in $x$, poses the question of whether one can extract the Sivers function 
within a full QCD scheme, as that defined in Ref.~\cite{Aybat:2011ge}. In this theoretical framework,
one must determine, based on the data, an input function which can be interpreted as the Sivers function
at a given initial scale, and the non-perturbative function $g_K$, which is responsible for the broadening 
of the TMDs as a function of the scale. Moreover, it requires a full knowledge of the unpolarized TMD PDF 
and FF, for which studies are still at a very early stage~\cite{Anselmino:2013lza,Signori:2013mda,Bacchetta:2017gcc}. 
Further complications arise from considerations as those discussed in Refs.~\cite{Boglione:2014oea,Boglione:2016bph}, 
where it has been suggested that at the kinematics of the current data, the errors of factorization may 
not be completely under control.
Note that when performing a fit on experimental data, it may happen that large theoretical errors are ``absorbed'' 
by the model-dependent parts of the TMDs, which can make the interpretation of the analysis results more problematic. 

For a reliable extraction of the Sivers function, it is also crucial to understand the extent to which 
different aspects associated with TMDs are visible in the data. Given the complications discussed above, 
it makes sense to use a bottom-up approach, that addresses questions that regard only the data and the information 
which can be inferred from them. 
In this paper, we will focus on these particular issues
and study the extent to which effects relevant to TMD physics
are likely to be observed in the existing sets of SIDIS experimental data.
For this analysis, in fact, we will model the Sivers function using a parameterization similar to that 
used in our past work~\cite{Anselmino:2008sga,Anselmino:2012aa,Anselmino:2016uie}, but 
we will relax the assumption that the Sivers functions should be parameterized in terms of the corresponding 
unpolarized TMD PDF. 
The chosen parametrization will be simple but flexible enough to  
allow for a realistic 
evaluation of the uncertainties affecting the extracted functions.

Finally, as the new COMPASS experimental data are separated in different $Q^2$ regions, it will 
be interesting to compare the results obtained by using different $Q^2$ evolution schemes. 
We will analyze the scale dependence of the Sivers function predicted by three assumptions: 
the no-evolution case, where the Sivers function does not depend at all on the scale $Q$,  
the collinear twist-three  approach, where the Sivers function varies with $Q$ only through the 
kinematic variable $x$, and a TMD-like scheme, in which the $Q^2$ evolution proceeds 
through a modification of the width of the Sivers $k_\perp$-distribution with varying $Q$.
The new functional form of the parameterization we have introduced, 
independent of the unpolarized TMD PDFs, is particularly suited to be applied to the full TMD-evolution 
scheme.

The paper is organized as follows. In Section~\ref{Strategy} we will describe the general framework used in our 
analysis and the parametrization which will be adopted. In Section~\ref{Extr} the two main best fits 
performed to extract the Sivers function will be presented and illustrated in detail, together with a thorough 
analysis of the corresponding  uncertainties and comparisons to the experimental measurements. 
In Section~\ref{new-deuterium} some results based on the error projections of a new run of the COMPASS II experiment 
with polarized deuterium targets~\cite{COMPASSII:2021} will be presented. 
The uncertainties obtained using present data and new projected errors will be compared.
In Section~\ref{evo} we will comment on how possible signals of scale dependence can be detected in the 
examined SIDIS data. 
Final remarks and conclusions will be drawn in Section~\ref{Concl}.

\section{\label{Strategy} General Strategy}

For the current study, we adopt a model for the Sivers function similar to that of 
Refs.~\cite{Anselmino:2005nn,Anselmino:2008sga,Anselmino:2011ch,Anselmino:2016uie}. 
We assume  a factorized form for the $x$ and $\kt$ dependences, and use a Gaussian 
model for the latter
\begin{equation}
\Delta ^N f_{q/p^\uparrow}(x,\kt) = 4 N_q x^{\alpha _q} (1-x)^{\beta  _q} 
\frac{M_p}{\langle k_\perp^2\rangle_S}\,
k_\perp\,
\frac{e^{-k_\perp^2/\langle k_\perp^2\rangle _S}}{\pi \langle k_\perp^2 \rangle_S}\,.
\label{eq:siv} 
\end{equation}
As mentioned above, the main difference between this parametrization and those used 
in previous analyses~\cite{Anselmino:2005nn,Anselmino:2008sga,Anselmino:2009st} is that in 
Eq.~\eqref{eq:siv} the $x$-dependent part of the 
Sivers function, for each flavour, is no longer parametrized in terms of the corresponding 
unpolarized PDF. 

In the past, when data were scarce and affected by rather large experimental uncertainties, 
this parametrization provided a useful input to allow for a successful extraction of the Sivers 
function even though the information contained in the experimental data was quite incomplete. 
It also had the advantage of ensuring the automatic fulfillment of the required positivity bounds.
In the current study, however, we relax this assumption in order to test in the most agnostic possible way 
aspects of the data related to TMD physics, like flavour separation and scale dependence. 
Our approach is also flexible enough
to allow for a realistic determination of the uncertainties in the extraction of the Sivers function.

Furthermore, in the new model the width of the Sivers function is not written in terms of the width of 
the unpolarized TMD PDF; instead, we parametrize the Sivers function directly in terms of its TMD width, 
$\langle k_\perp^2 \rangle_S$.
Note also that the parameterization of Eq.~\eqref{eq:siv} has been arranged in such a way that 
its first moment assumes a much simpler form, namely
\begin{equation}
\Delta ^N f^{(1)}_{q/p^\uparrow}(x) = 
\int d^2 {\bf k_\perp} \frac{k_\perp}{4 M_p} 
\Delta ^N f_{q/p^\uparrow}(x,k_\perp) =  N_q x^{\alpha _q} (1-x)^{\beta  _q} = - f_{1T}^{\perp(1)q}(x)\,,
\label{eq:first-mom2}
\end{equation}
where the rightmost equation provides the relation of the first moment with the Amsterdam notation.
For the unpolarized TMD PDFs and FFs we use the same functional forms as that adopted  
in Ref.~\cite{Anselmino:2016uie}, namely
\begin{align}
f_{q/p}(x,\kt) =& f_{q/p}(x)\,
\frac{e^{-k_\perp^2/\langle k_\perp^2\rangle}}{\pi \langle k_\perp^2 \rangle}\,,
\label{eq:unp-f}
\\
D_{h/q}(z,\pt) =& D_{h/q}(z)\,
\frac{e^{-\pt^2/\langle p_\perp^2\rangle}}{\pi \langle p_\perp^2 \rangle}\,,
\label{eq:unp-D}
\end{align}
where $f_{q/p}(x)$ and $D_{h/q}(z)$ are the usual unpolarized PDFs and FFs, which we will take from the
CTEQ6l~\cite{Stump:2003yu} and DSS~\cite{deFlorian:2007aj} leading order (LO) sets, 
respectively; $\langle k_\perp^2\rangle$ and $\langle p_\perp^2\rangle$ are the widths of the 
corresponding TMD distributions, which will be fixed according to the values extracted in 
Ref.~\cite{Anselmino:2013lza}, as we will explain in detail below. 
Although not explicitly indicated, our model for the unpolarized 
TMD PDFs and FFs depend on $Q^2$, according to Dokshitzer, Gribov, Lipatov, Altarelli, Parisi (DGLAP) 
equations~\cite{Gribov:1972ri,Altarelli:1977zs,Dokshitzer:1977sg}.

The SIDIS Sivers asymmetry, defined as
\be
A_{UT}^{\sin(\phi_h-\phi_S)}= 2\, 
\frac{ \int d\phi_S d\phi_h \,[d\sigma^\uparrow - d\sigma^\downarrow ]\sin(\phi_h-\phi_S) }
{ \int d\phi_S d\phi_h \, [d\sigma^\uparrow + d\sigma^\downarrow ]}=
\frac{F_{UT}^{\sin(\phi_h-\phi_S)}}{F_{UU}}\,,
\label{eq:siv-asy}
\ee
can be expressed, within this framework, through the following relations
\begin{align}
F_{UT}^{\sin(\phi_h-\phi_S)} (x,P_T,z)&=
2\,z\; \frac{P_T M_p}{{\langle P_T^2 \rangle}_S}
\frac{e^{-P_T^2/\langle P_T^2 \rangle_S}}{\pi \langle P_T^2\rangle_S}
\sum_q e^2_q 
 \Big(N_q x^{\alpha _q} (1-x)^{\beta  _q}  \Big) 
D_{h/q}(z) \,,
\label{eq:FUT} 
\\
F_{UU}(x,P_T,z)&=
\frac{e^{-P_T^2/\langle P_T^2 \rangle}}{\pi \langle P_T^2\rangle}
 \sum_q e^2_q \,f_{q/p}(x) \, D_{h/q}(z)\,,
\label{eq:FUU}\\
\langle P_T^2 \rangle=\langle \pt ^2 \rangle +& z^2 \avk
\,,\quad\quad
\langle P_T^2 \rangle_S=\langle \pt ^2 \rangle + z^2\langle \kt ^2 \rangle _S\,.
\label{eq:PTavg2}
\end{align}

We will examine different possible scenarios: 
our starting point is to set $\alpha_q=0$ and $\langle \kt ^2 \rangle _S=constant$  in Eq.~\eqref{eq:siv} 
and to consider only the contributions from $u$ and $d$ flavours. 
This provides a reference best fit that will be used as a baseline for comparison. 
Then, we analyze the data with different modifications of the above reference parametrization, each of them 
properly devised to address different aspects regarding the sensitivity
of the data to some chosen features. 

Specifically, we will investigate to which extent the present experimental data support the flavour separation 
of the Sivers function and, in turn, how we can estimate its uncertainties in the low-$x$ region, where the 
sea contributions are expected to become dominant. Moreover we will explore the sensitivity of the experimental 
measurements to $Q^2$ and $x$ correlations, to $Q^2$ dependence and, possibly, to TMD-evolution effects.

\section{\label{Extr} Extracting information from the Sivers asymmetry in SIDIS }

\subsection{\label{data} A closer look to data}

Our fits will include all experimental data presently available on the Sivers asymmetries in SIDIS processes: 
from the HERMES Collaboration for $\pi^\pm$, $\pi^0$ and $K^+$ SIDIS production off a proton target~\cite{Airapetian:2009ae}, 
from the COMPASS Collaboration for $\pi^\pm$, $K^0$ and  $K^+$ on LiD~\cite{Alekseev:2008aa} and for $h^\pm$ on NH$_3$ 
targets~\cite{Adolph:2016dvl} with $z>0.2$, which correspond to a very recent reanalysis of COMPASS 2010 measurements using a 
novel $Q^2$ binning, and finally from JLab data on $^3$He target~\cite{Qian:2011py}. 
We will not include the $K^-$ data, as they are mainly driven by the sea contributions of the Sivers function; as explained 
below in this analysis sea and valence will not be separated.  

For all experiments, these data are provided as functions of $x$, $P_T$ and $z$ kinematic variables, with the exception 
of JLab data which provides only $x$ dependent asymmetries.
We will not include the $z$-distributions, as in our model the $z$ dependence of the asymmetries is essentially fixed 
by the FFs, and it has essentially no sensitivity to our free parameters. 
In order to estimate the uncertainties in our extractions, we carefully explore the parameter space and consider 
a $2\sigma$ confidence level (C.L.), corresponding to a coverage probability of $95.4$\%. We then consistently accept 
parameter configurations that render a value of $\chi^2$ in the range $[\chi^2_{min},\chi^2_{min}+\Delta \chi^2]$. 
Note that the value of $\Delta\chi^2$ depends on the number of parameters considered and will be reported in the 
tables corresponding to each fit.

In order to extract the Sivers function, the first necessary ingredients are the unpolarized TMD PDFs and TMD FFs. 
This poses a big complication, since knowledge of these TMD functions from SIDIS data is very limited. 
In our study, we use the Gaussian functional forms of Eqs.~\eqref{eq:unp-f} and~\eqref{eq:unp-D} for the unpolarized TMDs, 
with the minimal parameters obtained in~\cite{Anselmino:2013lza}, where HERMES and COMPASS multiplicities were analyzed and fitted
within the same scheme adopted here. 
There, it was found that COMPASS and HERMES multiplicities could be well reproduced by a simple Gaussian 
model, like that of Eq.~\eqref{eq:FUU}, using only two free parameters $\langle \kt ^2 \rangle$ and $\langle \pt ^2 \rangle$, 
i.e. the widths of the TMD PDF and the TMD FF, respectively.
We stress that in Ref.~\cite{Anselmino:2013lza} data sets from the two experiments had to be fitted separately and no simultaneous
extraction was possible. In fact, it is likely that a simultaneous extraction can only be achieved by fully accounting for the highly non-trivial 
dynamics encoded in TMD-evolution equations, and properly dealing with the delicate interplay between perturbative and non-perturbative regimes. 
This is indeed a topic of ongoing research~\cite{Boglione:2014oea,Aidala:2014hva,Melis:2015ycg,Collins:2016hqq,Boglione:2016bph}.
For our current purposes all what is needed is that we consistently use the results of Ref.~\cite{Anselmino:2013lza} for each individual experiment. 
Furthermore, we use the unpolarized widths from the HERMES extraction for the JLab Sivers asymmetries, since these two experiments were shown 
to be compatible in Ref.~\cite{Anselmino:2013lza}. 

\begin{figure}[tp]
\centering
\includegraphics[width=7.0cm]{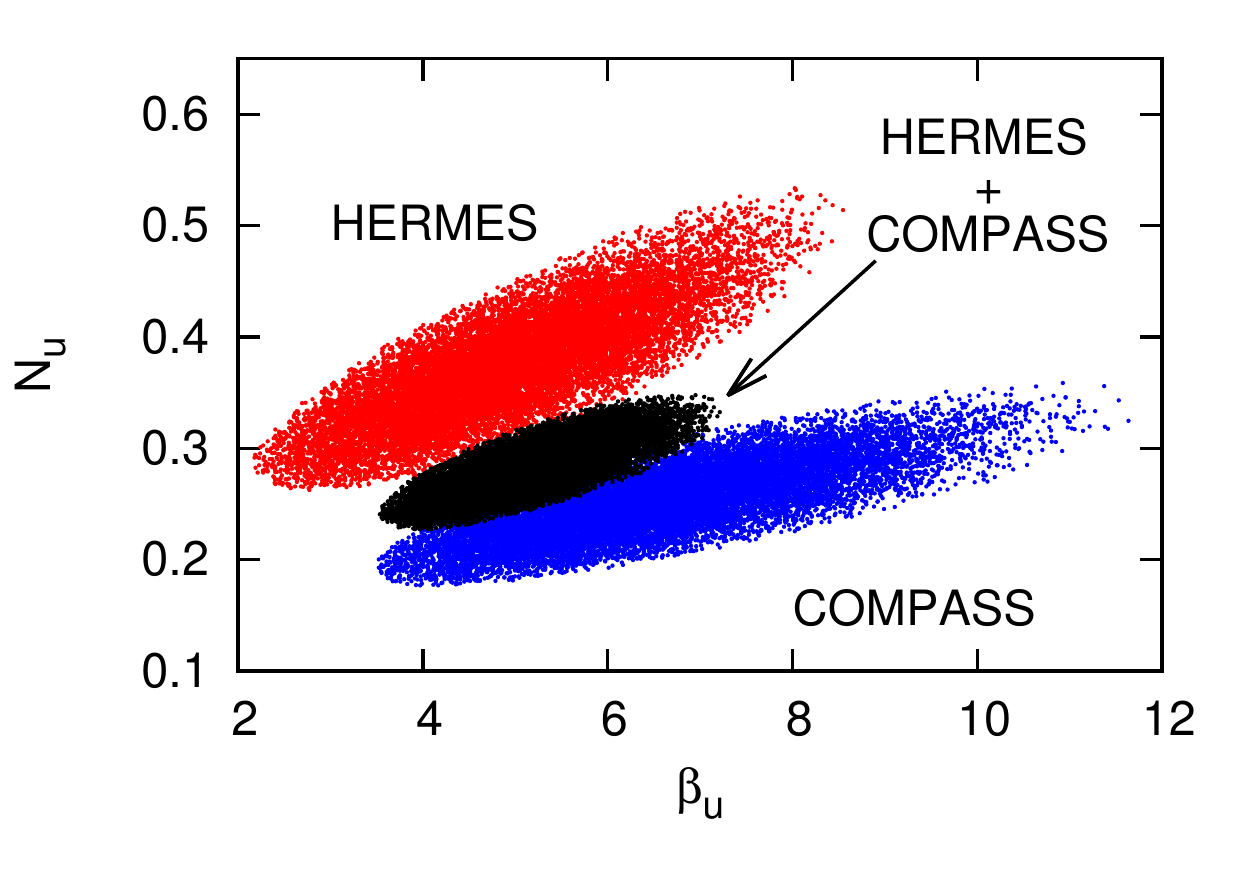} 
\includegraphics[width=7.0cm]{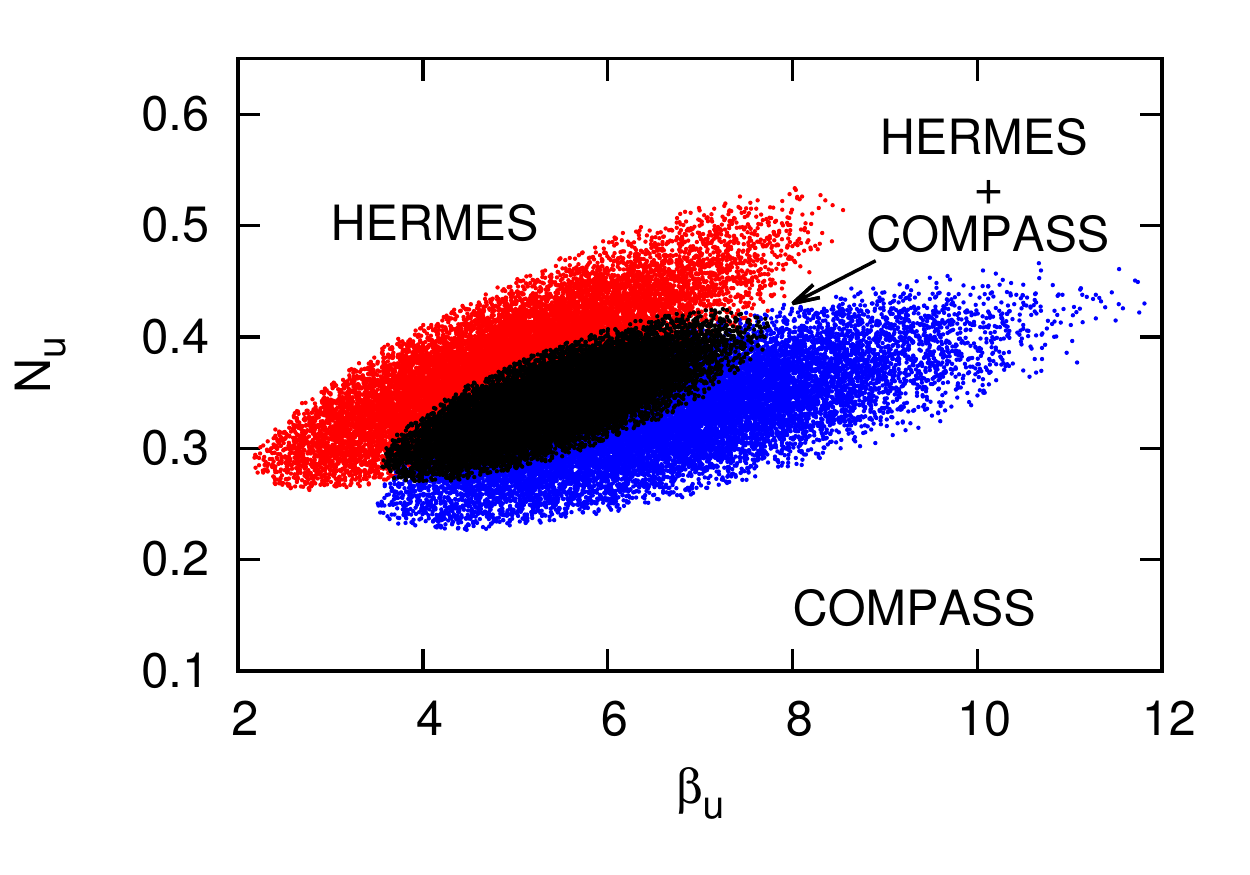}
\caption{Compatibility tests on $\pi^+$ production from a proton target using only the $u$-contribution (all others flavours being set to zero) 
of the Sivers function, as described in the text. In each panel, we show the scatter plot of the allowed 
values of $\beta_u$ and $N_u$, corresponding to a $2\sigma$ C.L., for three cases: HERMES data(red), COMPASS data(blue), HERMES+COMPASS(black).
Left panel: same unpolarized widths ($\langle \kt ^2 \rangle =0.57$ GeV$^2$ and $\langle \pt ^2 \rangle=0.12$ GeV$^2$ as obtained from HERMES 
multiplicities). Right panel: different unpolarized widths for each experiment ($\langle \kt ^2 \rangle =0.57$ GeV$^2$ and 
$\langle \pt ^2 \rangle=0.12$ GeV$^2$ for HERMES data, 
$\langle \kt ^2 \rangle =0.60$ GeV$^2$ and $\langle \pt ^2 \rangle=0.20$ GeV$^2$ for COMPASS data).
}
\label{fig:combined-fit}
\end{figure}

In order to illustrate this last point, we performed a simple test, where we evaluated the effects of using different 
Gaussian widths for the  computation of unpolarized HERMES and COMPASS cross sections, i.e. our asymmetry denominators. 
We compared two hypotheses:
$i$) using the same unpolarized widths for both HERMES and COMPASS asymmetries (namely the widths extracted from HERMES), 
$ii$) using different unpolarized widths corresponding to each experiment. 
In each case, we performed three fits on $\pi^+$ production from a proton target, considering only the $u$ contributions (all other flavours being set to zero): 
HERMES only, COMPASS only, HERMES+COMPASS simultaneously. 
Results are shown in Fig.~\ref{fig:combined-fit}, where scatter plots for the parameter space are displayed, at 
$2\sigma$ C.L.
The left panel shows that choosing the same unpolarized widths for the HERMES and COMPASS data sets 
results in fits that populate different regions of the parameter space (red and blue areas). 
In fact, the almost completely disjoint sets signal some tension. 
As a consequence, the combined fit (black area), although still giving a good value of the $\chi^2$, will have to
compromise, rendering values of the normalization parameter $N_u$ which end up being ``half way" between the 
red and the blue regions. 
In contrast, the regions in the parameter space explored in the right panel all overlap, visibly reducing 
the tension. 
This supports our choice to use the appropriate unpolarized Gaussian widths for each experimental set in our
analysis. 

This preliminary investigation illustrates how having a good knowledge of the unpolarized 
TMD distribution and fragmentation functions is of crucial importance for the analysis and extraction 
of any polarized observables. 
In this particular case, the fact that two different experimental data sets seem to point 
to different unpolarized Gaussian widths could be attributed to many different origins; possibly, 
a signal of TMD evolution effects. 
We will not get into this here, but clearly this is an issue which deserves further investigation.

\subsection{\label{Ref} Reference fit}

\begin{table}[tp]
\begin{tabular}{ccccc}
\hline
\hline
\multicolumn{5}{c}{n. of data points = 220}\\
\hline
\multicolumn{5}{c}{{\bf One flavour fits} (3 parameters)}\\
\hline
&&~\hspace*{1.5cm}~&$\chi^2_{\rm tot}$&$\chi^2_{\rm dof}$\\
&&$u$&408&1.88\\
&&$d$&914&4.21\\
\hline
\multicolumn{5}{c}{{\bf Two flavour fits} (5 parameters)}\\
\hline
&&~\hspace*{1.5cm}~&$\chi^2_{\rm tot}$&$\chi^2_{\rm dof}$\\
&&$u,\bar{u}$&266&1.24\\
&&$u,\bar{d}$&228&1.06\\
&&$\bm{u,d}$&{\bf 213}&0.99\\
\hline
\hline
\end{tabular}
\caption{
Comparison of minimal $\chi^2$ values obtained by fitting the Sivers asymmetries according to the model of Eqs.~\eqref{eq:siv}
to~\eqref{eq:PTavg2}, under different hypotheses for the flavour content of 
$F_{UT}^{\sin(\phi_h-\phi_S)}$. 
The left column indicates the flavour contribution considered in each fit (all others being set to zero).  
The top panel shows how the $u$ flavour contribution dominates the effects visible in the data.
The bottom panel shows the improvement on the description of the data when including 
one more flavour to the leading $u$ contribution. We highlight the chosen configuration for our study: 
$u$ and $d$ contributions only. Note that adding more parameters to disentangle the sea, would put the analysis
procedure at risk of over-fitting.
}
\label{tab:flavour-fits}
\end{table}

The baseline of our analysis is given by Eq.~\eqref{eq:siv}, in which we set $\alpha _u = \alpha _d = 0$, 
so that the first moment of the Sivers function simply reduces to  
\be
\Delta ^N f^{(1)}_{q/p^\uparrow}(x) = 
N_q (1-x)^{\beta  _q} \,.
\label{eq:first-mom-ref}
\ee
Furthermore, we assume the width of the Sivers function, $\langle k_\perp^2 \rangle_S$, to be independent
of other kinematic variables and of flavour. This introduces only one extra free parameter.

For all of the cases considered in this article, we will not attempt a flavour separation of sea and valence 
contributions. In fact, we have tested different hypotheses regarding the flavour content of $F_{UT}^{\sin(\phi_h-\phi_S)}$;
our results are shown in Table~\ref{tab:flavour-fits}, where the left column indicates which flavour 
component has been included in each fit (all other components being set to zero).
As it can be seen in the upper panel of Table~\ref{tab:flavour-fits},
the $u$ flavour Sivers function represents the leading contribution to the asymmetries.
The total $\chi ^2$ improves significantly if one more flavour is added to the fit, as shown on the 
lower panel of Table~\ref{tab:flavour-fits}. 
Any further addition of different flavour contributions will not improve the quality of the fit,
making convergence to the minimum more cumbersome and exposing us to the risk of over-fitting.
\begin{table}[b]
\begin{tabular}{l l r}
\hline
\hline
\multicolumn{3}{c}{{\bf Reference fit - no evolution}}\\
\hline
$\chi^2_{\rm tot}$ = 212.8      & ~ n. of points = 220  & ~\\
$\chi^2_{\rm dof}$ = 0.99       & ~ n. of free parameters = 5 & ~\\
$\Delta \chi^2$ = 11.3           & ~ & ~\\
\hline
HERMES & ~~$\langle k_\perp^2 \rangle = 0.57$ GeV$^2$ & ~~~~$\langle p_\perp^2 \rangle = 0.12$ GeV$^2$  \\
COMPASS & ~~$\langle k_\perp^2 \rangle = 0.60$ GeV$^2$ & ~~~~$\langle p_\perp^2 \rangle = 0.20$ GeV$^2$ \\
\hline
$N_u= 0.40 \pm 0.09$ & ~~$\beta_u=5.43 \pm 1.59$ & ~\\
$N_d=-0.63 \pm 0.23$ & ~~$\beta_d=6.45 \pm 3.64$ & ~\\
\multicolumn{3}{l}{$\langle k_\perp^2 \rangle _S = 0.30 \pm 0.15$ GeV$^2$}\\
\hline\hline
\end{tabular}
\caption{
Best fit parameters and $\chi^2$ values for the reference fit. 
The parameter errors correspond to $2\sigma$ C.L. 
Notice that these errors are well in agreement with the uncertainties on the 
free parameters shown in the scatter plots of Fig.~\ref{fig:para-space-ref}. }
\label{tab:ref}
\end{table}
%
\begin{figure}[t]
\centering
\includegraphics[width=7.0cm]{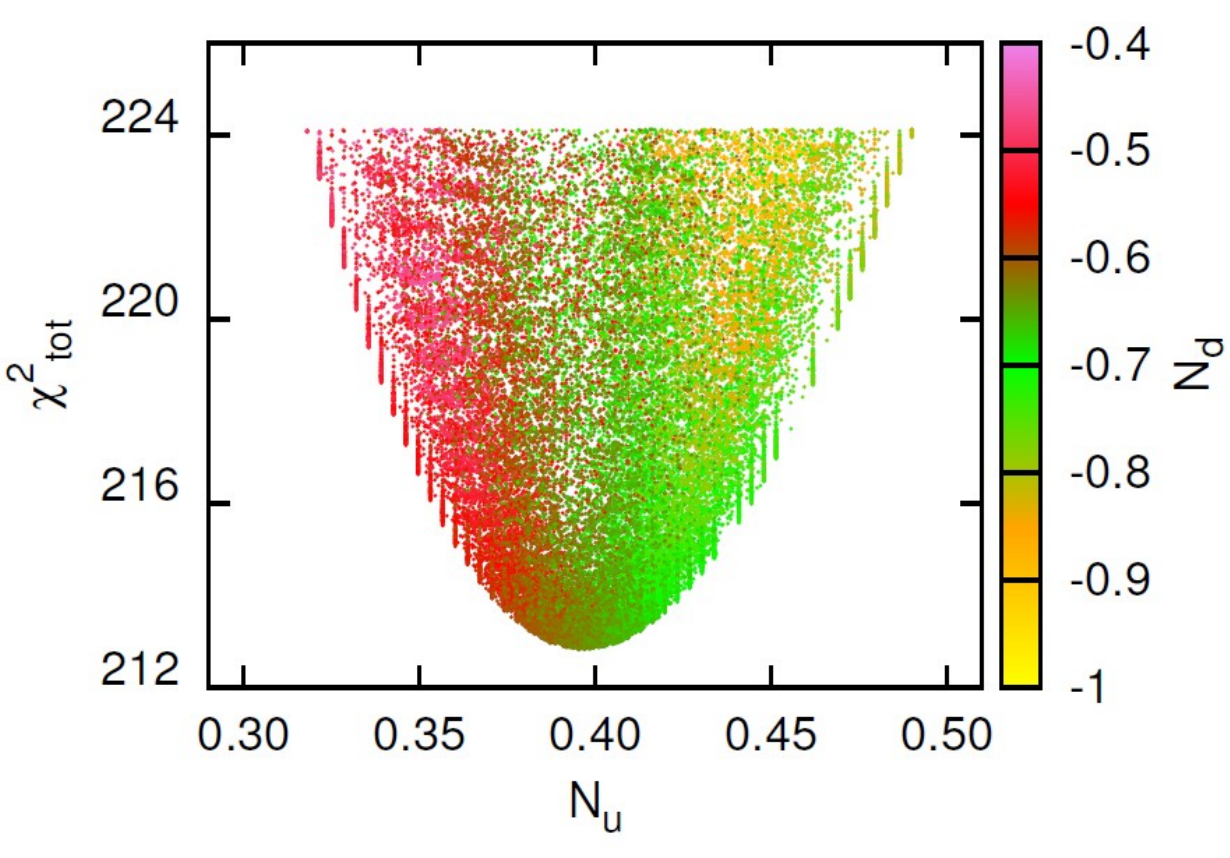} \hspace*{0.5cm}
\includegraphics[width=7.0cm]{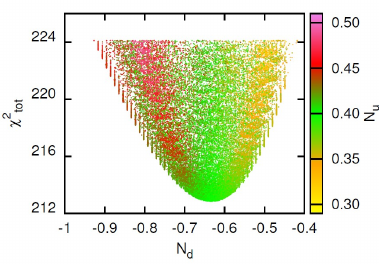} \\
\includegraphics[width=7.0cm]{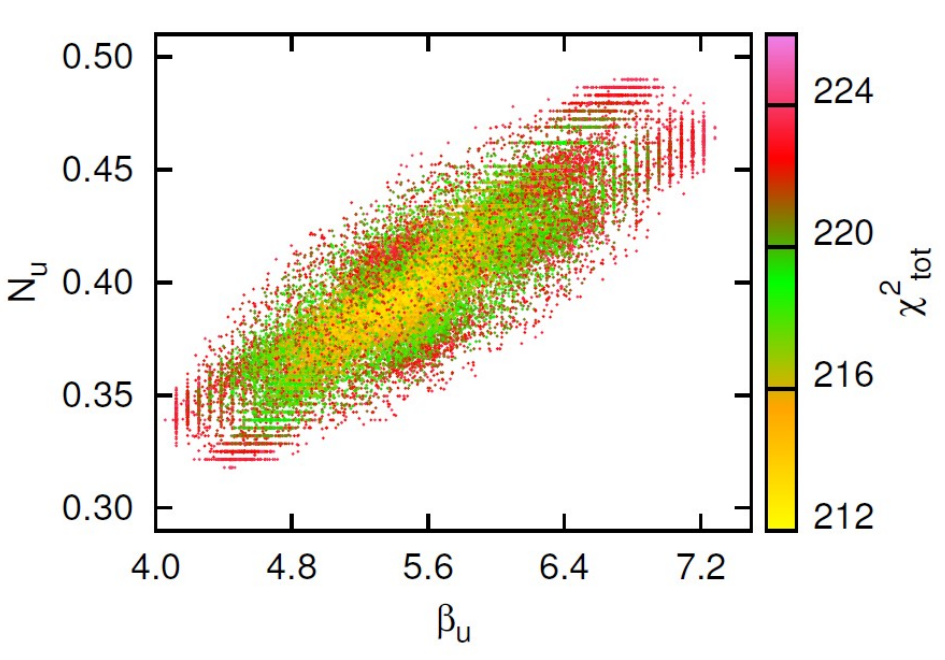} \hspace*{0.5cm}
\includegraphics[width=7.0cm]{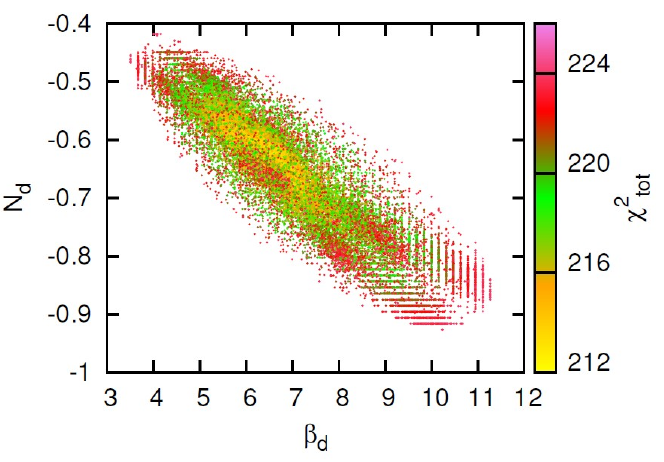}\\
\caption{Scatter plots representing the parameter space of the reference best fit. 
The shaded regions correspond to our estimate of $2\sigma$ C.L. error band.}
\label{fig:para-space-ref}
\includegraphics[width=7.0cm]{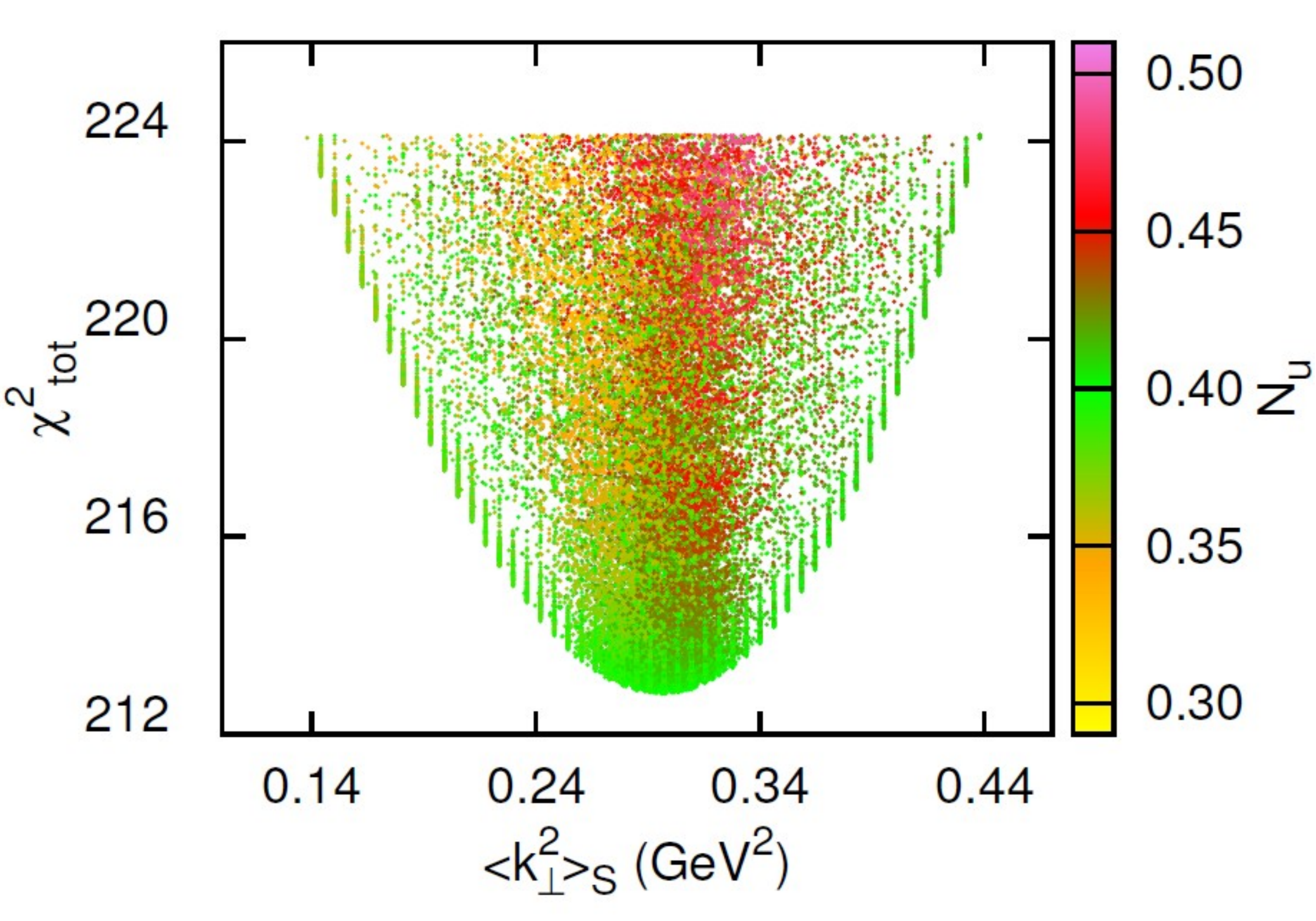}
\caption{Allowed values for the width of the Sivers function $\langle k_\perp^2\rangle_S$ as a function of $\chi^2_{\rm tot}$ . 
The region displayed corresponds to our estimate of $2\sigma$ C.L. error band.}
\label{fig:para-space-ref-width}
\end{figure}
For our analysis, we use the configuration that renders the smallest minimal $\chi^2$, 
i.e. we directly parametrize the total $u$ and $d$ flavours as follows:
\bea
 &&\Delta ^N f_{u/p^\uparrow}=
 \Delta ^N f_{u_v/p^\uparrow}+\Delta ^N f_{\bar{u}/p^\uparrow} = 
 4\,N_u (1-x)^{\beta _u}\,\frac{M_p}{\langle k_\perp^2\rangle_S}\,
 k_\perp\,
 \frac{e^{-k_\perp^2/\langle k_\perp^2\rangle _S}}{\pi \langle k_\perp^2 \rangle_S}\,,
 \label{eq:delta-f-u}
 \\
 &&\Delta ^N f_{d/p^\uparrow}=
 \Delta ^N f_{d_v/p^\uparrow}+\Delta ^N f_{\bar{d}/p^\uparrow} = 
 4\,N_d (1-x)^{\beta _d}\,\frac{M_p}{\langle k_\perp^2\rangle_S}\,
 k_\perp\,
 \frac{e^{-k_\perp^2/\langle k_\perp^2\rangle _S}}{\pi \langle k_\perp^2 \rangle_S}\,.
 \label{eq:delta-f-d}
\eea
This results in a fit with a total of 5 free parameters, $N_u$, $N_d$, $\beta_u$, $\beta_d$  and $\langle k_\perp^2\rangle_S$. 
We call this ``reference fit''. 
The role of the sea contributions, which are expected to be relevant at small $x$ where the behaviour of the 
Sivers function is mainly driven by the $\alpha _q$ parameters, will be addressed in Section~\ref{low-x}.

As explained above, the $\kt$ widths of the unpolarized TMDs are fixed according to the values extracted 
in Ref.~\cite{Anselmino:2013lza}, to make sure that the unpolarized 
cross sections appearing in the asymmetry denominator reproduce well the measured multiplicities for both 
HERMES and COMPASS experiments. In this first, simple fit no $Q^2$ evolution is applied to the Sivers function,
and the corresponding plots are labeled by ``no-evolution''; the extracted function will therefore represent 
the Sivers function at the average $Q^2$ scale of the experimental data.

Table~\ref{tab:ref} shows the values of the free parameters as determined by our best fit, together with the 
minimal values of the $\chi^2$ and the total number of data points included. 
The errors reported in Table~\ref{tab:ref} are MINUIT errors, corresponding to $2\sigma$ C.L., i.e. 
to a coverage probability of $95.4$\%.

The top panels of Fig.~\ref{fig:para-space-ref} show the $\chi^2 _{\rm tot}$ profiles as functions of the 
parameters $N_u$ and $N_d$. 
In these two plots the correlations between the parameters $N_u$ and $N_d$ are colour-coded: yellow corresponds to lowest, 
green to intermediate and purple to highest allowed values of $N_u$ (top left panel) and $N_d$ (top right panel).
It is evident that these profiles are quite well approximated by a quadratic function, confirming that the Hessian 
method adopted to evaluate the errors on the parameters is reliable. 
In fact, the errors reported in Table~\ref{tab:ref} are well in agreement with the uncertainties on the 
free parameters that can easily be inferred by looking at the scatter plots. 
The lower panels of Fig.~\ref{fig:para-space-ref} represent the correlations between the parameters 
$N_u$, $\beta_u$ (lower left) and $N_d$, $\beta_d$  (lower right). 
Here it is the corresponding $\chi^2_{\rm tot}$ which is colour-coded: yellow corresponds to the 
lowest $\chi^2_{\rm tot}$ values, green to intermediate and purple to the highest $\chi^2_{\rm tot}$ values. 
As expected from a fit consistent with a Hessian approximation, 
the correlations among parameters cover regions of reasonably regular, ellipsoidal shapes.  
Fig.~\ref{fig:para-space-ref-width} shows the $\chi^2_{\rm tot}$ profile of the Sivers $\kt$ width, $\avkS$, 
and its correlation with $N_u$ (color coded). 
Also in this case the uncertainties indicated by the scatter plots of the parameter space are perfectly consistent 
with the errors estimated by adopting the Hessian approximation, reported in Table~\ref{tab:ref}.

Plots showing the $u$ and $d$ Sivers functions and their estimated uncertainty bands, as extracted in this fit 
will be shown below. 
They will be extensively discussed in Section~\ref{low-x}.

The reference fit, with 5 free parameters, is able to reproduce all the existing 
SIDIS experimental measurements. Moreover, it provides a successful extraction of the Sivers function as well as a reliable estimate of 
the uncertainties, over the kinematic region covered by the bulk of experimental data (i.e. approximately $0.03 < x < 0.3$). 
Below this region, where only very few data points are present, the error bands from the reference fit are at risk of
being artificially small. 
In section~\ref{low-x}, 
we will consider the case where the $\alpha$ parameters
in Eq.~\eqref{eq:siv}, 
which regulate the low-$x$ behaviour of the Sivers function, 
may be different from zero. As we will discuss, this provides 
error bands that better reflect the amount of information which can be inferred from data.

\subsection{\label{low-x} Low-$x$ uncertainties}

Starting from the reference fit described above, which represents the basis for all further studies 
presented in this paper, we now move on to explore in more detail the low-$x$ kinematic region. 
To do this, we perform a different fit in which we allow our parametrization to become more flexible 
at small $x$ by including two extra parameters, $\alpha_u$ and $\alpha_d$, in the following way
\be
\Delta ^N f^{(1)}_{q/p^\uparrow}(x) = 
N_q x^{\alpha _q} (1-x)^{\beta  _q} \,.
\label{eq:first-mom-alpha}
\ee
This best fit will be referred to as the ``$\alpha$-fit''.
Table~\ref{tab:ref-alpha} shows the $\chi^2$ and the values of the parameters obtained in this case.
As it is immediately evident, the value of $\chi^2_{\rm dof}$ is unchanged, 
therefore the overall quality of the fit does not improve.
Moreover, the central values of the free parameters are extremely close to those obtained in the 
reference fit. 
This suggests that the experimental data currently used are not sensitive to the particular choice 
of value for $\alpha_u$ and $\alpha_d$. 
Consequently, further constraining the low-$x$ behaviour of the Sivers function seems at the moment unlikely. 

\begin{table}[t]
\begin{tabular}{l l r}
\hline
\hline
\multicolumn{3}{c}{{\bf $\bm{\alpha}$ fit - no evolution}}\\
\hline
$\chi^2_{\rm tot}$ = 211.5         & ~n. of points = 220 & ~\\
$\chi^2_{\rm dof}$ = 0.99          & ~n. of free parameters = 7  & ~\\
$\Delta \chi^2$ = 14.3      & ~ & ~\\
\hline
HERMES~~~~~~ $\langle k_\perp^2 \rangle = 0.57$ GeV$^2$ & ~~$\langle p_\perp^2 \rangle = 0.12$ GeV$^2$ & ~ \\
COMPASS~~~~ $\langle k_\perp^2 \rangle = 0.60$ GeV$^2$ & ~~$\langle p_\perp^2 \rangle = 0.20$ GeV$^2$ & ~\\
\hline
$N_u= 0.40 \pm 0.09$ & ~~ $\beta_u=5.93 \pm 3.86$ & $\alpha_u=0.073 \pm 0.46$~~ \\
$N_d=-0.63 \pm 0.23$ & ~~ $\beta_d=5.71 \pm 7.43$ & $\alpha_d=-0.075 \pm 0.83$  \\
$\langle k_\perp^2 \rangle _S = 0.30 \pm 0.15$ GeV$^2$ & ~ & ~\\
\hline\hline
\end{tabular}
\caption{Best fit parameters and $\chi^2$ values corresponding to the $\alpha$-fit. 
Notice that, despite the presence of two extra parameters 
w.r.t. the reference fit presented in Table~\ref{tab:ref}, the value of $\chi^2_{\rm tot}$ 
remains practically unchanged.
However, the uncertainty on the free parameters increases considerably. 
This generates much larger uncertainty bands in the low-$x$ region, as shown in Fig.~\ref{fig:ref-vs-alpha}.
}
\label{tab:ref-alpha}
\end{table}
%
%
%
%
\begin{figure}[t]
\centering
\includegraphics[width=7.0cm]{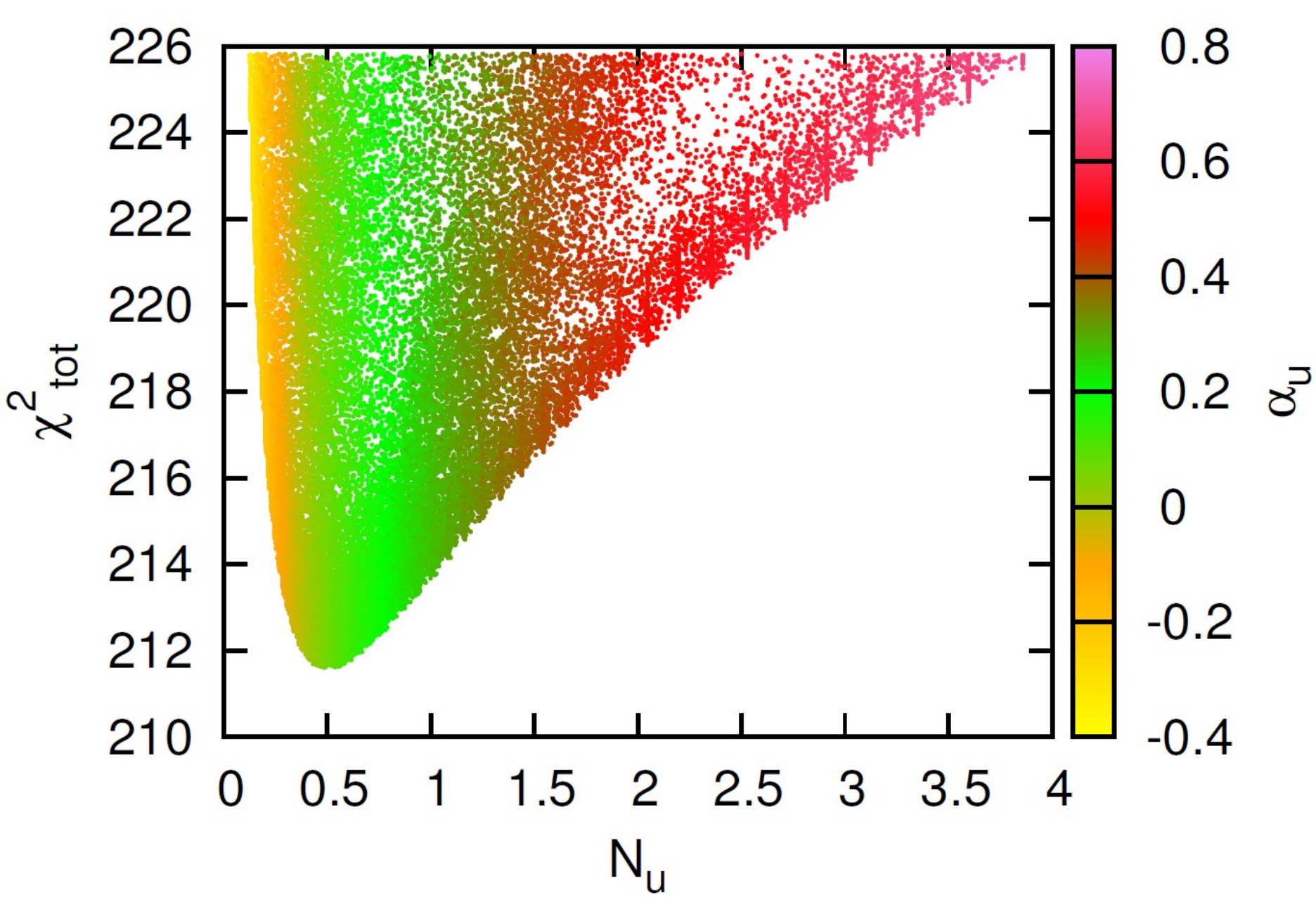} \hspace*{0.5cm} 
\includegraphics[width=7.0cm]{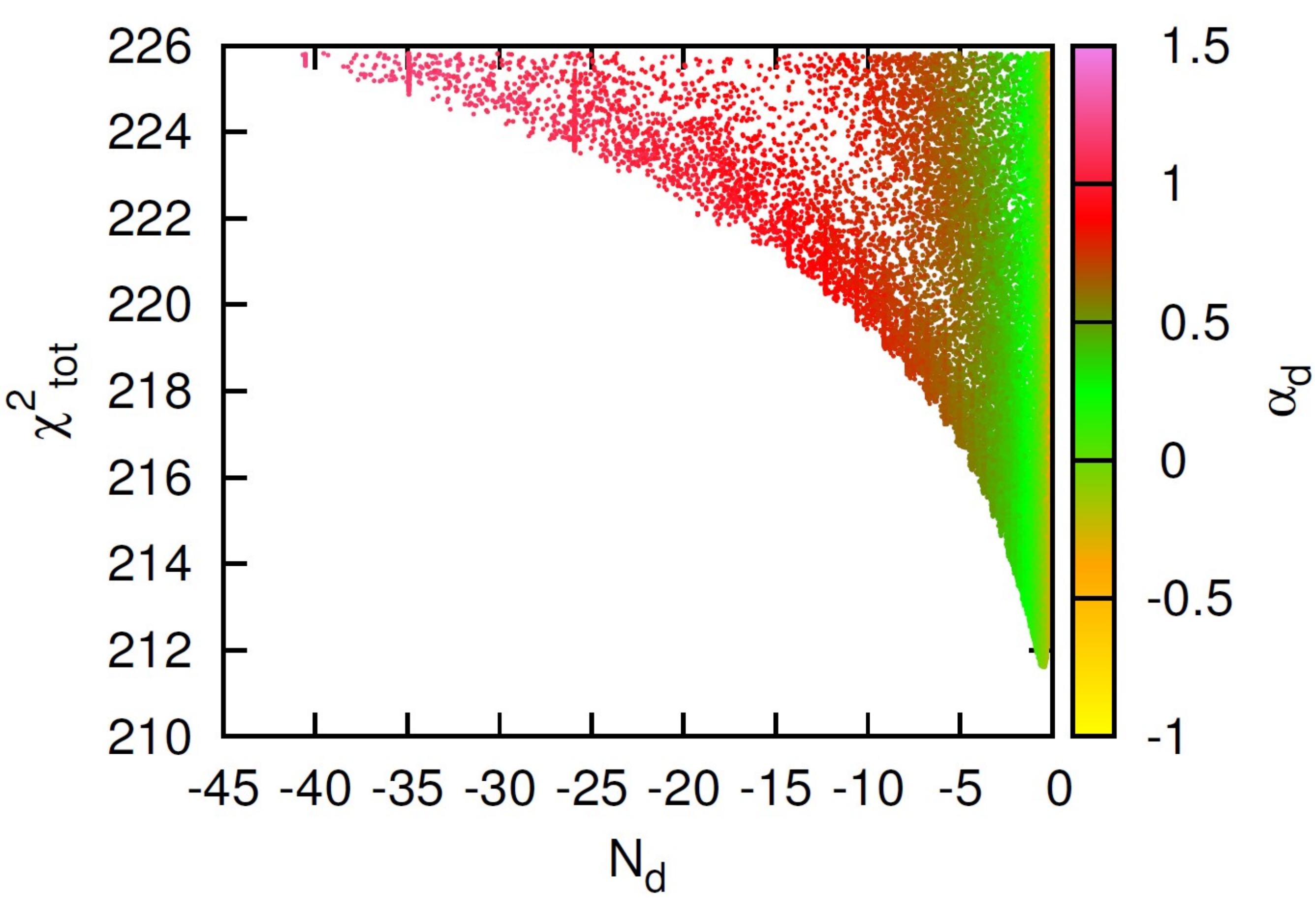} \\
\includegraphics[width=7.0cm]{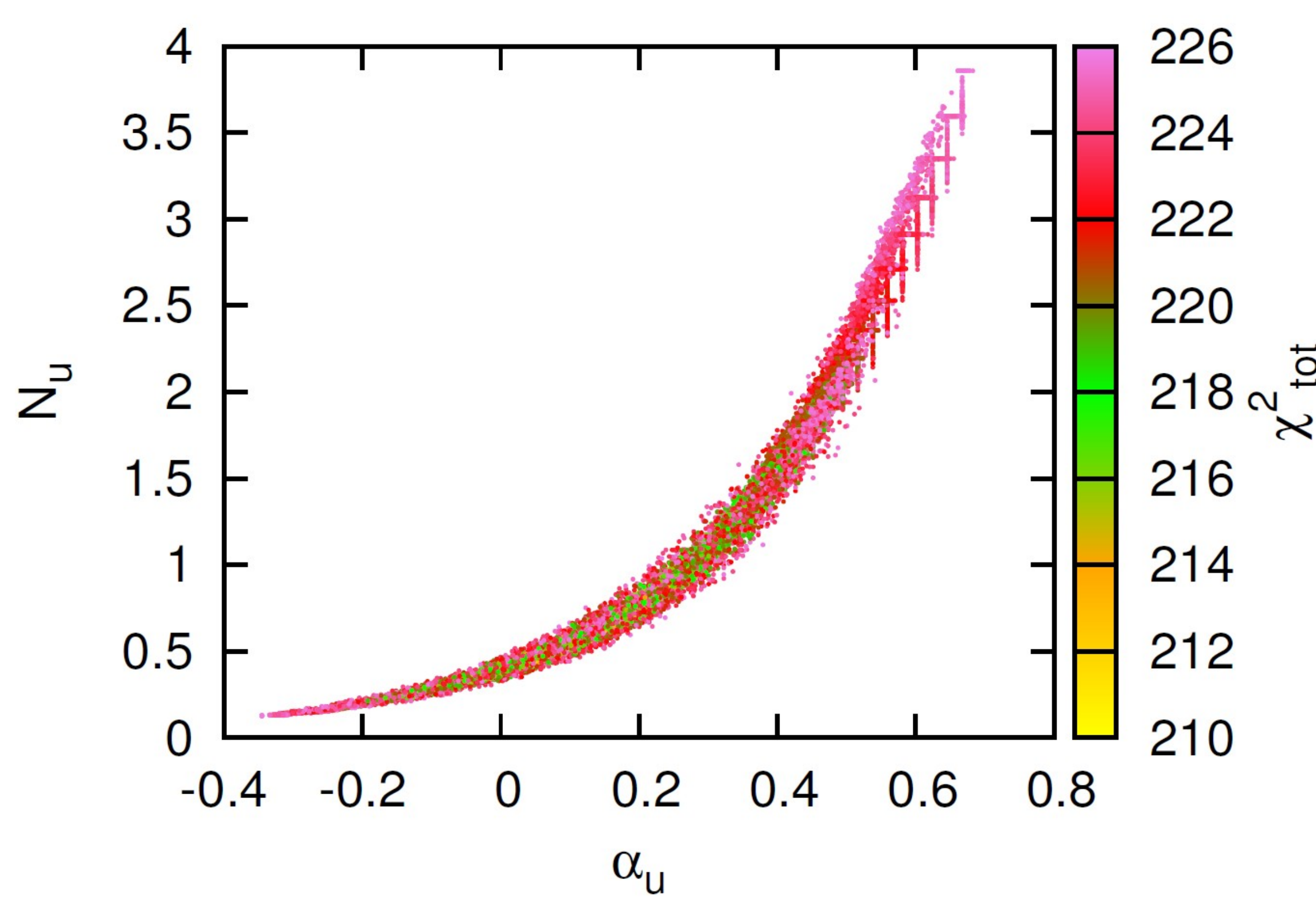} \hspace*{0.5cm}
\includegraphics[width=7.0cm]{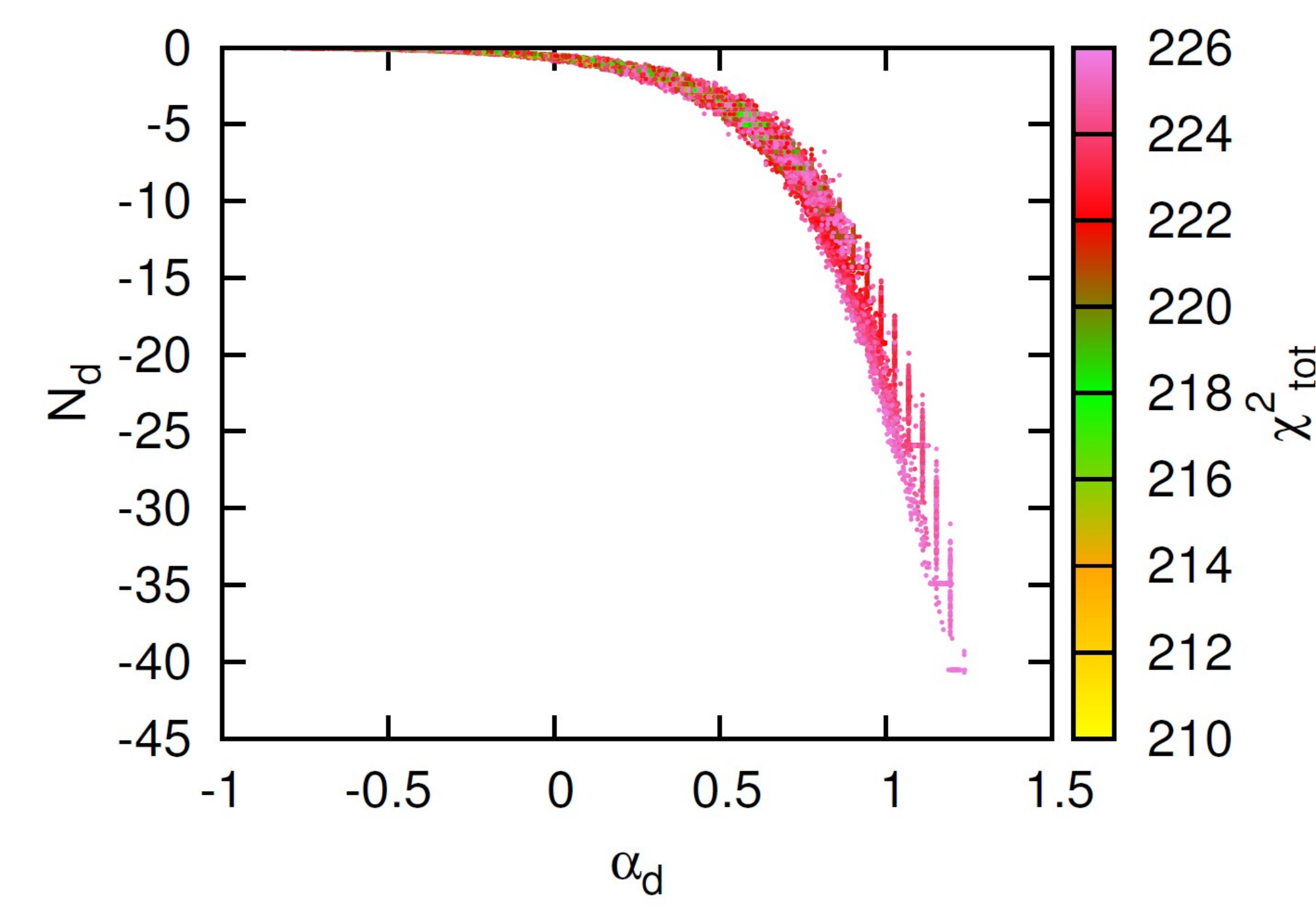} \\
\caption{Parameter space scatter plots for the $\alpha$-fit, 
which includes the $\alpha_u$ and $\alpha_d$ free parameters.
The regions displayed correspond to our estimate of $2\sigma$ C.L. error band.
Notice that the uncertainties on the parameters which can be inferred from the scatter plots are much larger 
than the errors reported in Table~\ref{tab:ref-alpha}.}
\label{fig:para-space-alpha}
\end{figure}
%
%
%
%
\begin{figure}[t]
\centering
\includegraphics[width=7.0cm]{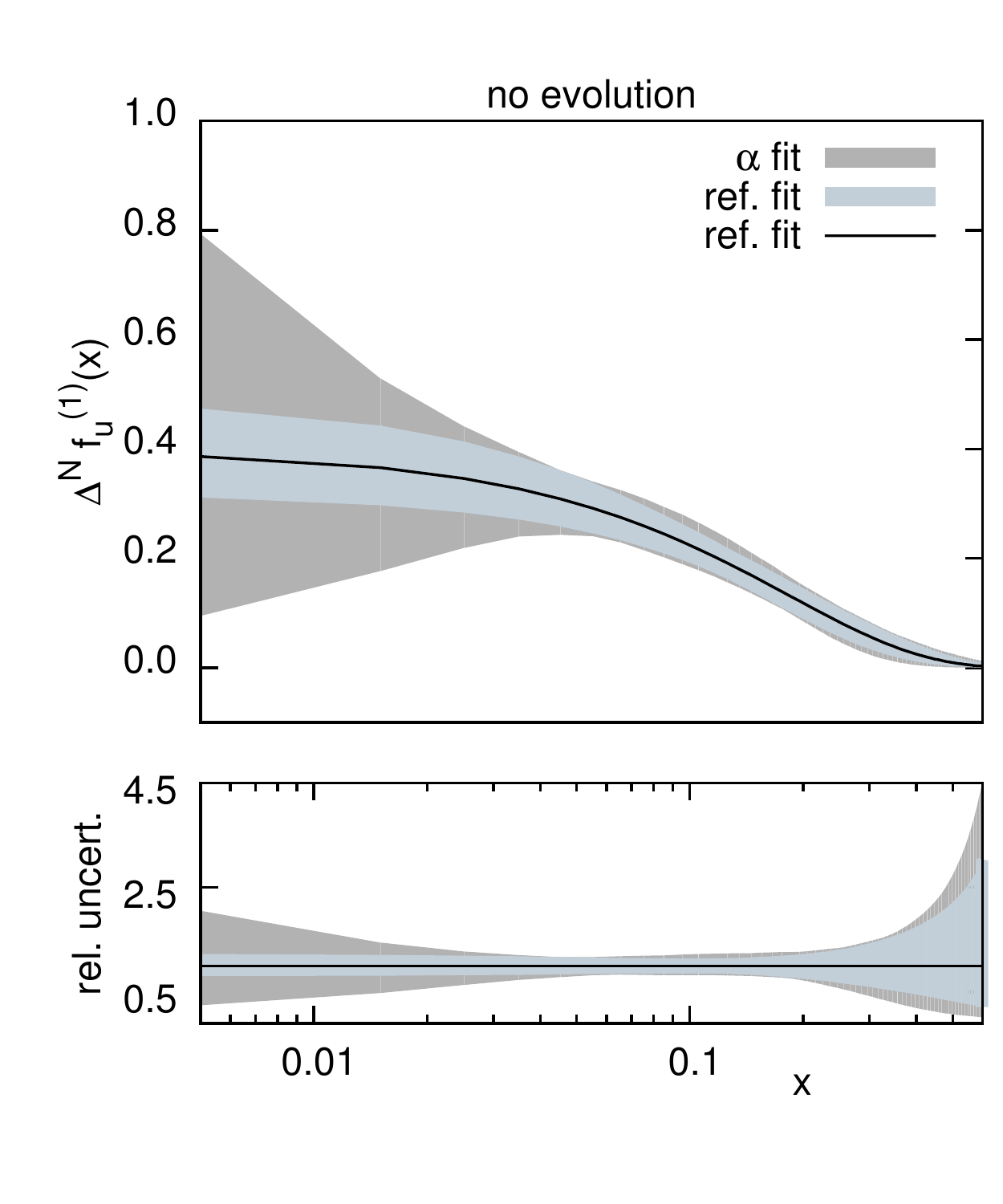}
\includegraphics[width=7.0cm]{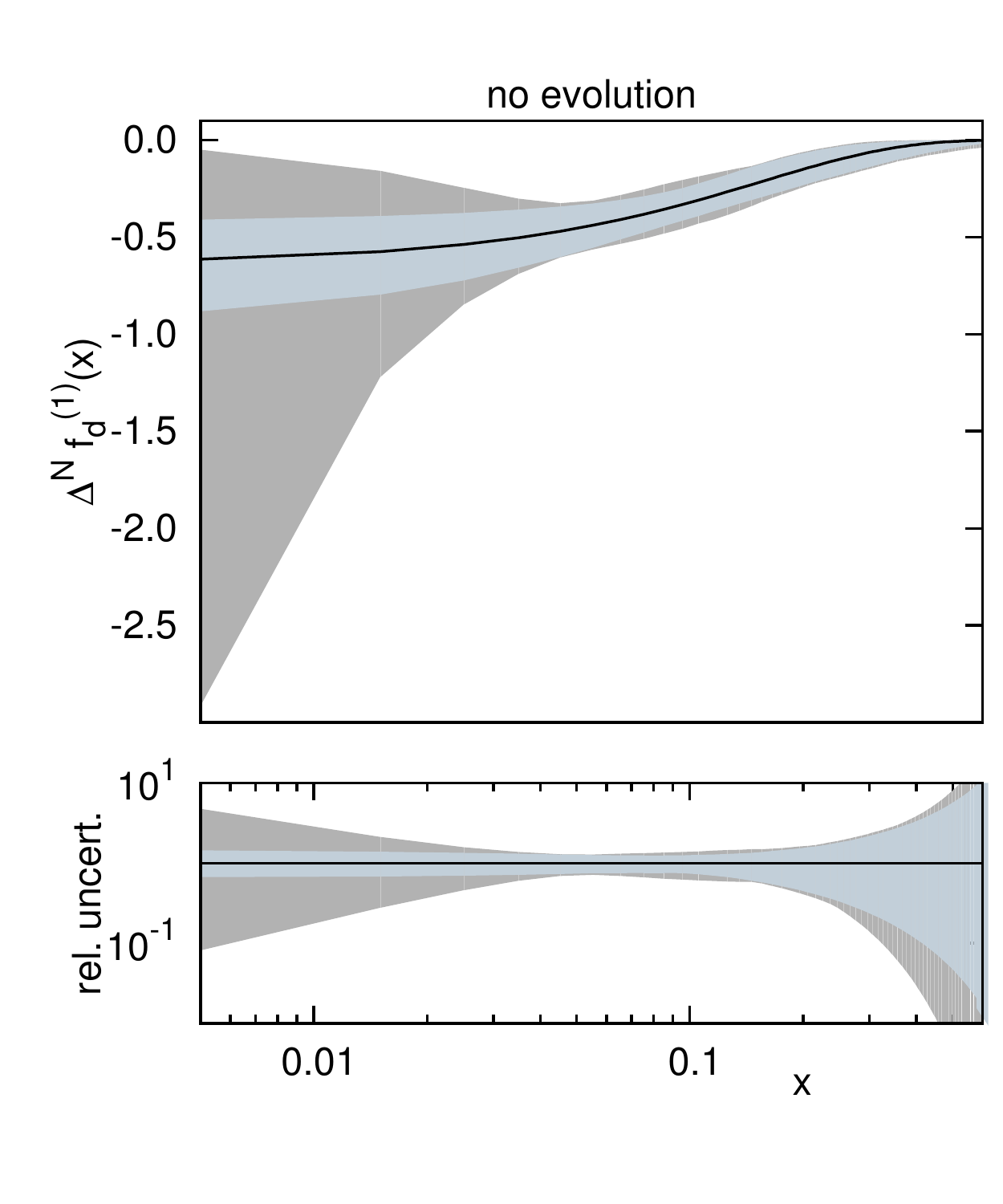}\\
\includegraphics[width=7.0cm]{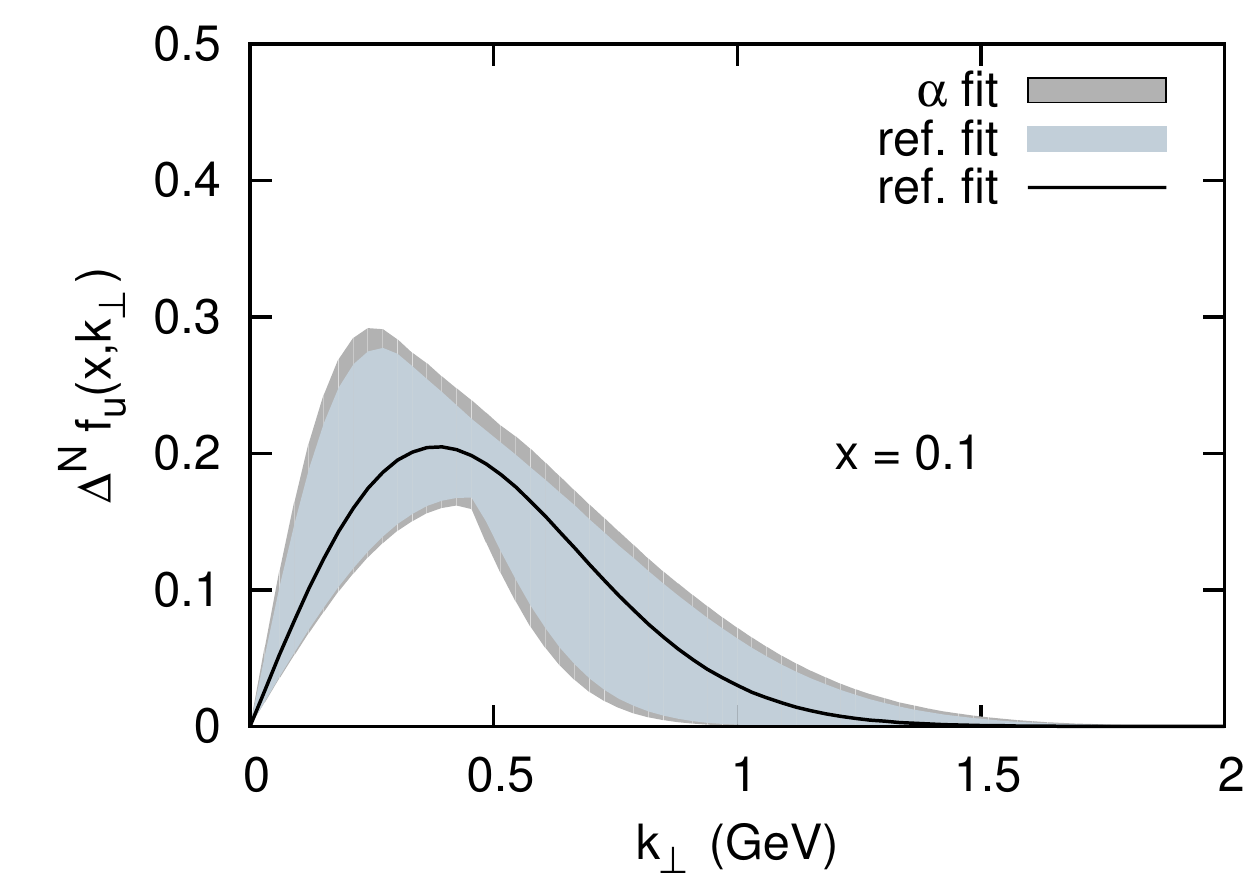}
\includegraphics[width=7.0cm]{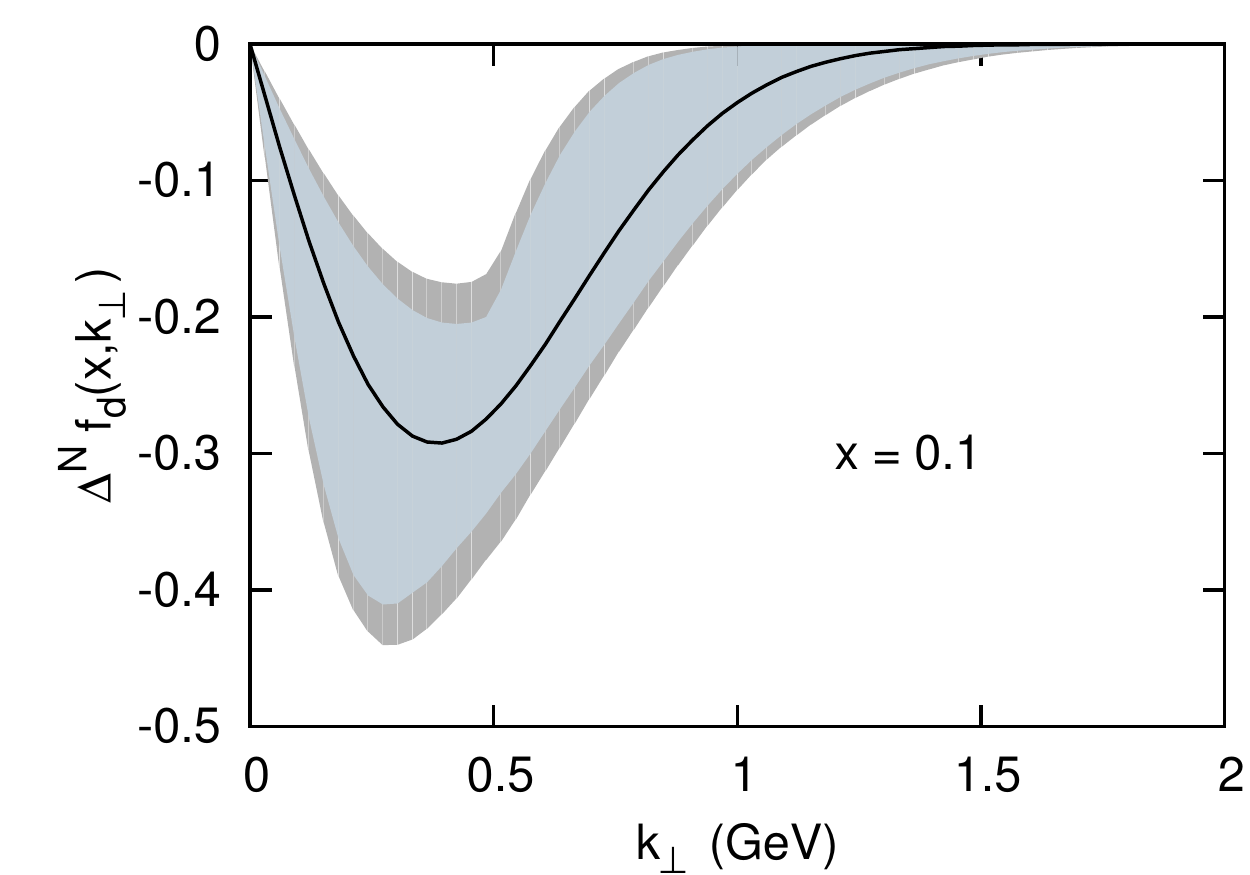}
\caption{
The extracted Sivers distributions for $u = u_v + \bar u$ and  
$d = d_v + \bar d$. Upper panels: the first moments of the 
Sivers function, Eqs.~\eqref{eq:first-mom-ref} and~\eqref{eq:first-mom-alpha}, are shown versus $x$. 
Middle panel: relative uncertainties, 
given by the ratio between the upper/lower border of the uncertainty bands and the 
best-fit curve for the reference fit. 
Lower panel: the Sivers functions, Eqs.~(\ref{eq:siv}), is shown versus $\kt$, at $x=0.1$. 
Here we have no $Q^2$ dependence. 
The shaded bands correspond to our estimate of $2\sigma$ C.L.
In all panels, the light blue bands correspond to the uncertainties of the reference fit 
(only $N_{u(d)}$ and $\beta_{u(d)}$ free parameters), while the large grey bands correspond to the 
uncertainties for the fit which includes also the $\alpha_u$ and $\alpha_d$ parameters.
}
\label{fig:ref-vs-alpha}
\end{figure}
%
\begin{figure}[htp]
\centering
\includegraphics[width=7.0cm]{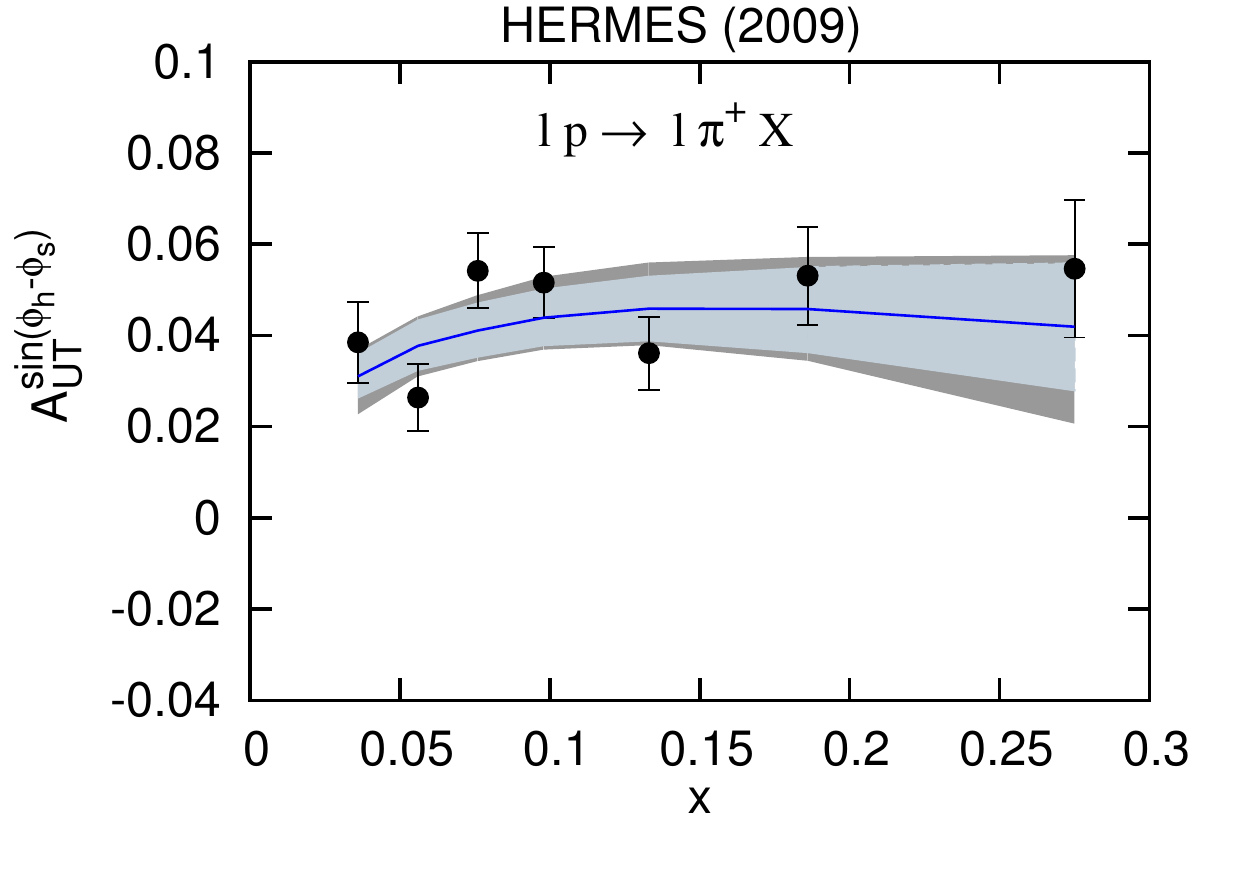}
\includegraphics[width=7.0cm]{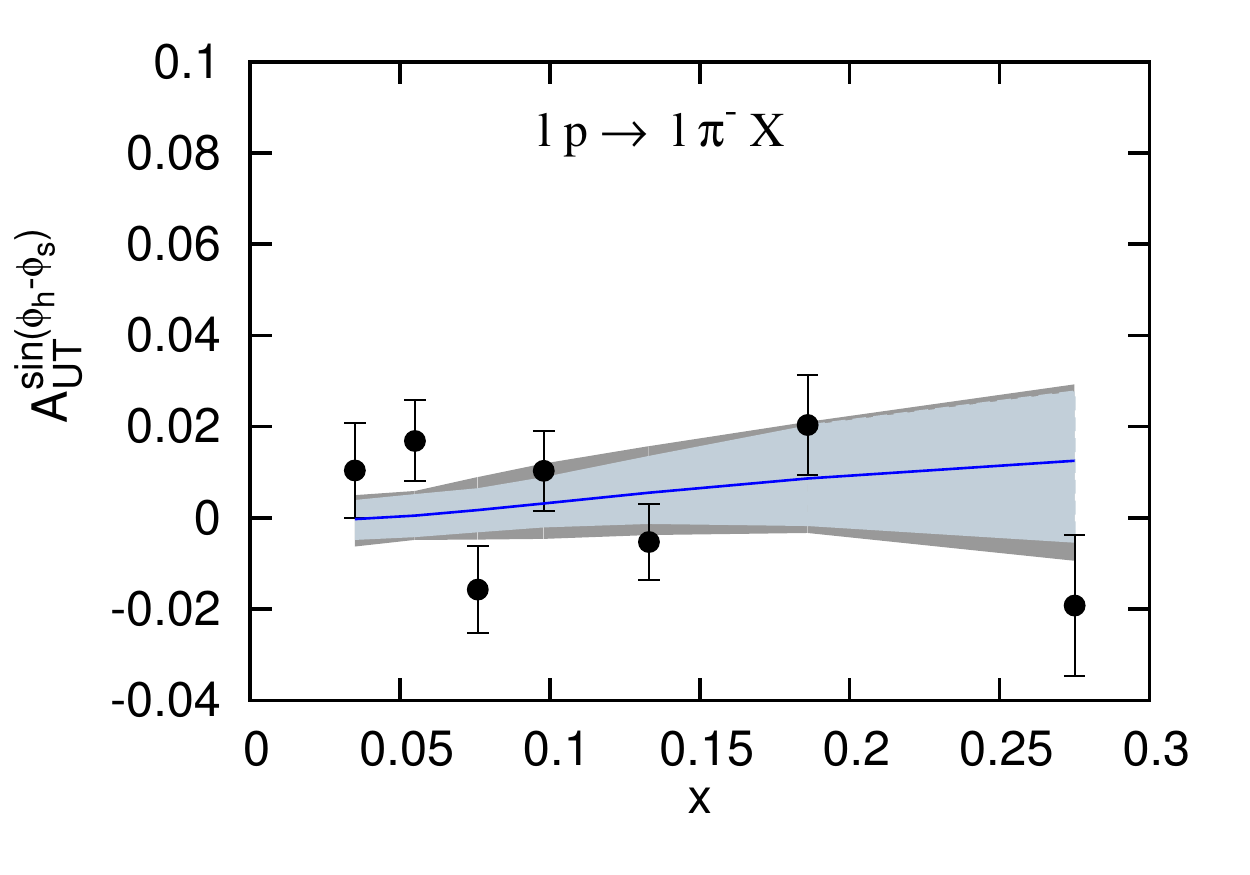} 
\includegraphics[width=7.0cm]{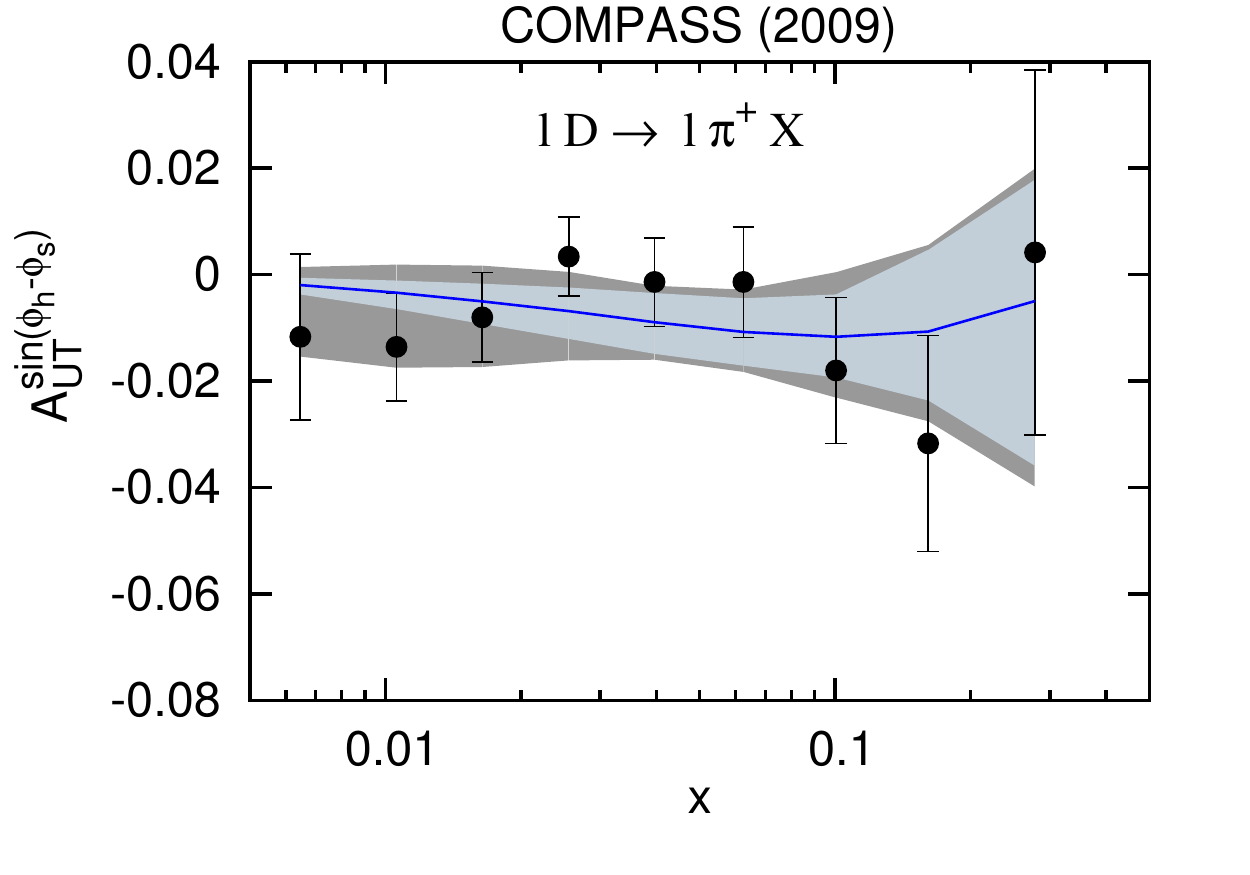}
\includegraphics[width=7.0cm]{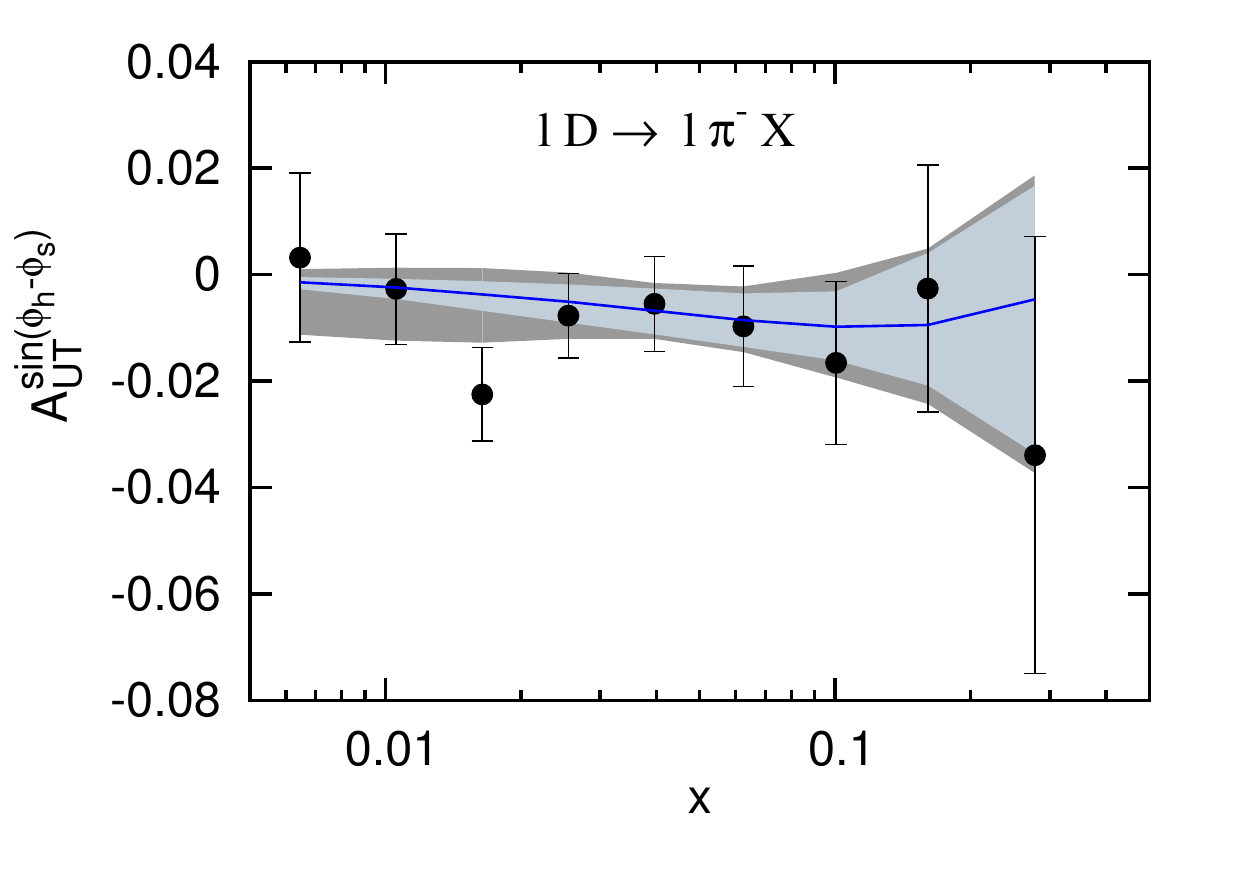} 
\includegraphics[width=7.0cm]{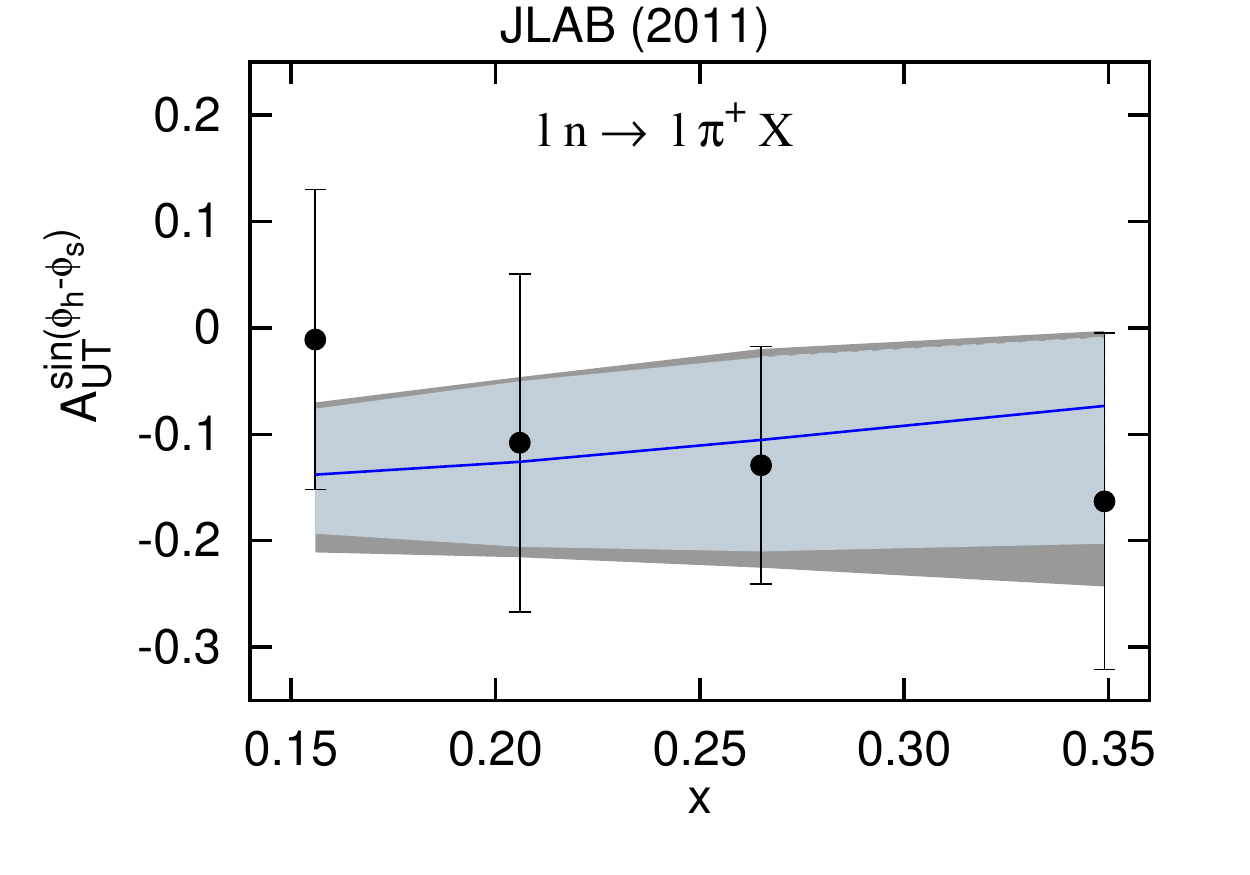}
\includegraphics[width=7.0cm]{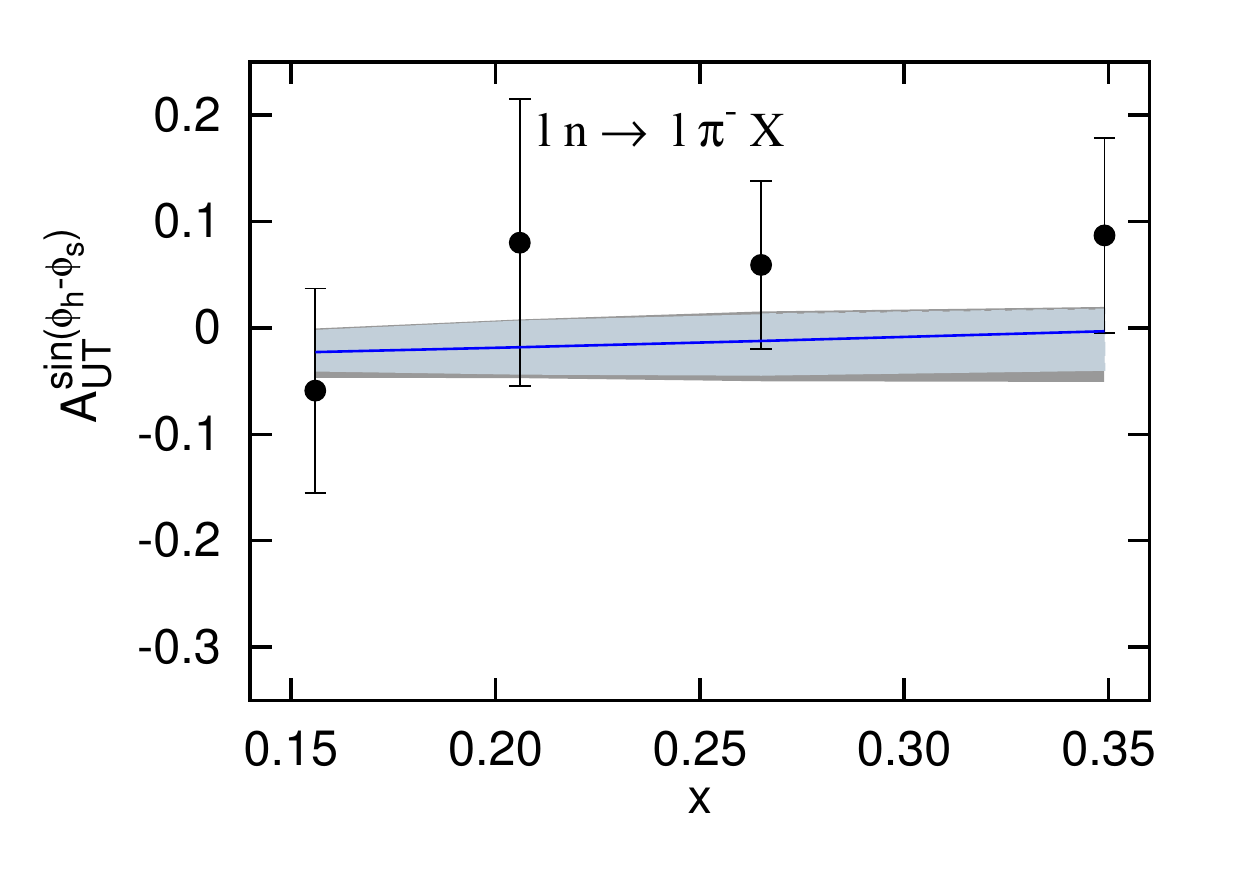} 
\caption{The results obtained from the reference fit and the $\alpha$-fit are compared to 
the HERMES measurements of the SIDIS Sivers asymmetry for $\pi^\pm$ production 
off a proton target~\cite{Airapetian:2009ae} (upper panels), 
to the COMPASS measurements of the SIDIS Sivers asymmetry on a LiD 
target~\cite{Alekseev:2008aa} for $\pi^\pm$ production (middle panels),
and to the JLab data for $\pi^\pm$ production on a $^3$He target~\cite{Qian:2011py} (bottom panel). 
Here we show the $x$ dependence only. 
The shaded region corresponds to our estimate of $2\sigma$ C.L. error band.
The light-blue bands correspond to the uncertainties of the reference fit 
(only $N_{u(d)}$ and $\beta_{u(d)}$ free parameters), while the (larger) gray bands correspond to the 
uncertainties of the $\alpha$-fit, which includes also the $\alpha_{u(d)}$  parameters.}
\label{fig:old-data}
\end{figure}
%
%
%
%
%
%
\begin{figure}[htp]
\centering
\includegraphics[width=7.0cm]{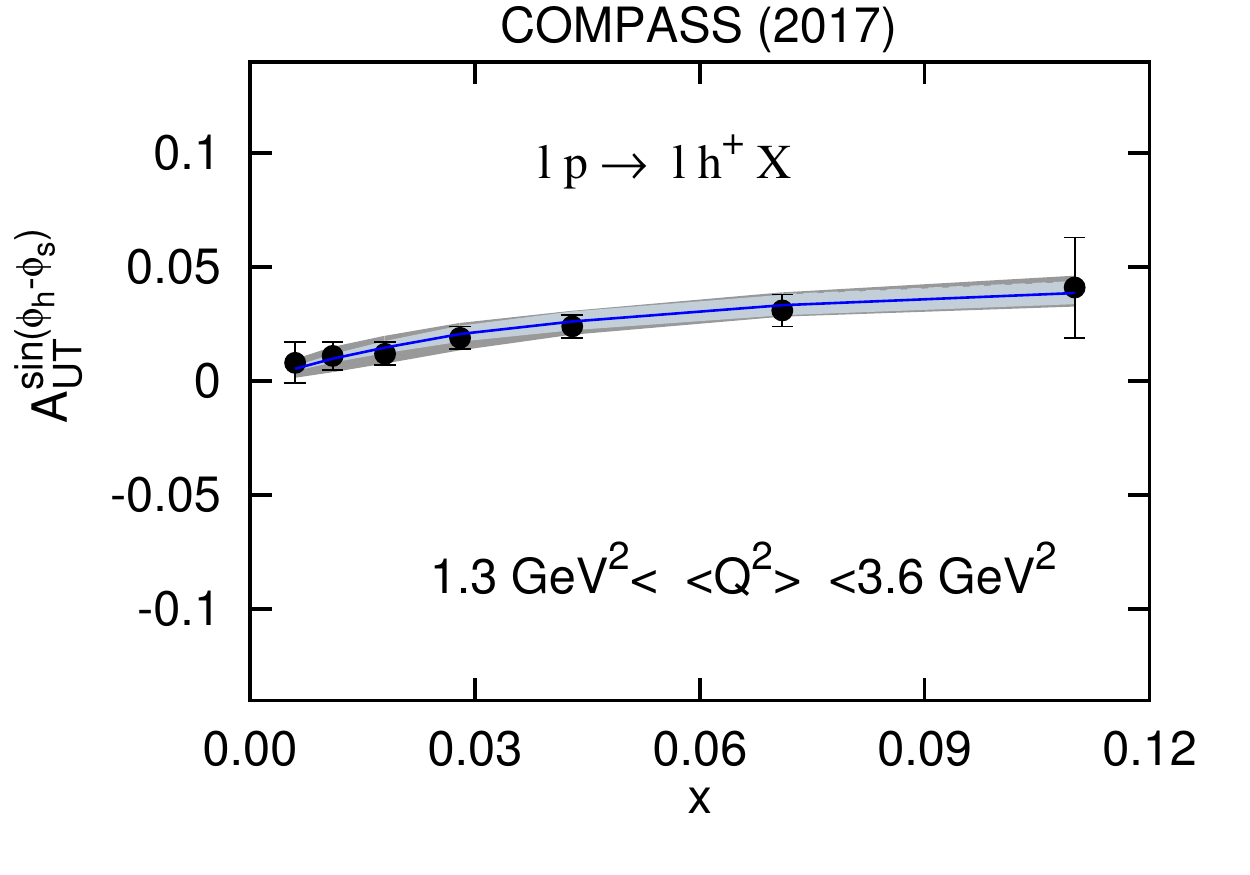}
\includegraphics[width=7.0cm]{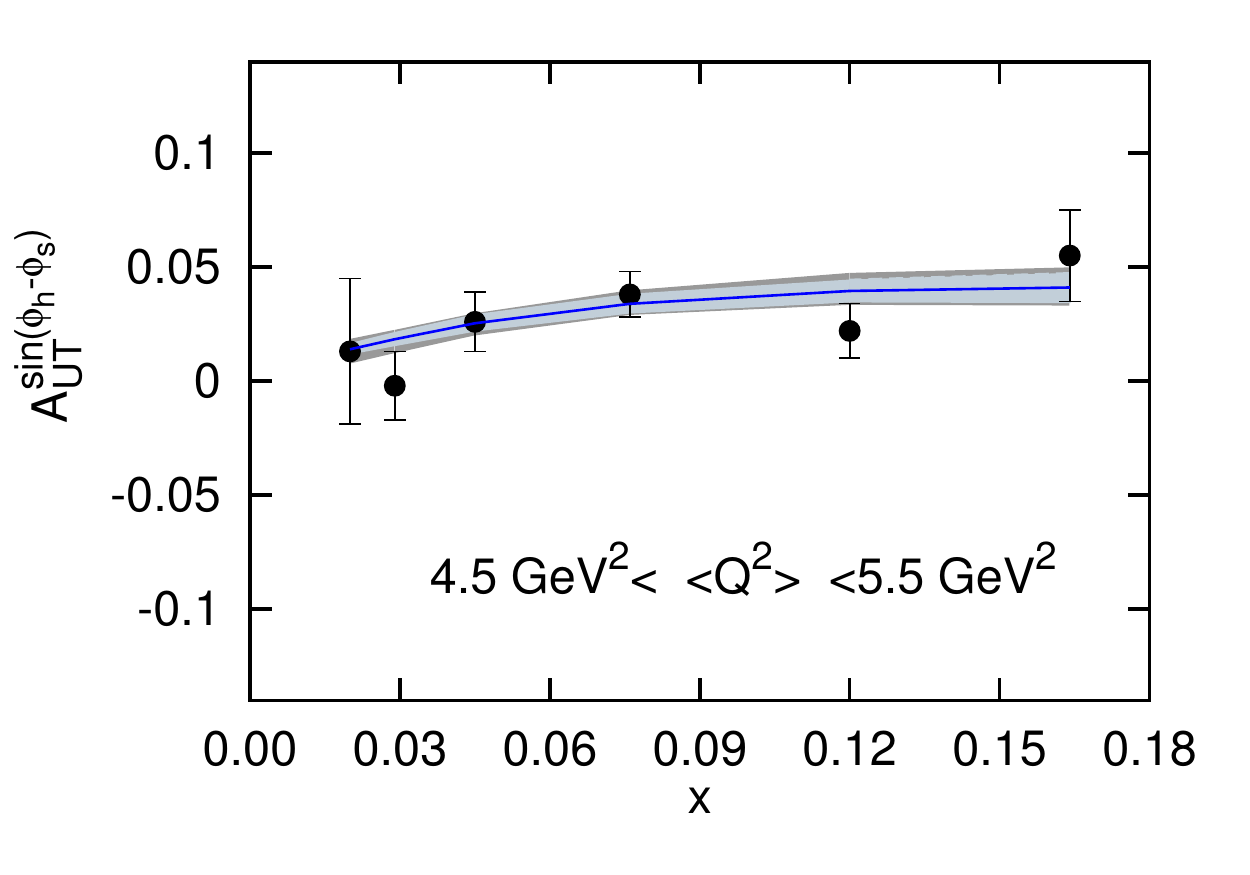}\\
\includegraphics[width=7.0cm]{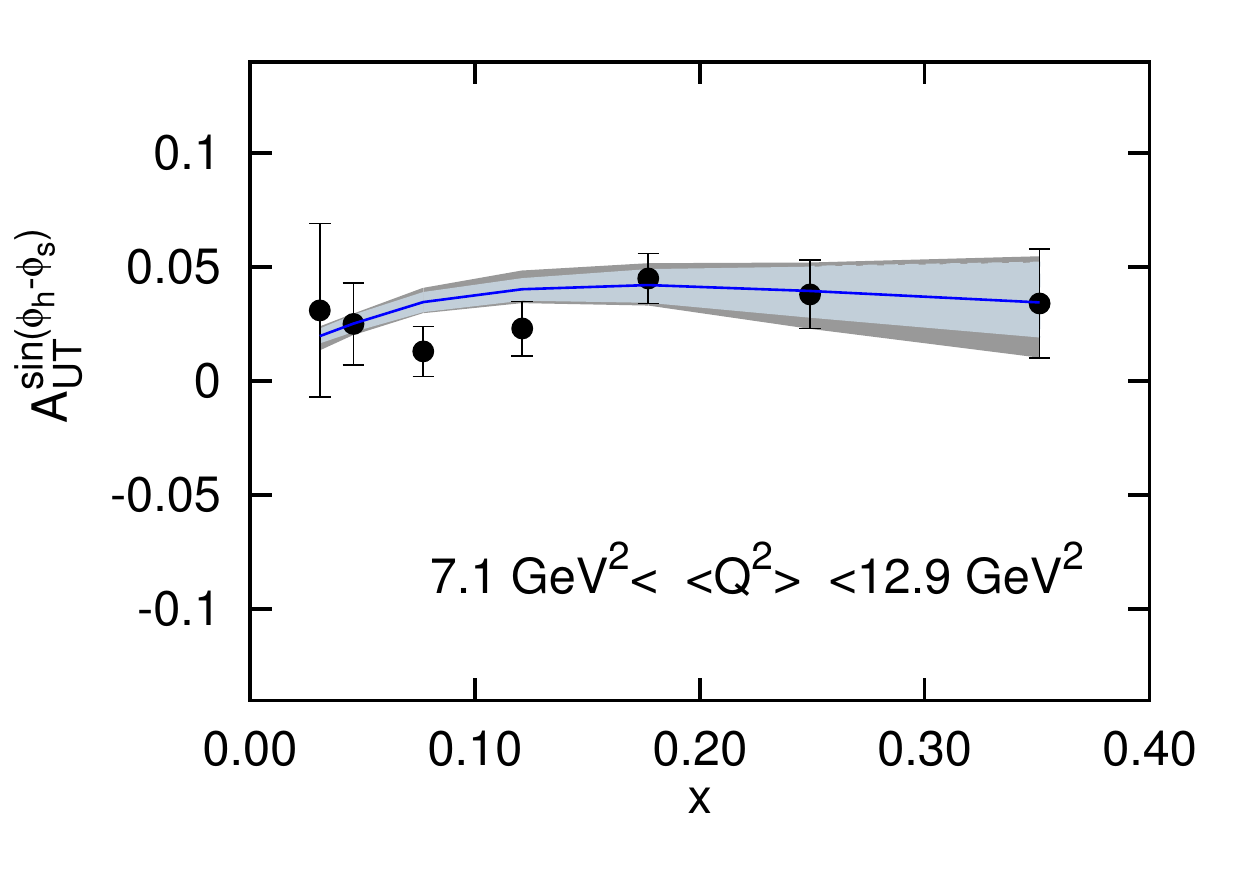}
\includegraphics[width=7.0cm]{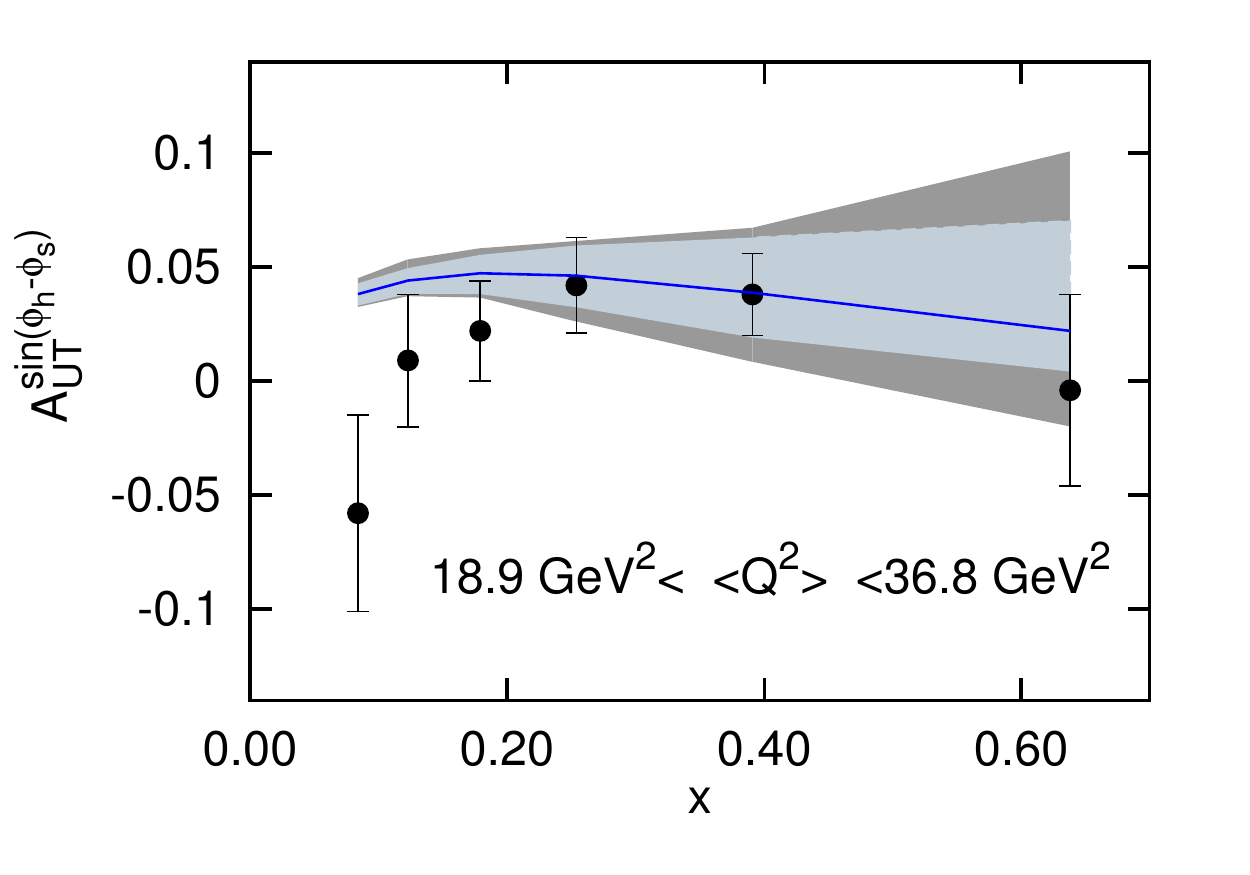}\\
\includegraphics[width=7.0cm]{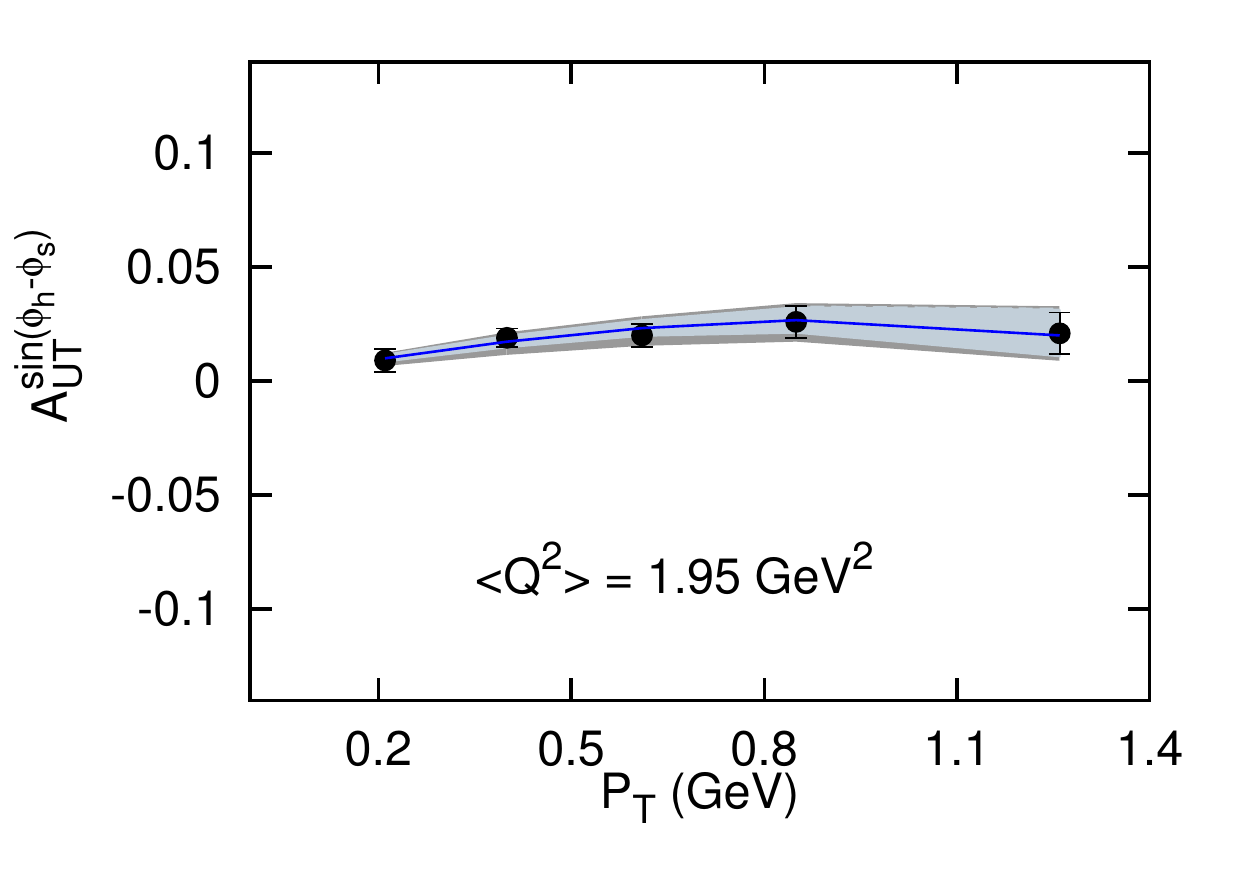}
\includegraphics[width=7.0cm]{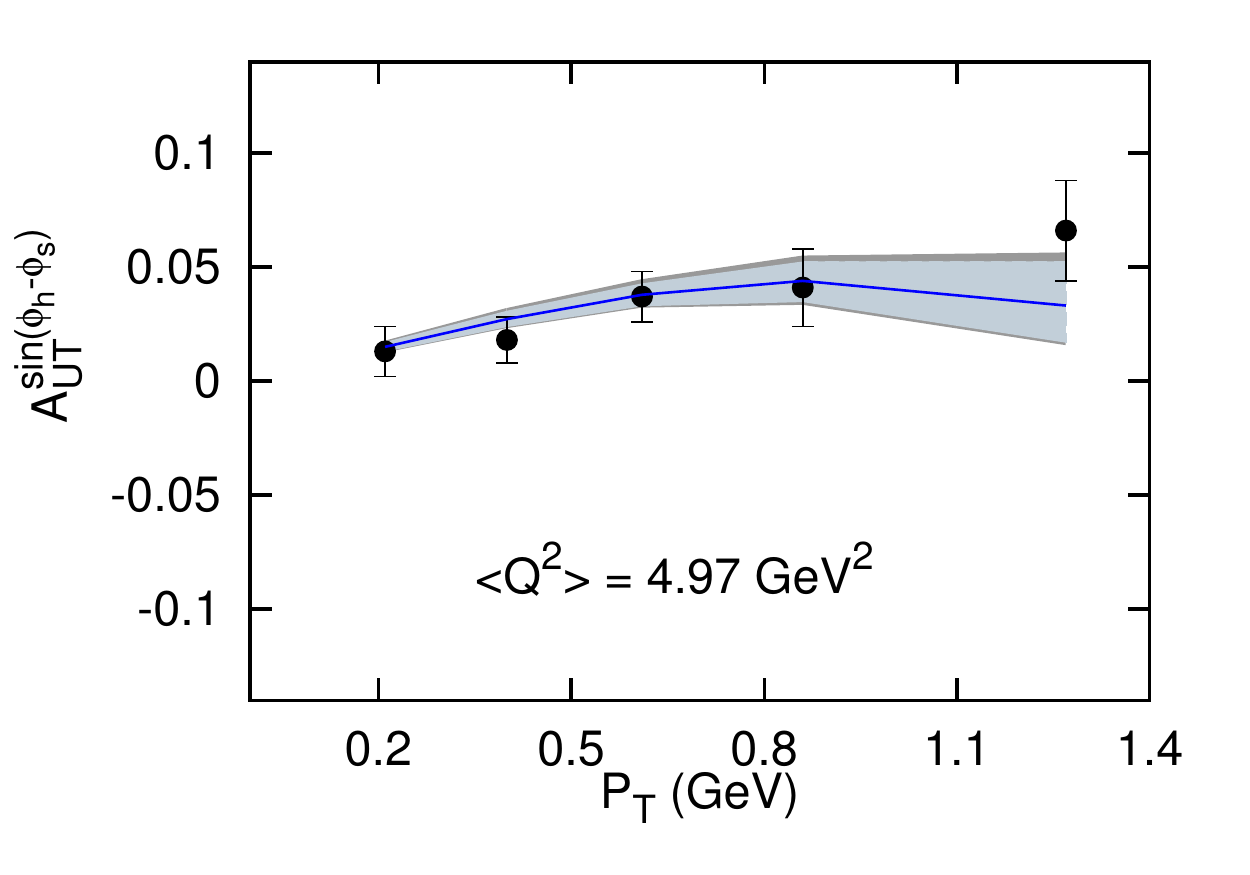}\\
\includegraphics[width=7.0cm]{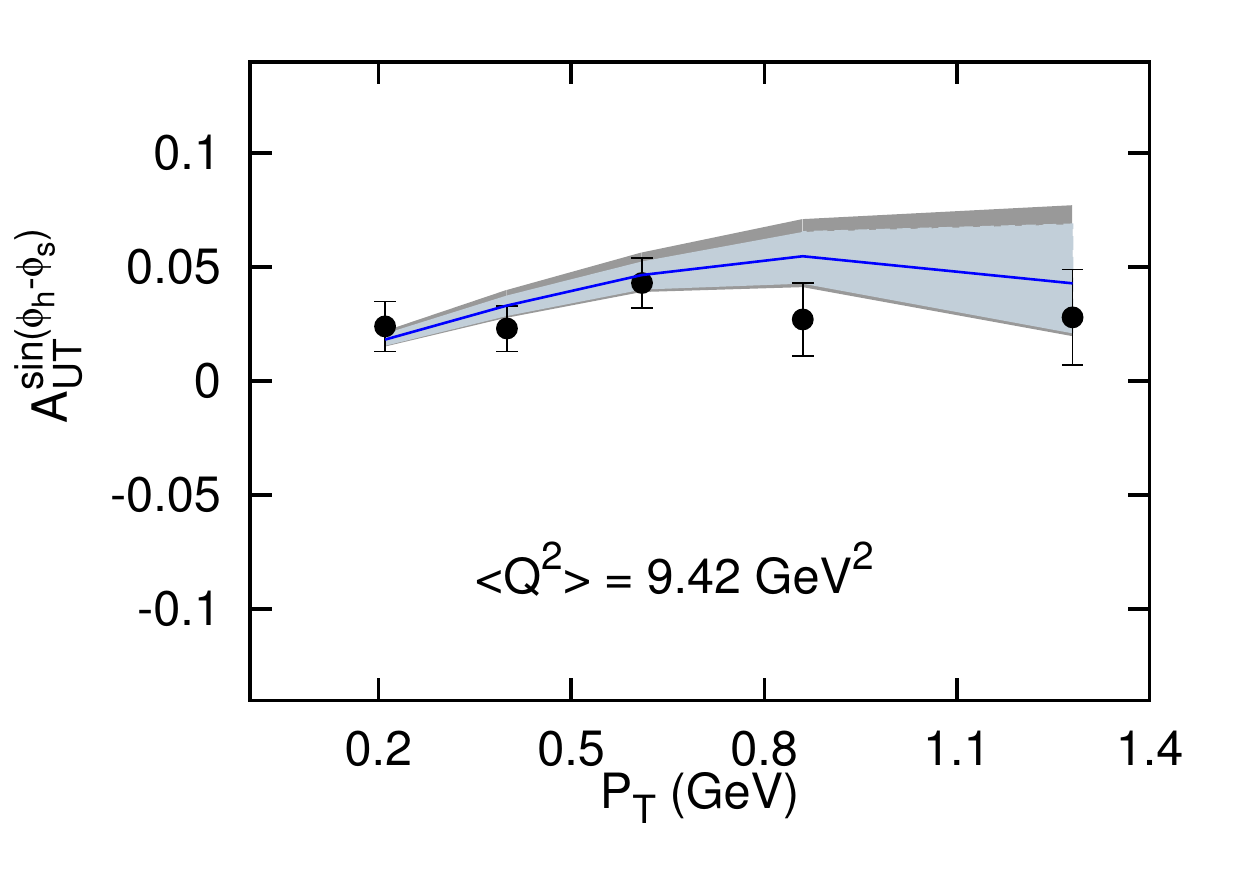}
\includegraphics[width=7.0cm]{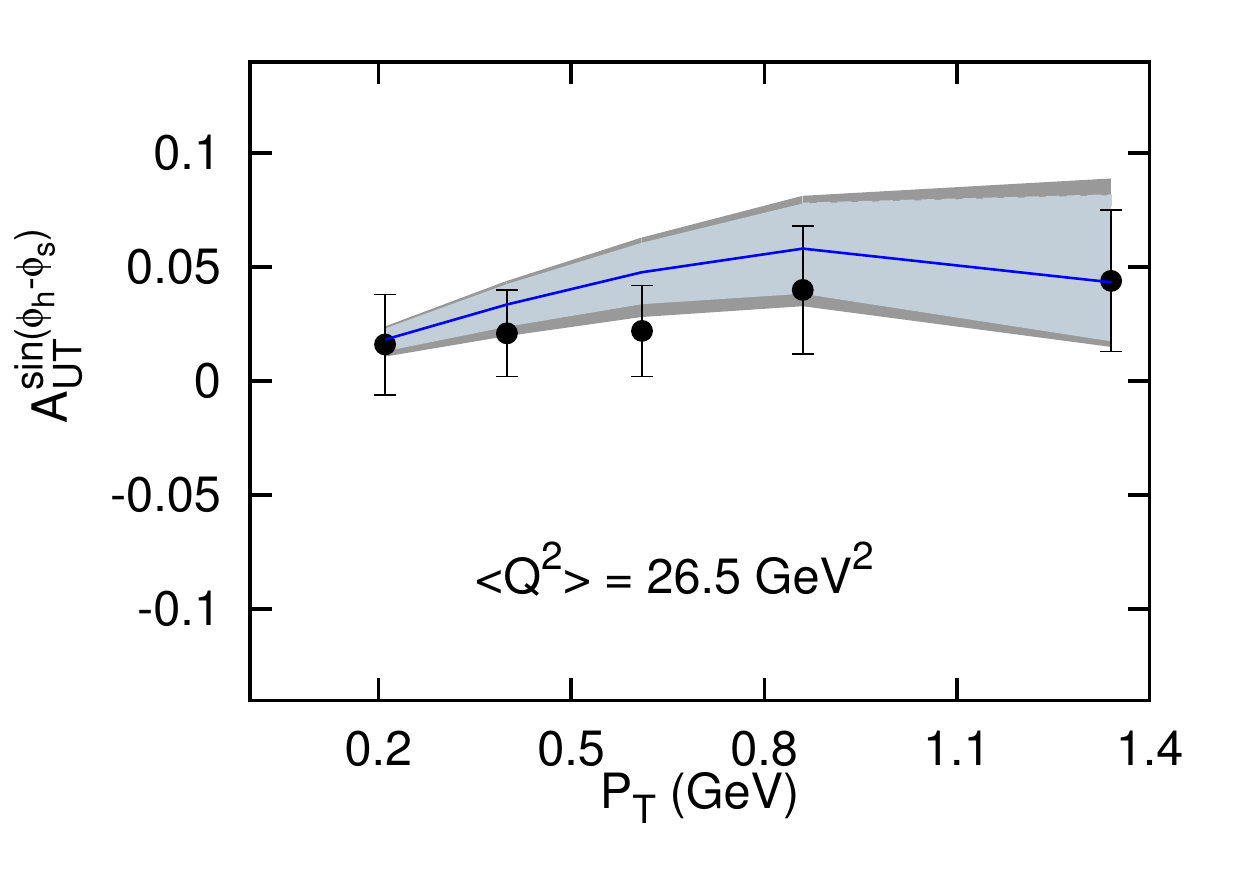}
\caption{The results obtained from the reference fit and the $\alpha$-fit are compared to the COMPASS Collaboration measurements of 
the SIDIS Sivers asymmetry on a NH$_3$ target~\cite{Adolph:2016dvl} for $h^+$ production. 
We show the $x$ and $P_T$ dependences, 
the $z$ depedences are not included in the fit. 
The shaded regions correspond to our estimate of $2\sigma$ C.L. error band.
The light-blue bands correspond to the uncertainties of the reference fit 
(only $N_{u(d)}$ and $\beta_{u(d)}$ free parameters), while the (larger) gray bands correspond to the 
uncertainties of the $\alpha$-fit which includes also the $\alpha_{u(d)}$ parameters.
}
\label{fig:compass17}
\end{figure}
%
\begin{figure}[htp]
\centering
\includegraphics[width=7.0cm]{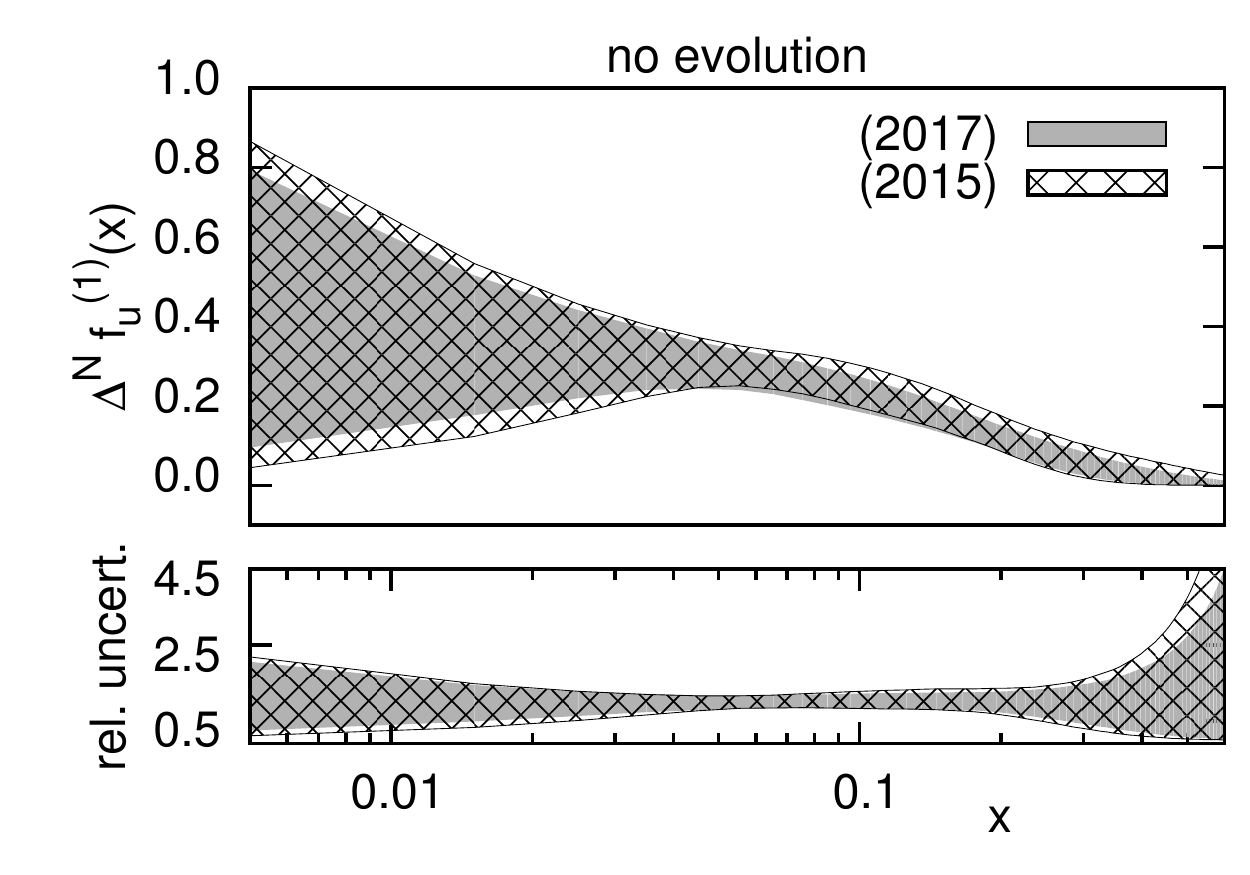}\hspace{0.5cm}
\includegraphics[width=7.0cm]{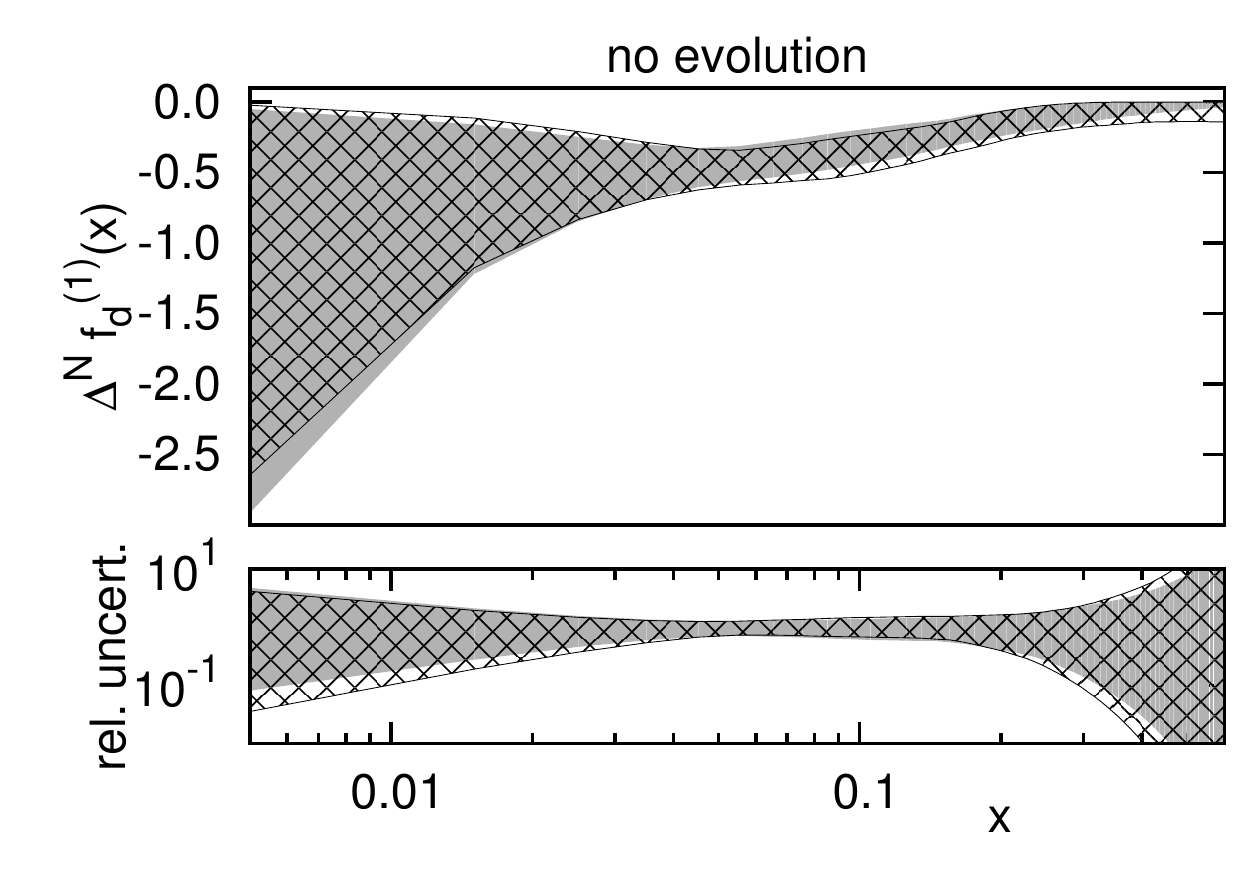}
\caption{Uncertainties on the first moments of the Sivers distribution for $u = u_v + \bar u$ (left panel) and  
$d = d_v + \bar d$ (right panel). The shaded 
bands correspond to our estimate of $2\sigma$ C.L. error for $\alpha$-fit case.
In both panels, the gray bands correspond to the fit which includes the 
new COMPASS-2017 data~\cite{Adolph:2016dvl} for $h^+$ and $h^-$ 
production off NH$_3$ target, while the meshed areas correspond to the uncertainties obtained 
when the COMPASS-2015 data from the 
older analysis~\cite{Adolph:2014zba} are included. The lower panels show the 
relative errors, given by the ratio between the upper/lower border of the uncertainty bands and the 
best-fit curve for the reference fit. }
\label{fig:first-mom-17-vs-15}
\end{figure}
%
One can notice, however, a sizable increase of the parameter errors. This 
has an effect on the uncertainty bands of the extracted Sivers function, as shown in 
Fig.~\ref{fig:ref-vs-alpha}, where the light-blue bands correspond to 
the reference fit, while the wider, gray bands refer to the ``$\alpha$-fit''. 

The modifications on the parameter space induced by adding the two $\alpha$ parameters is shown 
in detail in Fig.~\ref{fig:para-space-alpha}.
The top panels show the $\chi^2_{\rm tot}$ profiles as functions of the parameters $N_u$ (top-left) and 
$N_d$ (top-right). 
Contrary to the reference fit, these profiles are very far from resembling quadratic functions, and 
therefore the Hessian method adopted to evaluate the errors on the parameters cannot 
be trusted. The MINUIT errors reported in Table~\ref{tab:ref}, in fact, largely underestimate 
the uncertainties on the free parameter determination: by looking at the plots in Fig.~\ref{fig:para-space-alpha}, 
one can easily see that, to $2\sigma$ C.L., $N_u$ can go as low as $0.1$ and as 
large as $4.0$, over a very asymmetric range. 
Similarly for $N_d$, which can span over an even larger range, from $0$ to $-45$, 
on an extremely asymmetric range. A clear indication, however, is given on the sign: $N_u$ 
is positive and $N_d $ is negative, signaling a preference of the data for a positive $u$ 
and a negative $d$ Sivers function.

In the upper panels of Fig.~\ref{fig:para-space-alpha} the correlations ($N_u$,$\alpha_u$) and 
($N_d$, $\alpha_d$) are colour-coded: the very 
evident structure in bands of the same colour points to extremely strong correlations. 
This becomes even more explicit in the bottom panels of Fig.~\ref{fig:para-space-alpha}, 
where we show $N_u$ vs. $\alpha_u$ (bottom left) and $N_d$ vs. $\alpha_d$ (bottom right). 
In these scatter plots, the expected ellipsoidal shapes are replaced by 
very thin and stretched distributions, which indicate that an extremely large number of equally good 
fits can be obtained provided $N$, $\alpha$ (and $\beta$) are in the appropriate ratio among each other. 
In other words, even very large values of the $\alpha$ and $\beta$ parameters can result in an
acceptable $\chi^2$, provided the corresponding $N$ parameter is adequately large in size. 
Conversely, low values of $\alpha$ and $\beta$ are also equally appropriate if $N$ is small enough. 
The strong correlations introduced by $\alpha$, in fact, make it cumbersome to find a good fit by 
 a simple minimization procedure.
  
Nonetheless, the study of the parameter space including the $\alpha$ parameters allows for a more 
realistic estimate of the uncertainty bands in the small-$x$ region.
This is shown in Fig.~\ref{fig:ref-vs-alpha}, where the gray shaded areas 
represent the uncertainty bands corresponding to the $\alpha$-fit, while the light-blue bands represent 
the uncertainties corresponding to the reference best fit. Clearly, the two fits have very similar bands 
over the region $0.03<x<0.3$, while the $\alpha$-fit uncertainties 
grow larger outside this range, where experimental data are less dense. 
Notice that the Sivers width, $\avkS$, is not significantly affected by this strong broadening of the uncertainty 
bands: its central value remains the same (see Tables~\ref{tab:ref} and~\ref{tab:ref-alpha}), while 
error bands show no significant change, 
as it is clearly evident in the bottom panels of Fig.~\ref{fig:ref-vs-alpha}, where the $u$ and 
$d$ Sivers functions are plotted vs. $\kt$ at  
$x=0.1$. 

Fig.~\ref{fig:old-data} shows the results obtained from this reference fit compared to older data, 
from HERMES-proton (top panels) and COMPASS-deuterium (middle panels), 
which have historically been present in our previous fits, together with JLab-neutron measurements 
(bottom panels).


The bands corresponding to the reference best fit are shown in light-blue. 
The enlargement of the gray bands for the $\alpha$-fit provides a more sensible estimate of the uncertainties 
at low $x$. 
%
In fact, as seen in the central panels of Fig.~\ref{fig:old-data}, the agreement of the light blue bands 
with the deuteron data seem to deteriorate at small $x$, while the gray bands corresponding to the 
$\alpha$-fit improve the compatibility with these experimental measurements. 
Note that, since separating valence and sea contributions is not possible with the current data, the effect 
on the uncertainties introduced by allowing $\alpha\neq0$ also reflects our ignorance about the sea contributions. 

This supports the need to learn more about the Sivers function 
in the low-$x$ region and, in turn, about its sea contributions. 
In fact, this is one of the main tasks of the future 
Electron-Ion-Collider (EIC)~\cite{Accardi:2012qut}, 
which is planned to be built in the next few years in the USA.
Besides the clear benefits of an EIC to resolve the sea of the Sivers function, 
we stress the importance of the deuteron target measurements as those performed 
by COMPASS~\cite{Alekseev:2008aa}.
Recall that the $\alpha$-fit uncertainties only have a significant impact in the 
description of these data in the low-$x$ regime, where the reference fit delivered 
uncomfortably small uncertainties. 
Thus, an improvement on the statistics for these measurements may prove very 
useful in constraining the low-$x$ regime, 
rendering information about the sea distributions (see section~\ref{new-deuterium}).


In Fig.~\ref{fig:compass17} we compare the results obtained from the reference fit   
to the newest COMPASS data on the SIDIS Sivers asymmetry on a NH$_3$ target~\cite{Adolph:2016dvl},  
for $h^+$ 
production, binned in four values of $Q^2$ (the average value of $Q^2$ 
corresponding to the bin is indicated on each panel).
Only the $x$ and the $P_T$ dependences are shown, as the $z$ dependences are not included in the fit. 
However, we have checked that all $z$-distributions are always successfully reproduced.
The shaded regions correspond to our estimate of 95.4\% C.L. error band.

Finally, we compare the results obtained from our fit which uses 
the newly re-analyzed data by the COMPASS Collaboration~\cite{Adolph:2016dvl} to those obtained 
using the older set of COMPASS-proton data~\cite{Adolph:2014zba}. These newly released data 
sets belong to the same measurements (2010 run), 
but they differ in the way they are binned.
In fact, in their more recent analysis the data are binned in $x$ and $P_T$ as well as in four bins 
of $Q^2$, the same bins in which the Drell-Yan analysis is being carried out. 
As shown in Fig.~\ref{fig:first-mom-17-vs-15}, some reduction of the uncertainty bands is 
obtained when using the COMPASS-2017 data set, indicating that an increased degree of 
information is reached by applying the new binning. An important feature of the new binning 
is the separation of different ranges of $Q^2$. This, for the first time, 
allows to explore the possibility of scale dependence in the Sivers function. 
We will address this in Section~\ref{evo}.


\subsection{\label{large-x} Large-$x$ uncertainties}

The study of large-$x$ uncertainties is indeed very delicate. 
At present, as shown in this analysis, the Sivers function is largely unconstrained
in the range from $x \sim 0.3$ up to $x = 1.0$. 

In this region the Sivers function should approach zero with the only theoretical constraint 
given by the positivity bound, 
$|\Delta^N f_{q/p^\uparrow}| \le 2 f_{q/p}$, which should hold for any flavour $q$ 
and at every value of $x$ and $\kt$.
Notice, however, that also the integrated unpolarized PDFs are largely undetermined at 
large-$x$, undermining the significance of any phenomenological application of the 
positivity bound itself. 

To make the large-$x$ uncertainties more visible, in the middle panel of Fig.~\ref{fig:ref-vs-alpha},  
we show the bands corresponding to the relative uncertainties, i.e. the ratios 
between the upper/lower border of the uncertainty bands and the 
best-fit curve for the reference fit, at each value of $x$ and for any given flavour. 

\begin{figure}[tp]
\centering
\includegraphics[width=7.0cm]{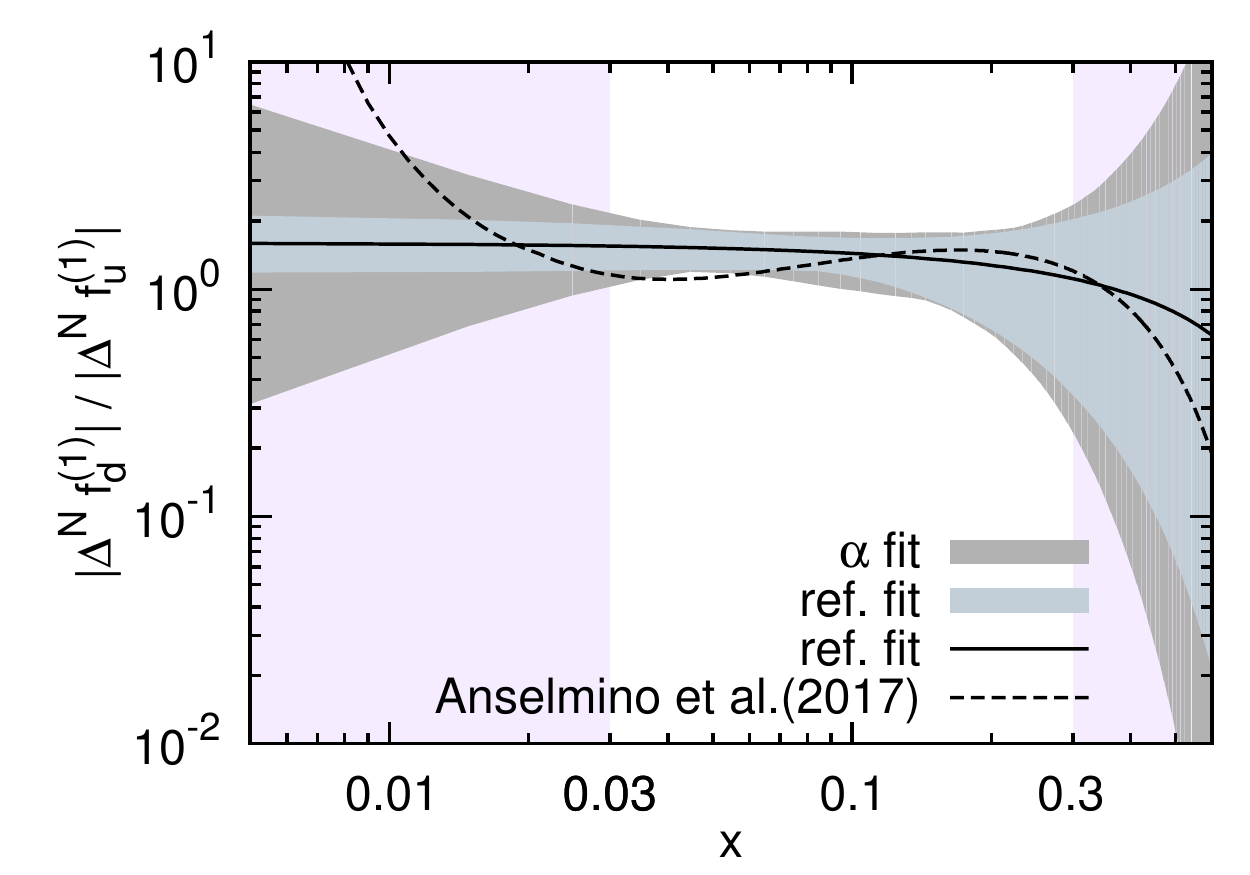}
\caption{Flavour ratio for the first moments of the Sivers TMD distributions, 
$|\Delta ^N f_{d/p^\uparrow}^{(1)}| / |\Delta ^N f_{u/p^\uparrow}^{(1)}|$, 
as extracted in the reference fit ($\alpha=0$) 
(light-blue bands) and in the $\alpha$-fit (gray bands).
The solid line shows the central line for the ratio, according to our reference fit. For comparison, we include the 
corresponding central line according to the extraction of Ref.~\cite{Anselmino:2016uie}. 
The low-$x$ and large-$x$ regions, where experimental information is scarse, are highlighted in purple.
}
\label{fig:ratio-moment-du}
\end{figure}

In a similar way, in Fig.~\ref{fig:ratio-moment-du}, we show the ratio 
$|\Delta ^N f_{d/p^\uparrow}^{(1)}| / |\Delta ^N f_{u/p^\uparrow}^{(1)}|$. For comparison, we also display the 
central line of the previous extraction of the Sivers functions from Ref.~\cite{Anselmino:2016uie}. 
As expected, in the range $0.03<x<0.3$, the agreement between the two extractions is acceptable. 
As one goes outside of this region, however, the two extractions exhibit 
more distinct behaviours. While this does not compromise compatibility at large $x$, 
as both central lines fall within the error bands, differences at low $x$ are more dramatic. 
This should serve as a warning that model dependence has an important effect outside the bounds 
of experimental information.

Future measurements at JLab12~\cite{Dudek:2012vr} will be able to shed 
some light on the large-$x$ kinematic region.
These measurements will give a crucial contribution in the extraction of the 
unpolarized TMD parton distribution functions as well as the Sivers functions  
in the large-$x$ range, and will give us a much clearer signature for the 
flavour separation of the valence contributions.

\newpage

\section{\label{new-deuterium} Comments on the precision of deuteron target data}

As mentioned above, the $d$ Sivers function is poorly determined by 
the existing Sivers data, in spite of the fact that it should be constrained by the
identification of the final state hadrons.
Possibly this can be traced back to the large $u$-dominance of SIDIS on proton targets. 

The COMPASS collaboration has recently proposed a new run of their SIDIS measurements on a 
deuteron target, with increased statistics and precision~\cite{COMPASSII:2021}. 
It is therefore interesting to evaluate the impact of such a measurement on the
extraction of the first moment of the $u$ and $d$ Sivers function, using the projected errors for 
the proposed 2021 deuteron run, as reported in Ref.~\cite{COMPASSII:2021}.

\begin{figure}[tbp]
\centering
\includegraphics[width=7.0cm]{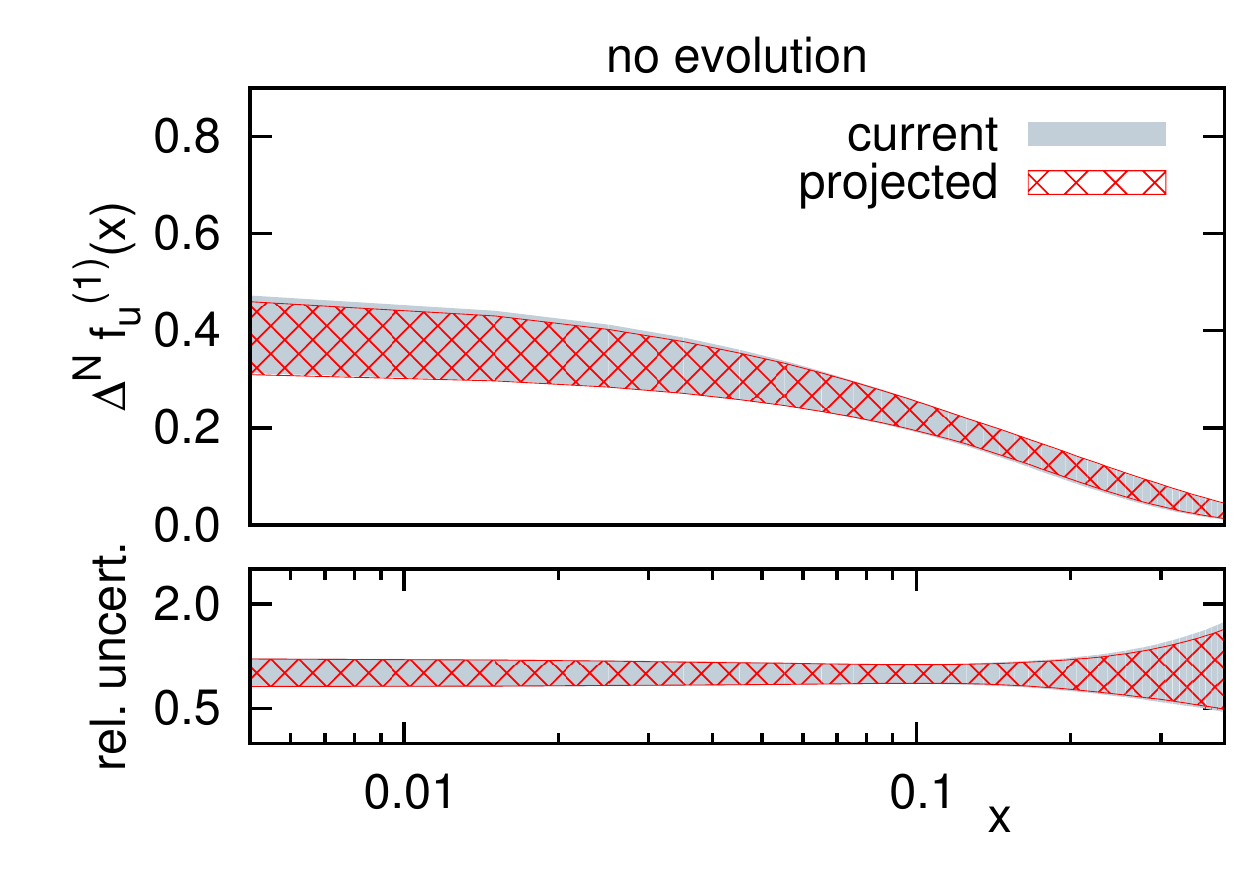}\hspace{0.5cm}
\includegraphics[width=7.0cm]{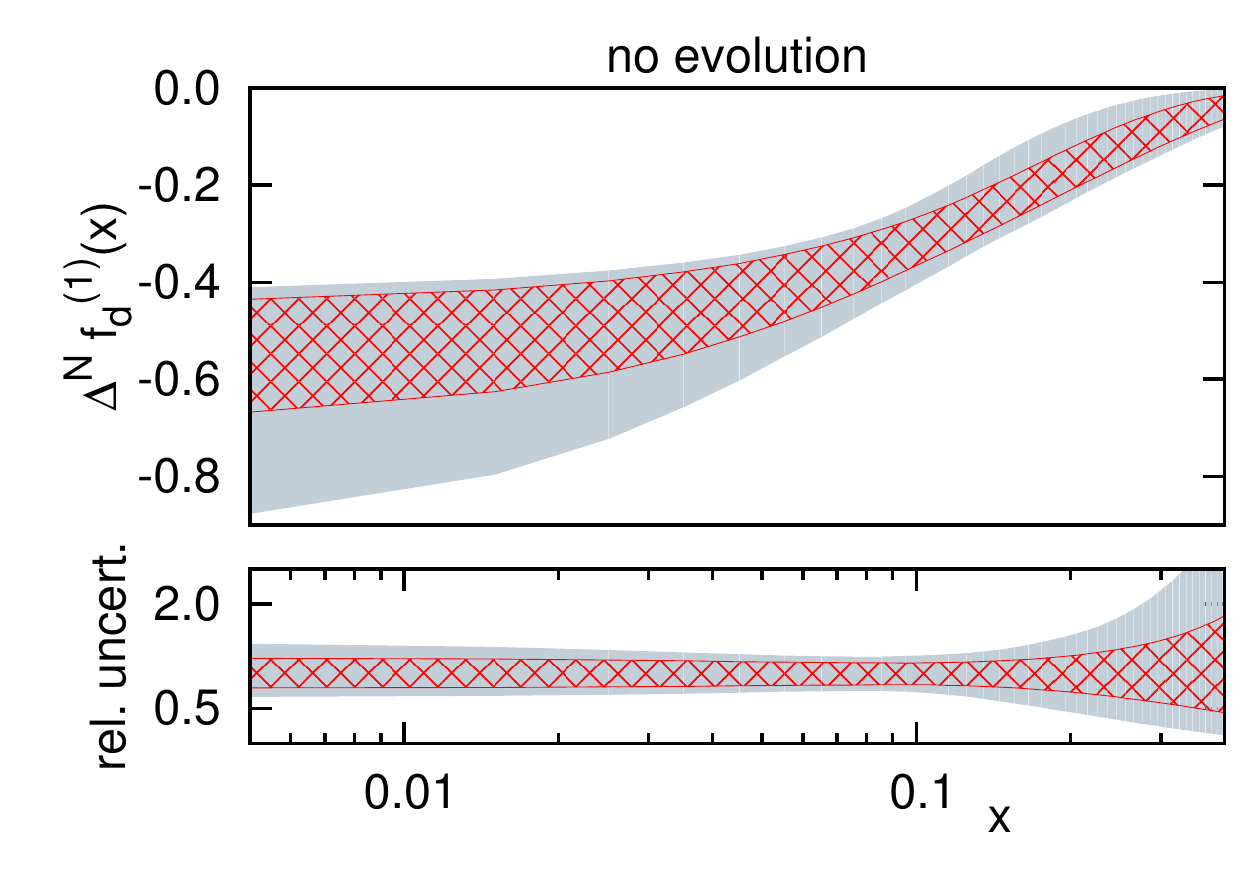}
\caption{
First moment of the extracted Sivers distributions for $u = u_v + \bar u$ (left panel) and  
$d = d_v + \bar d$ (right panel). The shaded 
bands correspond to our estimate of $2\sigma$ C.L. error. 
The light-blue bands show the uncertainties corresponding to our reference fit (see Table~\ref{tab:ref}). 
The red (meshed) bands correspond to the uncertainties estimated by using the same model, with 
the projected experimetal errors of the future COMPASS run on deuteron target~\cite{COMPASSII:2021}.
}
\label{fig:first-mom-deuterium}
\end{figure}

Our results are shown in Fig.~\ref{fig:first-mom-deuterium}. 
The $2\sigma$ error bands marking the 95.45\% C.L. for the first moments of the 
$u$ and $d$ Sivers functions, obtained with the reference fit (here labeled by ``current") 
are shown in light-blue. 
The bands obtained when adding to the data set the projected errors on the 
asymmetries of the new deuteron run are shown in red, and labeled ``projected''.
The plots for the first moments (bottom panels) show the relative uncertainty,
i.e. the ratio between the upper/lower border of the uncertainty bands and the 
best-fit curve for the reference fit.
As expected, the new deuteron run will have a small impact on the $u$-quark first
moment of the Sivers function. On the contrary, the reduction in the error band for the
first moment of the Sivers function for the $d$-quark is considerable, and is about
a factor $2$ for $x < 0.1$. 

This new COMPASS run will therefore lead to a remarkable improvement of our knowledge 
on the other flavour contributions of the Sivers function,
besides the already well constrained $u$.

\section{\label{evo} Signals of Scale dependence}

In all the results presented above, no $Q^2$ dependence of the Sivers function was considered.
An important aspect of the new COMPASS binning is that it separates different ranges of $Q^2$,
which poses the question of whether one can distinguish different assumptions about scale dependence.
To test how this scale dependence can affect our analysis, we will consider two different 
approaches which will serve us as comparisons: on one side, we will adopt a collinear, 
twist-3 evolution scheme based on Refs.~\cite{Kang:2008ey,Vogelsang:2009pj,Braun:2009mi,Kang:2012em}; 
on the other side, we will apply to the Sivers function a TMD-like $Q^2$ evolution similar 
to that described in Ref.~\cite{Anselmino:2012aa}.

In the collinear higher-twist evolution framework, the correlation between spin and transverse momentum 
is included into the higher-twist collinear parton distributions or fragmentation functions. 
These functions have no probabilistic interpretation:
they are generated as quantum interferences between a collinear active quark state in the 
scattering amplitude and a collinear quark-gluon composite state in its complex conjugate amplitude.
There are no intrinsic $\kt$ in this case, which are integrated over, and the evolution 
in $Q^2$ occurs only through $x$. In other words, twist-3 PDFs and FFs evolve in $Q^2$ by changing 
shape in $x$. 

In the TMD factorization approach, spin asymmetries are generated by spin and transverse momentum 
correlations between the identified hadron and the active parton.
This correlations are embedded in the TMD parton distribution or fragmentation functions, 
which can be interpreted as probability densities. Here the $Q^2$ evolution affects the $x$ 
dependence as well as the shape in $\kt$.

%
%
\begin{table}[t]
\begin{tabular}{l l r}
\hline
\hline
\multicolumn{3}{c}{{\bf Collinear twist-3 evolution }}\\
\hline
$\chi^2_{\rm tot}$ = 201.5      & ~ n. of points = 220  & ~\\
$\chi^2_{\rm dof}$ = 0.94       & ~ n. of free parameters = 5 & ~\\
$\Delta \chi^2$ = 11.3           & ~ & ~\\
\hline
HERMES & ~~$\langle k_\perp^2 \rangle = 0.57$ GeV$^2$ & ~~~~$\langle p_\perp^2 \rangle = 0.12$ GeV$^2$  \\
COMPASS & ~~$\langle k_\perp^2 \rangle = 0.60$ GeV$^2$ & ~~~~$\langle p_\perp^2 \rangle = 0.20$ GeV$^2$ \\
\hline
$N_u= 0.39 \pm 0.08$ & ~~$\beta_u=3.55 \pm 1.26$ & ~\\
$N_d=-0.65 \pm 0.27$ & ~~$\beta_d=4.77 \pm 3.41$ & ~\\
\multicolumn{3}{l}{$\langle k_\perp^2 \rangle _S = 0.33 \pm 0.14$ GeV$^2$}\\
\hline\hline
\end{tabular}
\caption{
Best fit parameters and $\chi^2$ values for the collinear twist-3 evolution case. 
The parameter errors correspond to $2\sigma$ C.L. 
Notice the reduced value of $\chi^2_{tot}$ w.r.t. that of the reference fit in Table~\ref{tab:ref}.
}
\label{tab:twist3}
\end{table}
%
\begin{figure}[b]
\centering
\includegraphics[width=7.0cm]{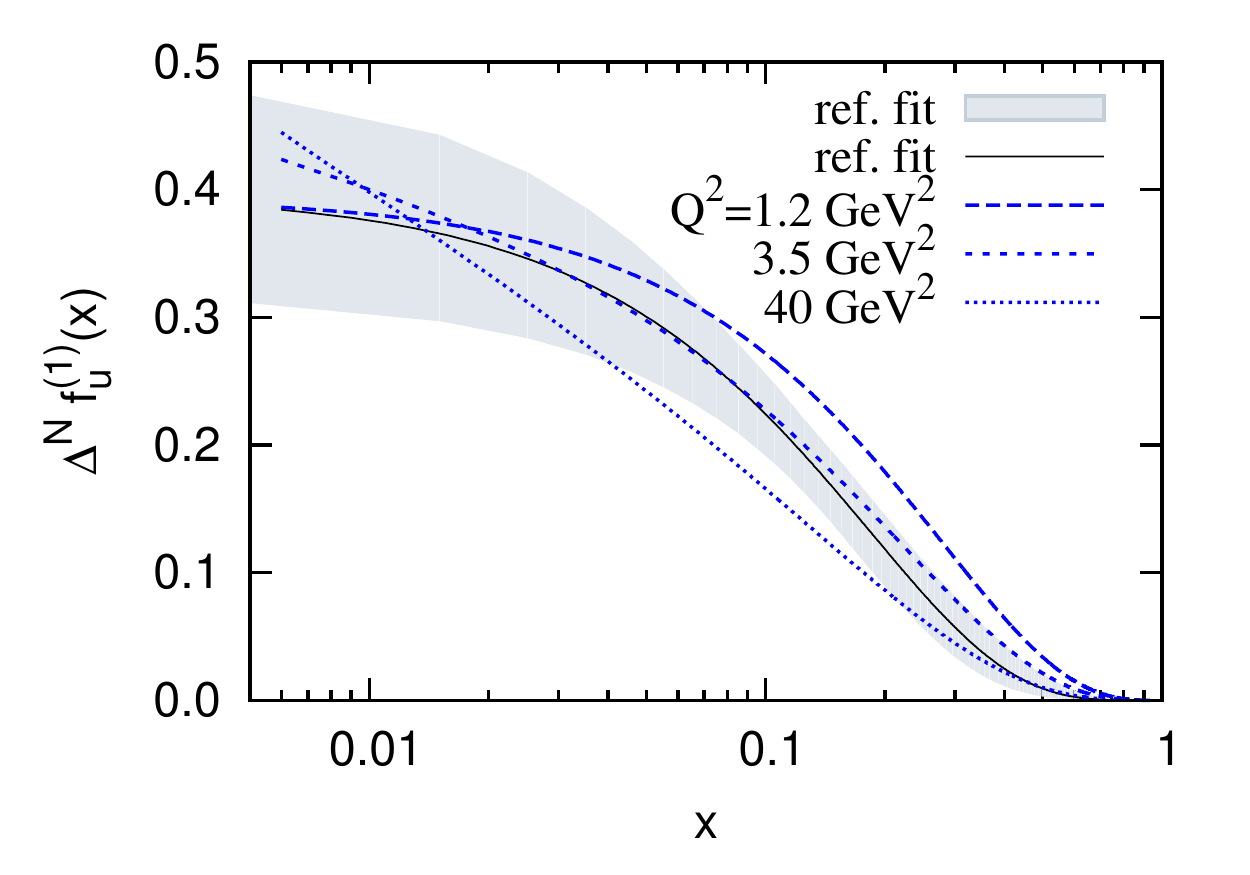}
\includegraphics[width=7.0cm]{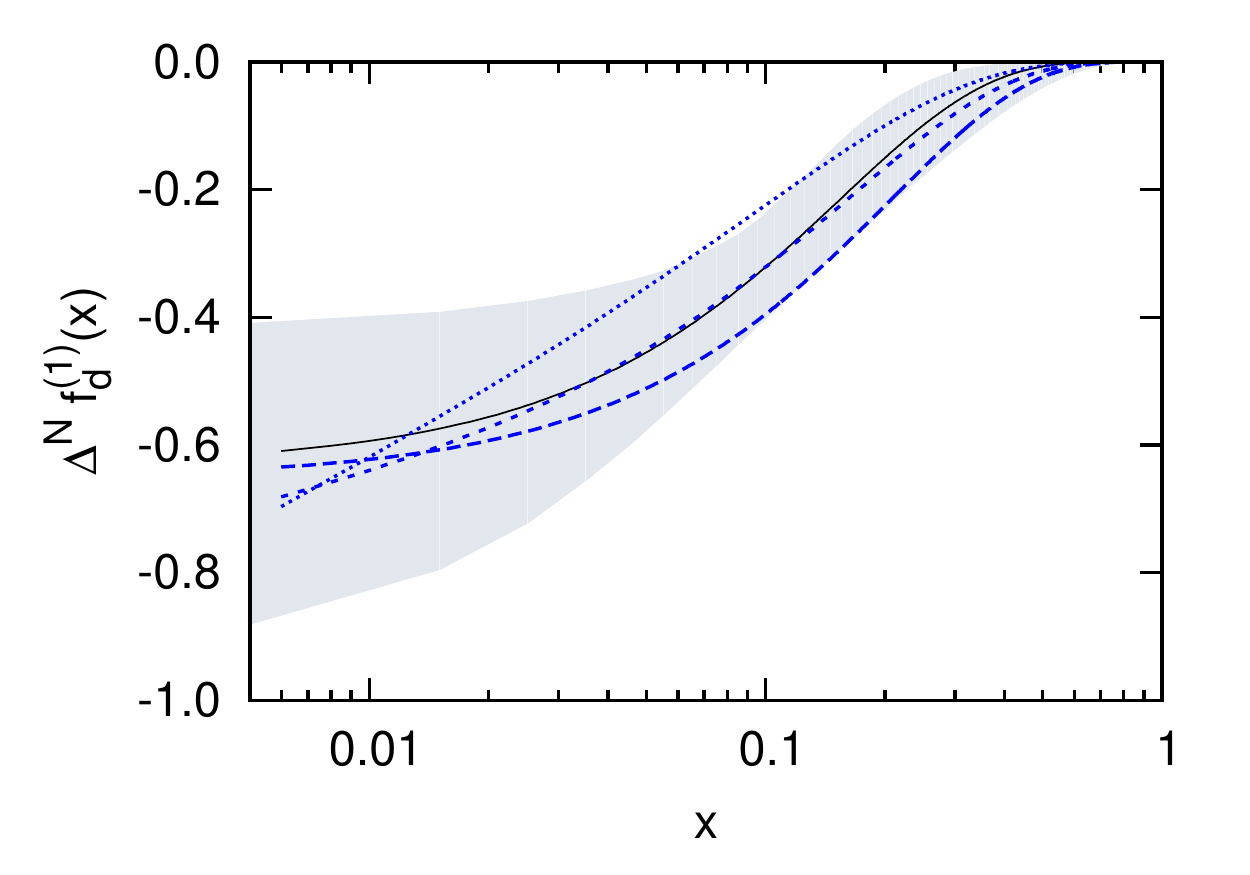}
\caption{Extracted first moments of the Sivers function for $u = u_v + \bar u$ (left panel) and  
$d = d_v + \bar d$ (right panel). 
The results corresponding to the reference fit with no $Q^2$ evolution (solid, black line) are compared 
to those obtained by applying a collinear twist-3 evolution (blue lines), as described in the text, for three values of $Q^2$: 
$1.2\,{\rm GeV}^2$ (long-dashed), $3.5\,{\rm GeV}^2$ (short-dashed) and $40\,{\rm GeV}^2$ (dotted). 
The bands correspond to the reference fit with no $Q^2$ evolution.
}
\label{fig:first-mom-twist-3}
\end{figure}
%
%
Although they are defined in different contexts, TMD and collinear quark-gluon correlation functions 
are closely related to each other. 
In particular, the first $\kt$-moment of the Sivers function is related to the collinear, 
twist-3 quark-gluon correlation function $T_{q,F}(x, x)$~\cite{Kang:2012em}.
As the evolution equations for $T_{q,F}(x, x)$ are known, we can adopt them in our study to render 
the $Q^2$ dependence of the Sivers function, from the initial scale $Q_0^2 = 1.2$ GeV$^2$ 
(which coincides with the lowest $Q^2$ of the experimental data included 
in our best fit) to the $Q^2$ corresponding to each specific data point at which the asymmetry 
is evaluated. 
To implement the collinear twist-3 evolution we use the HOPPET code~\cite{Salam:2008qg}, 
appropriately modified to include the kernels corresponding to the Sivers 
function~\cite{Prokudin:hoppet}.
Notice that while we do not include off diagonal terms in the twist-3 evolution case,
this approximation is enough for our purposes: we will test whether 
the existing data can distinguish between an approach with no evolution (reference fit) and
another where some scale dependence appears in the first moment of the Sivers function.

In Fig.~\ref{fig:first-mom-twist-3} the $u$ and $d$ Sivers first moments extracted in the 
reference best fit with no $Q^2$ evolution (solid, black lines) are compared to those obtained by applying 
the collinear, twist-3 evolution described above (blue lines). Three values of $Q^2$ are shown: 
$Q^2=1.2$ GeV$^2$ (long-dashed) and $Q^2=40$ GeV$^2$ (dotted), the lowest and largest $Q^2$ values of the COMPASS 
measurements and $Q^2=3.5$ GeV$^2$ (short-dashed), which is approximately the mean value 
of the full data sample.
The corresponding values of the $\chi^2$ and best fit parameters are presented in Table~\ref{tab:twist3}.

The Sivers functions extracted in the reference fit are very similar to those obtained using 
the twist-3 evolution scheme at the experiment average value, $Q^2=3.5$  GeV$^2$; 
in fact, they are very similar in the region 
where data constraints are stronger, $0.03<x<0.3$. Instead, they grow progressively 
apart when $Q^2$ is varied 
to reach its lowest and largest limits. 
For the $u$ flavour, the $Q^2$ variation of the first moment due to the collinear twist-3 evolution is actually 
larger than the uncertainty band corresponding to the no-evolution case. 
This gives a positive message 
about the precision of the data, in particular about the new binning of COMPASS asymmetries:
signals of collinear evolution could possibly be observed in the experimental data. 
This does not happen for the $d$ flavour which, as we have already pointed out, 
is affected by a larger uncertainty. The right panel of Fig.~\ref{fig:first-mom-twist-3} clearly shows that the 
error band corresponding to $\Delta ^N f ^{(1)} _{d/p^\uparrow}$ is of the same size (or even larger at small $x$) 
of the variation induced by  $Q^2$ evolution.
Notice that in this case the whole $Q^2$ scaling occurs only through $x$, leaving the $\kt$ part 
of the Sivers function unchanged.   

Finally we turn to the discussion on TMD evolution effects. To extract the Sivers function 
within a full TMD-scheme, one needs to
exploit an ``input function'', i.e. the value of the Sivers function at the initial $Q^2$ scale.
Then, a TMD factorization scheme as that discussed in Ref.~\cite{Aybat:2011ge}, and successively 
implemented in Refs.~\cite{Aybat:2011ta,Anselmino:2012aa}, can be applied to compute 
the Sivers function at any larger value of $Q^2$. 
%
%
\begin{table}[b]
\begin{tabular}{l l r}
\hline
\hline
\multicolumn{3}{c}{{\bf $\bm{Q^2}$-dependent $\bm{\langle k_\perp^2\rangle_S}$ fit}}\\
\hline
$\chi^2_{\rm tot}$ = 212.8         & ~n. of points = 220 & ~\\
$\chi^2_{\rm dof}$ = 0.99          & ~n. of free parameters = 6  & ~\\
$\Delta \chi^2$ = 12.9      & ~ & ~\\
\hline
%
%
%
HERMES~~~~~~ $\langle k_\perp^2 \rangle = 0.57$ GeV$^2$ & ~~$\langle p_\perp^2 \rangle = 0.12$ GeV$^2$ & ~ \\
COMPASS~~~~ $\langle k_\perp^2 \rangle = 0.60$ GeV$^2$ & ~~$\langle p_\perp^2 \rangle = 0.20$ GeV$^2$ & ~\\
\hline
$N_u= 0.40 \pm 0.09$ & ~~ $\beta_u=5.42 \pm 1.70$ & ~ \\
$N_d=-0.63 \pm 0.26$ & ~~ $\beta_d=6.45 \pm 3.89$ & ~ \\
\hline
$\langle k_\perp^2 \rangle _S= g_1 + g_2 \log\left(Q^2/Q^2_0\right)$&&\\
$g_1 = 0.28 \pm 0.29$ GeV$^2$ & $g_2 = 0.01 \pm 0.20$ GeV$^2$ & ~\\
\hline\hline
\end{tabular}
\caption{
Best fit parameters and $\chi^2$ values for the $Q^2$-dependent $\langle k_\perp^2 \rangle_S$ case, in which our reference model 
is modified according to Eq.~\eqref{eq:Q-width}. 
The parameter errors correspond to $2\sigma$ C.L. 
Notice that there is \emph{no} reduction in the value of $\chi^2_{tot}$ w.r.t. that of the reference fit in Table~\ref{tab:ref}.
}
\label{tab:tmdevo}
\end{table}
%
%
One should keep in mind that, within this approach, TMD parton densities change their shape 
in $\kt$ as $Q^2$ varies: 
in particular, their $\kt$-distributions broaden and dilute as $Q^2$ increases. 
While the TMDs themselves (and their first moments) experience variations in their 
$x$-distributions too as $Q^2$ increases, in the azimuthal asymmetries these effects 
are expected to roughly cancel in the ratio.
One complication of this type of analysis is the limited knowledge of the unpolarized functions 
at the kinematics of the available Sivers asymmetries. In fact, recent studies  have
suggested that the errors of factorization, at these kinematics, may not be under 
control~\cite{Boglione:2014oea,Boglione:2016bph,Collins:2016hqq,Moffat:2017sha}. This 
may affect all of the measurements in SIDIS. 

\begin{figure}[t]
\centering
\includegraphics[width=7.0cm]{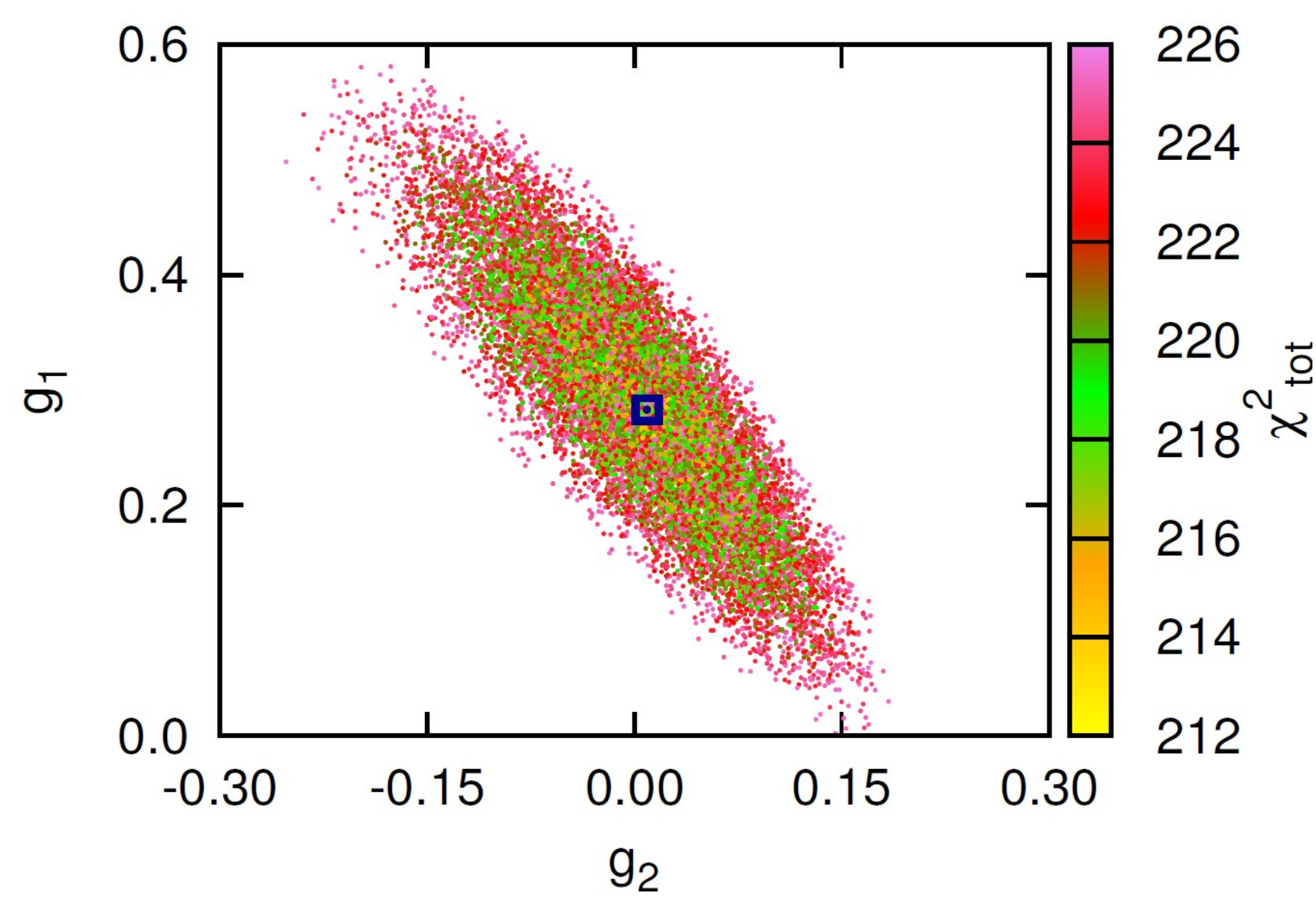}
\caption{Scatter plots showing the correlation between the $g_1$ and $g_2$ free parameters of Eq.~\eqref{eq:Q-width}. 
The  $\chi ^2_{tot}$ corresponding to this fit is colour-coded: yellow corresponds to its lowest values while red and purple to the 
highest accepted values.
}
\label{fig:para-space-g2}
\end{figure}
%
\begin{figure}[b]
\centering
\includegraphics[width=7.0cm]{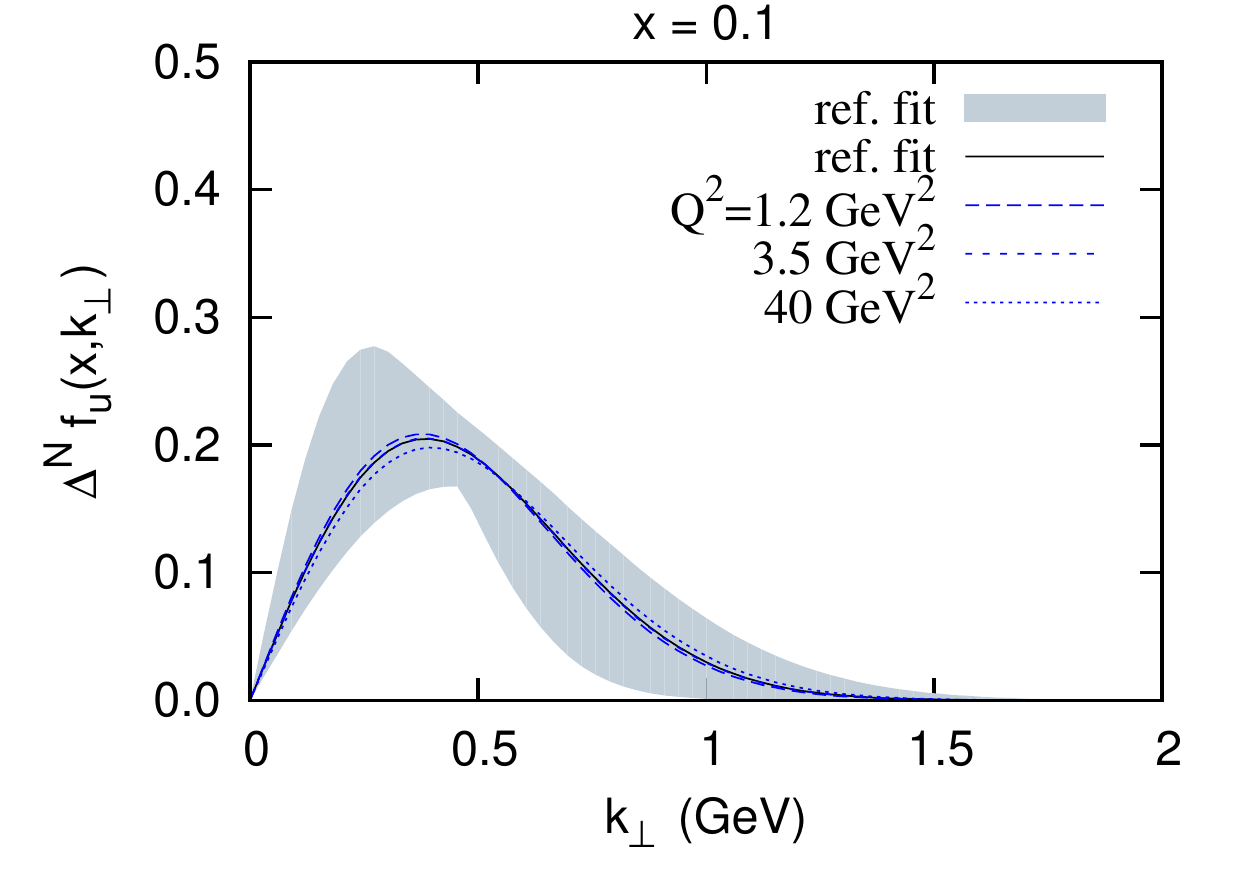}
\includegraphics[width=7.0cm]{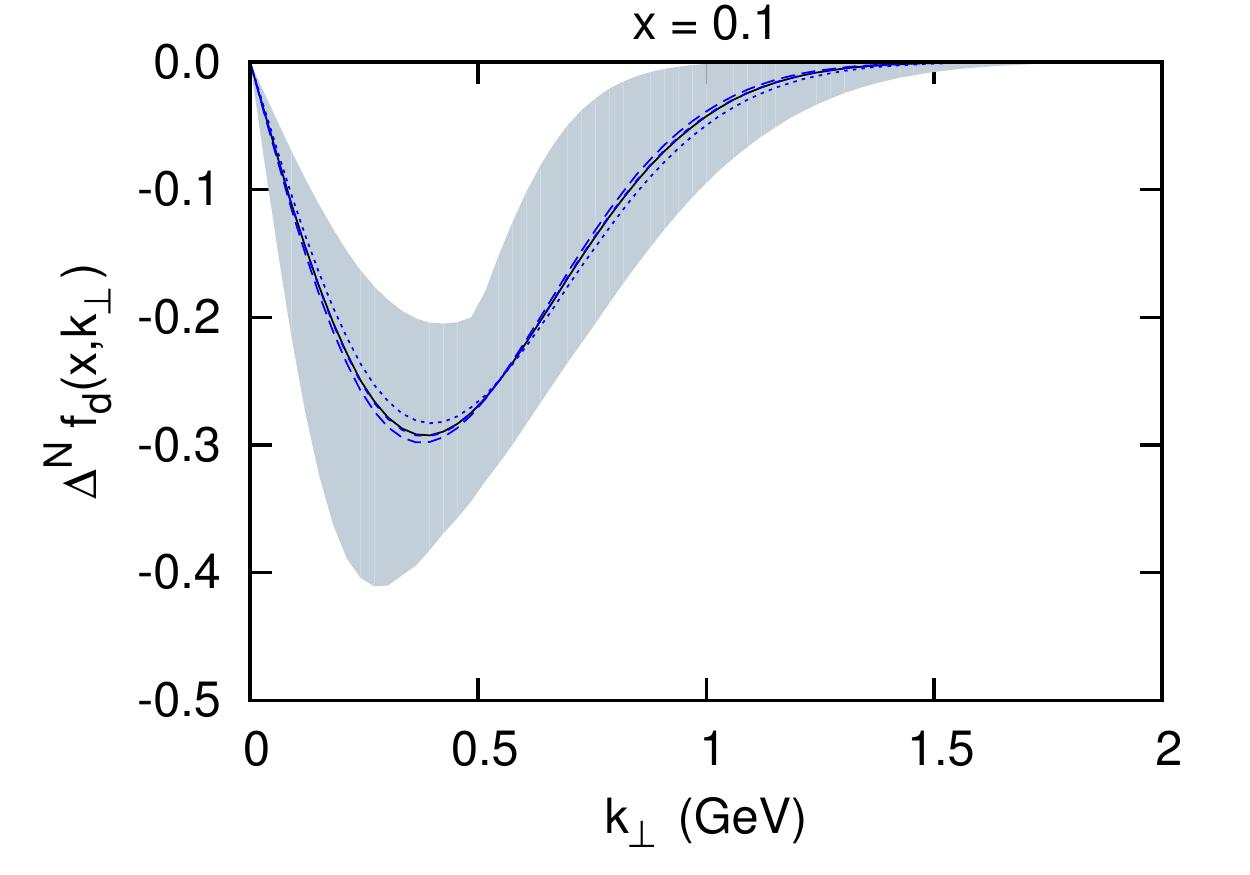}
\caption{
Extracted Sivers distributions for $u = u_v + \bar u$ (left panel) and  
$d = d_v + \bar d$ (right panel). The central line of our reference fit (solid black), is compared
the Sivers
function, extracted with the modified model according to Eq.~\eqref{eq:Q-width}, 
at a fixed value of $x=0.1$ (blue lines), 
for three values of $Q^2$: $1.2\,{\rm GeV}^2$(long-dashed), $3.5\,{\rm GeV}^2$(short-dashed) and $40\,{\rm GeV}^2$(dotted).
For comparison, we show  the error bands corresponding to the reference fit with no $Q^2$ evolution. 
}
\label{fig:log-dep-width}
\end{figure}

While waiting for further studies to clarify this situation, 
one may ask questions regarding the data and assess to what extent
TMD-evolution effects are visible.
To address this, we consider a modification of our model, where we allow for the width of the 
Sivers function , $\langle \kt ^2 \rangle_S$, 
to become a function of $Q^2$, according to
\be
\langle \kt ^2 \rangle_S = g_1 + g_2 \ln \frac{Q^2}{Q_0^2}\,,
\label{eq:Q-width}
\ee
where  $g_1$ and $g_2$ are two free parameters to be determined by a best fit, 
and $Q_0=1$ GeV. 


The particular choice of Eq.~\eqref{eq:Q-width}, is intended to mimic the 
main feature of the scale dependence of TMDs, the broadening of the $k_\perp$-distribution 
with variations of $Q^2$. In the full TMD definition, this is partly regulated by 
the non-perturbative, universal function $g_K$ (see, for instance, Eq.~(44) in \cite{Aybat:2011ge}). While 
a one to one correspondence between $g_K$ and our parameter $g_2$ cannot be made, it serves as a proxy
to study the sensitivity of the data to TMD effects.

The values of the $\chi^2$ and of the best fit parameters obtained within this model are presented 
in Table~\ref{tab:tmdevo}. Notice that there is \emph{no} reduction in the value of $\chi^2_{tot}$ 
w.r.t. that of the reference fit in Table~\ref{tab:ref}, although one extra parameter is added 
to the fit.

Fig.~\ref{fig:para-space-g2} shows the correlation between $g_1$ and $g_2$ resulting from our
analysis; the 
$\chi ^2$ is colour-coded: yellow corresponds to its lowest values while red and purple to the 
highest accepted values.
As it is clearly indicated by this plot, both the parameters $g_1$ and $g_2$ are affected by a 
rather large uncertainty. In particular, the central value of $g_1$ remains quite close to that 
extracted in the reference fit while its error increases significantly. The central value of 
$g_2$, instead, turns out to be extremely small, but again affected by a very large uncertainty.
Also in this case the two parameters are strongly correlated, and equally good description of 
the data can be obtained by using rather large and positive values of $g_2$ provided $g_1$ is 
sufficiently small. Paradoxically, even negative values of $g_2$ are acceptable if $g_1$ is 
allowed to grow large and positive, in such a way that the combination 
$(g_1 + g_2 \ln \frac{Q^2}{Q_0^2})$ remains overall positive. 

Fig.~\ref{fig:log-dep-width} shows the effect of TMD evolution on the $k_\perp$-distributions 
of the Sivers functions for $u$ and $d$ flavours, where blue lines represent the Sivers function at 
a fixed value of $x=0.1$, for three different scales of the data, and the light-blue shaded 
bands correspond to the uncertainties for the reference fit (no evolution).
As expected, the small best fit value of $g_2$ 
renders virtually no visible effect in the Sivers function. 

Notice that, as in all of our analysis, the widths of the unpolarized TMDs  
are allowed to be different for each experiment, in accordance with the best description achieved 
within the approach of Ref.~\cite{Anselmino:2013lza}. While more refinements are possible within the 
same gaussian model, for instance, to include $Q^2$ dependence as that of Eq.~\eqref{eq:Q-width}, 
this is unlikely to change the main result of this section, due to the large uncertainties 
on our $g_2$ parameter. We remark that while the differences on the unpolarized widths may be 
attributed to TMD-evolution, this remains as of today an open question. It is quite possible
for other effects to play a role (see for instance Refs.~\cite{Boglione:2016bph,Boglione:2017jlh,Collins:2018teg}).
Regardless of the poor knowledge on the unpolarized TMDs at the kinematics of the Sivers asymmetries,
the central point is that additional $Q^2$ dependence, introduced via $g_2$, does not render 
a result significantly different from that of the reference fit.

These results suggest that in a full TMD analysis, the current Sivers asymmetries will  
probably not constrain strongly the function $g_K$. 
On the other hand, due the large uncertainties on $g_2$, good compatibility between the 
Sivers asymmetries and extracted values of $g_K$ from other observables are likely to be achieved.

\section{\label{Concl} Conclusions}

In this paper we have performed a novel extraction of the Sivers function from SIDIS asymmetry measurements. 
We have exploited all available SIDIS data from HERMES~\cite{Airapetian:2009ae}, 
JLab~\cite{Qian:2011py} and COMPASS-deuteron~\cite{Alekseev:2008aa}, including the new re-analysis of the 2010 run of the 
COMPASS-proton experiment~\cite{Adolph:2016dvl}.

The increased statistics and precision of these new sets of data, together with a finer binning in $Q^2$ as well as in $x$, 
has allowed 
a critical re-analysis of the extraction procedure and its uncertainties. To do so we have adopted a 
simple and transparent parametric form of the Sivers function, as given in Eq.~\eqref{eq:siv}. 
The aim of this new approach is to attempt an extraction of the Sivers TMD based, as much as possible, on the sole 
information provided by experimental data. 
In this framework, it has also been possible to perform a very detailed and accurate study of the parameter space, to provide a 
reliable estimate of the uncertainties which affect the extracted functions, shedding light on the subtle interplay among experimental 
errors, theoretical uncertainties and model-dependent constraints.

With our particular choice of parametrization, see Eq.~\eqref{eq:siv},
we started by assessing how much the measured SIDIS asymmetries could tell us about the flavour content of the Sivers function, 
and on its separation into valence and sea contributions.
We found that the existing data can resolve unambiguously  the total $u$-flavour (valence + sea) contribution, 
while leaving all other flavours largely undetermined (see Table~\ref{tab:flavour-fits}).
We associate to the total $d$-flavour
the additional contribution needed to describe the data, but further investigations 
possibly with more precise data are necessary.
From the statistical point of view, we found that a good
configuration was given by a parametric form that considered the contribution of total $u$ and $d$ flavours of the 
Sivers function, as in Eqs.~\eqref{eq:delta-f-u} and~\eqref{eq:delta-f-d}.
Any attempt to separate valence from sea contributions, namely $u$ from  $\bar u$ and  $d$ from  $\bar d$, 
resulted in a decrease on the quality of the fit, due to a lack of information in the 
experimental data presently available.

For this analysis we have performed two best fits: the first one was a very basic fit, which we referred to as the ``reference fit'', 
based on the most simple parametric form which could reproduce the main features of the Sivers function; the second fit included 
two extra free parameters, to make the parametrization more flexible in the small-$x$ region, in such a way that possible sea 
contributions to the $u$ and $d$ flavours could be accounted for, at least partially. 
Although we could not separate sea from valence contributions within the Sivers first moments, this approach allowed us to obtain a 
much more realistic estimate of the uncertainties affecting the extracted functions at small values of $x$. 
Issues related to the large-$x$ regime, where uncertainties becomes extremely large due to the absence of 
experimental data, were also discussed.  
Drawing well-founded conclusions on the low-$x$ and large-$x$ kinematic regimes will only become possible when 
new experimental information will become available 
from dedicated experiments which are presently being planned, like the EIC~\cite{Accardi:2012qut}, or have just started to run, 
like the newly upgraded JLab12~\cite{Dudek:2012vr}.

A considerable part of our work was devoted to the study of scale dependence effects. 
We considered and compared 3 different scenarios:  no-evolution, collinear twist-3 evolution and TMD-evolution.

Collinear twist-3 evolution, which proceeds only through $x$ while not affecting the $k_\perp$ dependence of the Sivers 
function, was found to be quite fast. 
In fact, when spanning the range of $\langle Q^2 \rangle$ values covered by the experimental data 
($1.2$ GeV$^2 < \langle Q^2\rangle < 40$ GeV$^2$) the extracted Sivers function shows variations that are larger than the 
error band for the reference fit. This suggests that the data can help to determine some scale dependence on the first moment 
of the Sivers function. These results justify a cautious optimism in the 
possibility of observing this kind of scale dependence in the SIDIS asymmetry experimental measurements. 

Signals of TMD evolution, which instead affects mostly the $k_\perp$ dependence of the Sivers function 
(effects involving the $x$-dependence are expected to roughly cancel in the asymmetry ratios) 
turned out to be  more elusive.
Our attempts to estimate them resulted in a rather poor determination of the $g_2$ parameter, 
which regulates the logarithmic variation of the $k_\perp$ width with $Q^2$, and is intended
to mimic the behaviour of the non-perturbative function $g_K$ defined in the full TMD approach of~\cite{Aybat:2011ge}. 
Our best fit delivered a very small value of $g_2$, with a large uncertainty.
This does not mean that TMD evolution is slow. In fact, within the large uncertainty bands corresponding to this extraction, 
there is room for quite a large variety of different $Q^2$ behaviours.
Unfortunately, the available experimental information is presently too limited to determine $g_2$ 
with a satisfactory precision. While further constraints from a full TMD analysis may help to ease this uncertainty,
it is unlikely that, for instance,
$g_K$ can be constrained via the Sivers asymmetries. However, compatibility with information on $g_K$ from some other 
data sets, such as SIDIS multiplicities, can probably be easily accomplished, as evidenced by our large uncertainties in
$g_2$.

Finally, we comment on the role of the unpolarized TMDs in the extraction of the Sivers function. 
As shown in Fig.~\ref{fig:combined-fit},
different assumptions about these functions can alter results significantly. Differently from previous 
analyses, our choice for the unpolarized functions is based on an approach that better describes 
most of the available unpolarized SIDIS data. This consideration actually releases the tension on the model
of the Sivers function, otherwise encountered when trying to simultaneously fit COMPASS and HERMES data.
Among other complications this issue may raise, we realized that this kind of tension can 
reduce the statistical significance of the analysis, since it increases the minimal $\chi^2$ values. 
This can, for instance, lead to inadvertently over-fit the data by adding more parameters in order 
to reduce an artificially large $\chi^2$. This type of complications make it evident how critical it is to 
obtain a better knowledge of the unpolarized functions, not only for this but also for any other
SIDIS asymmetries.

In conclusion, this type of analysis is an essential step which, after the first decade of pioneering studies, 
may lead us toward a new phase of high precision TMD physics. 
Our bottom-up approach to extract the Sivers function, while carefully
keeping track of error estimation and of the sensitivity of the data to different TMD effects, 
may assume a relevant role 
as new measurements, with ever increasing statistics and precision, are becoming available 
from dedicated experiments.

 \acknowledgments
 \noindent
We are very grateful to M. Anselmino for his support and for his valuable contributions to this project, 
and to A. Prokudin for his crucial insights and for his help in the implementation of the collinear 
twist-3 evolution of the Sivers function. 
We thank F. Murgia for useful discussions and for contributing to the early stages of this work. 
%


\begin{thebibliography}{45}
\expandafter\ifx\csname natexlab\endcsname\relax\def\natexlab#1{#1}\fi
\expandafter\ifx\csname bibnamefont\endcsname\relax
  \def\bibnamefont#1{#1}\fi
\expandafter\ifx\csname bibfnamefont\endcsname\relax
  \def\bibfnamefont#1{#1}\fi
\expandafter\ifx\csname citenamefont\endcsname\relax
  \def\citenamefont#1{#1}\fi
\expandafter\ifx\csname url\endcsname\relax
  \def\url#1{\texttt{#1}}\fi
\expandafter\ifx\csname urlprefix\endcsname\relax\def\urlprefix{URL }\fi
\providecommand{\bibinfo}[2]{#2}
\providecommand{\eprint}[2][]{\url{#2}}

\bibitem[{\citenamefont{Sivers}(1990)}]{Sivers:1989cc}
\bibinfo{author}{\bibfnamefont{D.~W.} \bibnamefont{Sivers}},
  \bibinfo{journal}{Phys. Rev.} \textbf{\bibinfo{volume}{D41}},
  \bibinfo{pages}{83} (\bibinfo{year}{1990}).

\bibitem[{\citenamefont{Sivers}(1991)}]{Sivers:1990fh}
\bibinfo{author}{\bibfnamefont{D.~W.} \bibnamefont{Sivers}},
  \bibinfo{journal}{Phys. Rev.} \textbf{\bibinfo{volume}{D43}},
  \bibinfo{pages}{261} (\bibinfo{year}{1991}).

\bibitem[{\citenamefont{Brodsky et~al.}(2002)\citenamefont{Brodsky, Hwang, and
  Schmidt}}]{Brodsky:2002cx}
\bibinfo{author}{\bibfnamefont{S.~J.} \bibnamefont{Brodsky}},
  \bibinfo{author}{\bibfnamefont{D.~S.} \bibnamefont{Hwang}}, \bibnamefont{and}
  \bibinfo{author}{\bibfnamefont{I.}~\bibnamefont{Schmidt}},
  \bibinfo{journal}{Phys. Lett.} \textbf{\bibinfo{volume}{B530}},
  \bibinfo{pages}{99} (\bibinfo{year}{2002}), \eprint{hep-ph/0201296}.

\bibitem[{\citenamefont{Collins}(2002)}]{Collins:2002kn}
\bibinfo{author}{\bibfnamefont{J.~C.} \bibnamefont{Collins}},
  \bibinfo{journal}{Phys. Lett.} \textbf{\bibinfo{volume}{B536}},
  \bibinfo{pages}{43} (\bibinfo{year}{2002}), \eprint{hep-ph/0204004}.

\bibitem[{\citenamefont{Adamczyk et~al.}(2016)}]{Adamczyk:2015gyk}
\bibinfo{author}{\bibfnamefont{L.}~\bibnamefont{Adamczyk}} \bibnamefont{et~al.}
  (\bibinfo{collaboration}{STAR}), \bibinfo{journal}{Phys. Rev. Lett.}
  \textbf{\bibinfo{volume}{116}}, \bibinfo{pages}{132301}
  (\bibinfo{year}{2016}), \eprint{1511.06003}.

\bibitem[{\citenamefont{Huang et~al.}(2016)\citenamefont{Huang, Kang, Vitev,
  and Xing}}]{Huang:2015vpy}
\bibinfo{author}{\bibfnamefont{J.}~\bibnamefont{Huang}},
  \bibinfo{author}{\bibfnamefont{Z.-B.} \bibnamefont{Kang}},
  \bibinfo{author}{\bibfnamefont{I.}~\bibnamefont{Vitev}}, \bibnamefont{and}
  \bibinfo{author}{\bibfnamefont{H.}~\bibnamefont{Xing}},
  \bibinfo{journal}{Phys. Rev.} \textbf{\bibinfo{volume}{D93}},
  \bibinfo{pages}{014036} (\bibinfo{year}{2016}), \eprint{1511.06764}.

\bibitem[{\citenamefont{Anselmino et~al.}(2017)\citenamefont{Anselmino,
  Boglione, D'Alesio, Murgia, and Prokudin}}]{Anselmino:2016uie}
\bibinfo{author}{\bibfnamefont{M.}~\bibnamefont{Anselmino}},
  \bibinfo{author}{\bibfnamefont{M.}~\bibnamefont{Boglione}},
  \bibinfo{author}{\bibfnamefont{U.}~\bibnamefont{D'Alesio}},
  \bibinfo{author}{\bibfnamefont{F.}~\bibnamefont{Murgia}}, \bibnamefont{and}
  \bibinfo{author}{\bibfnamefont{A.}~\bibnamefont{Prokudin}},
  \bibinfo{journal}{JHEP} \textbf{\bibinfo{volume}{04}}, \bibinfo{pages}{046}
  (\bibinfo{year}{2017}), \eprint{1612.06413}.

\bibitem[{\citenamefont{Aghasyan et~al.}(2017)}]{Aghasyan:2017jop}
\bibinfo{author}{\bibfnamefont{M.}~\bibnamefont{Aghasyan}} \bibnamefont{et~al.}
  (\bibinfo{collaboration}{COMPASS}), \bibinfo{journal}{Phys. Rev. Lett.}
  \textbf{\bibinfo{volume}{119}}, \bibinfo{pages}{112002}
  (\bibinfo{year}{2017}), \eprint{1704.00488}.

\bibitem[{\citenamefont{Adolph et~al.}(2017)}]{Adolph:2016dvl}
\bibinfo{author}{\bibfnamefont{C.}~\bibnamefont{Adolph}} \bibnamefont{et~al.}
  (\bibinfo{collaboration}{COMPASS}), \bibinfo{journal}{Phys. Lett.}
  \textbf{\bibinfo{volume}{B770}}, \bibinfo{pages}{138} (\bibinfo{year}{2017}),
  \eprint{1609.07374}.

\bibitem[{\citenamefont{Aybat et~al.}(2012{\natexlab{a}})\citenamefont{Aybat,
  Collins, Qiu, and Rogers}}]{Aybat:2011ge}
\bibinfo{author}{\bibfnamefont{S.~M.} \bibnamefont{Aybat}},
  \bibinfo{author}{\bibfnamefont{J.~C.} \bibnamefont{Collins}},
  \bibinfo{author}{\bibfnamefont{J.-W.} \bibnamefont{Qiu}}, \bibnamefont{and}
  \bibinfo{author}{\bibfnamefont{T.~C.} \bibnamefont{Rogers}},
  \bibinfo{journal}{Phys. Rev.} \textbf{\bibinfo{volume}{D85}},
  \bibinfo{pages}{034043} (\bibinfo{year}{2012}{\natexlab{a}}),
  \eprint{1110.6428}.

\bibitem[{\citenamefont{Anselmino et~al.}(2014)\citenamefont{Anselmino,
  Boglione, Gonzalez~Hernandez, Melis, and Prokudin}}]{Anselmino:2013lza}
\bibinfo{author}{\bibfnamefont{M.}~\bibnamefont{Anselmino}},
  \bibinfo{author}{\bibfnamefont{M.}~\bibnamefont{Boglione}},
  \bibinfo{author}{\bibfnamefont{J.~O.} \bibnamefont{Gonzalez~Hernandez}},
  \bibinfo{author}{\bibfnamefont{S.}~\bibnamefont{Melis}}, \bibnamefont{and}
  \bibinfo{author}{\bibfnamefont{A.}~\bibnamefont{Prokudin}},
  \bibinfo{journal}{JHEP} \textbf{\bibinfo{volume}{04}}, \bibinfo{pages}{005}
  (\bibinfo{year}{2014}), \eprint{1312.6261}.

\bibitem[{\citenamefont{Signori et~al.}(2013)\citenamefont{Signori, Bacchetta,
  Radici, and Schnell}}]{Signori:2013mda}
\bibinfo{author}{\bibfnamefont{A.}~\bibnamefont{Signori}},
  \bibinfo{author}{\bibfnamefont{A.}~\bibnamefont{Bacchetta}},
  \bibinfo{author}{\bibfnamefont{M.}~\bibnamefont{Radici}}, \bibnamefont{and}
  \bibinfo{author}{\bibfnamefont{G.}~\bibnamefont{Schnell}},
  \bibinfo{journal}{JHEP} \textbf{\bibinfo{volume}{11}}, \bibinfo{pages}{194}
  (\bibinfo{year}{2013}), \eprint{1309.3507}.

\bibitem[{\citenamefont{Bacchetta et~al.}(2017)\citenamefont{Bacchetta,
  Delcarro, Pisano, Radici, and Signori}}]{Bacchetta:2017gcc}
\bibinfo{author}{\bibfnamefont{A.}~\bibnamefont{Bacchetta}},
  \bibinfo{author}{\bibfnamefont{F.}~\bibnamefont{Delcarro}},
  \bibinfo{author}{\bibfnamefont{C.}~\bibnamefont{Pisano}},
  \bibinfo{author}{\bibfnamefont{M.}~\bibnamefont{Radici}}, \bibnamefont{and}
  \bibinfo{author}{\bibfnamefont{A.}~\bibnamefont{Signori}},
  \bibinfo{journal}{JHEP} \textbf{\bibinfo{volume}{06}}, \bibinfo{pages}{081}
  (\bibinfo{year}{2017}), \eprint{1703.10157}.

\bibitem[{\citenamefont{Boglione et~al.}(2015)\citenamefont{Boglione,
  Gonzalez~Hernandez, Melis, and Prokudin}}]{Boglione:2014oea}
\bibinfo{author}{\bibfnamefont{M.}~\bibnamefont{Boglione}},
  \bibinfo{author}{\bibfnamefont{J.~O.} \bibnamefont{Gonzalez~Hernandez}},
  \bibinfo{author}{\bibfnamefont{S.}~\bibnamefont{Melis}}, \bibnamefont{and}
  \bibinfo{author}{\bibfnamefont{A.}~\bibnamefont{Prokudin}},
  \bibinfo{journal}{JHEP} \textbf{\bibinfo{volume}{02}}, \bibinfo{pages}{095}
  (\bibinfo{year}{2015}), \eprint{1412.1383}.

\bibitem[{\citenamefont{Boglione
  et~al.}(2017{\natexlab{a}})\citenamefont{Boglione, Collins, Gamberg,
  Gonzalez-Hernandez, Rogers, and Sato}}]{Boglione:2016bph}
\bibinfo{author}{\bibfnamefont{M.}~\bibnamefont{Boglione}},
  \bibinfo{author}{\bibfnamefont{J.}~\bibnamefont{Collins}},
  \bibinfo{author}{\bibfnamefont{L.}~\bibnamefont{Gamberg}},
  \bibinfo{author}{\bibfnamefont{J.~O.} \bibnamefont{Gonzalez-Hernandez}},
  \bibinfo{author}{\bibfnamefont{T.~C.} \bibnamefont{Rogers}},
  \bibnamefont{and} \bibinfo{author}{\bibfnamefont{N.}~\bibnamefont{Sato}},
  \bibinfo{journal}{Phys. Lett.} \textbf{\bibinfo{volume}{B766}},
  \bibinfo{pages}{245} (\bibinfo{year}{2017}{\natexlab{a}}),
  \eprint{1611.10329}.

\bibitem[{\citenamefont{Anselmino
  et~al.}(2009{\natexlab{a}})\citenamefont{Anselmino, Boglione, D'Alesio,
  Kotzinian, Melis, Murgia, Prokudin, and Turk}}]{Anselmino:2008sga}
\bibinfo{author}{\bibfnamefont{M.}~\bibnamefont{Anselmino}},
  \bibinfo{author}{\bibfnamefont{M.}~\bibnamefont{Boglione}},
  \bibinfo{author}{\bibfnamefont{U.}~\bibnamefont{D'Alesio}},
  \bibinfo{author}{\bibfnamefont{A.}~\bibnamefont{Kotzinian}},
  \bibinfo{author}{\bibfnamefont{S.}~\bibnamefont{Melis}},
  \bibinfo{author}{\bibfnamefont{F.}~\bibnamefont{Murgia}},
  \bibinfo{author}{\bibfnamefont{A.}~\bibnamefont{Prokudin}}, \bibnamefont{and}
  \bibinfo{author}{\bibfnamefont{C.}~\bibnamefont{Turk}},
  \bibinfo{journal}{Eur. Phys. J.} \textbf{\bibinfo{volume}{A39}},
  \bibinfo{pages}{89} (\bibinfo{year}{2009}{\natexlab{a}}), \eprint{0805.2677}.

\bibitem[{\citenamefont{Anselmino et~al.}(2012)\citenamefont{Anselmino,
  Boglione, and Melis}}]{Anselmino:2012aa}
\bibinfo{author}{\bibfnamefont{M.}~\bibnamefont{Anselmino}},
  \bibinfo{author}{\bibfnamefont{M.}~\bibnamefont{Boglione}}, \bibnamefont{and}
  \bibinfo{author}{\bibfnamefont{S.}~\bibnamefont{Melis}},
  \bibinfo{journal}{Phys. Rev.} \textbf{\bibinfo{volume}{D86}},
  \bibinfo{pages}{014028} (\bibinfo{year}{2012}), \eprint{1204.1239}.

\bibitem[{\citenamefont{Agarwala et~al.}(2018)}]{COMPASSII:2021}
\bibinfo{author}{\bibfnamefont{J.}~\bibnamefont{Agarwala}} \bibnamefont{et~al.}
  (\bibinfo{collaboration}{COMPASS Collaboration}),
  \bibinfo{journal}{CERN-SPSC-2017-034,SPSC-P-340-ADD-1}
  (\bibinfo{year}{2018}).

\bibitem[{\citenamefont{Anselmino et~al.}(2005)\citenamefont{Anselmino,
  Boglione, D'Alesio, Kotzinian, Murgia, and Prokudin}}]{Anselmino:2005nn}
\bibinfo{author}{\bibfnamefont{M.}~\bibnamefont{Anselmino}},
  \bibinfo{author}{\bibfnamefont{M.}~\bibnamefont{Boglione}},
  \bibinfo{author}{\bibfnamefont{U.}~\bibnamefont{D'Alesio}},
  \bibinfo{author}{\bibfnamefont{A.}~\bibnamefont{Kotzinian}},
  \bibinfo{author}{\bibfnamefont{F.}~\bibnamefont{Murgia}}, \bibnamefont{and}
  \bibinfo{author}{\bibfnamefont{A.}~\bibnamefont{Prokudin}},
  \bibinfo{journal}{Phys. Rev.} \textbf{\bibinfo{volume}{D71}},
  \bibinfo{pages}{074006} (\bibinfo{year}{2005}), \eprint{hep-ph/0501196}.

\bibitem[{\citenamefont{Anselmino et~al.}(2011)\citenamefont{Anselmino,
  Boglione, D'Alesio, Melis, Murgia, Nocera, and Prokudin}}]{Anselmino:2011ch}
\bibinfo{author}{\bibfnamefont{M.}~\bibnamefont{Anselmino}},
  \bibinfo{author}{\bibfnamefont{M.}~\bibnamefont{Boglione}},
  \bibinfo{author}{\bibfnamefont{U.}~\bibnamefont{D'Alesio}},
  \bibinfo{author}{\bibfnamefont{S.}~\bibnamefont{Melis}},
  \bibinfo{author}{\bibfnamefont{F.}~\bibnamefont{Murgia}},
  \bibinfo{author}{\bibfnamefont{E.~R.} \bibnamefont{Nocera}},
  \bibnamefont{and} \bibinfo{author}{\bibfnamefont{A.}~\bibnamefont{Prokudin}},
  \bibinfo{journal}{Phys. Rev.} \textbf{\bibinfo{volume}{D83}},
  \bibinfo{pages}{114019} (\bibinfo{year}{2011}), \eprint{1101.1011}.

\bibitem[{\citenamefont{Anselmino
  et~al.}(2009{\natexlab{b}})\citenamefont{Anselmino, Boglione, D'Alesio,
  Melis, Murgia, and Prokudin}}]{Anselmino:2009st}
\bibinfo{author}{\bibfnamefont{M.}~\bibnamefont{Anselmino}},
  \bibinfo{author}{\bibfnamefont{M.}~\bibnamefont{Boglione}},
  \bibinfo{author}{\bibfnamefont{U.}~\bibnamefont{D'Alesio}},
  \bibinfo{author}{\bibfnamefont{S.}~\bibnamefont{Melis}},
  \bibinfo{author}{\bibfnamefont{F.}~\bibnamefont{Murgia}}, \bibnamefont{and}
  \bibinfo{author}{\bibfnamefont{A.}~\bibnamefont{Prokudin}},
  \bibinfo{journal}{Phys. Rev.} \textbf{\bibinfo{volume}{D79}},
  \bibinfo{pages}{054010} (\bibinfo{year}{2009}{\natexlab{b}}),
  \eprint{0901.3078}.

\bibitem[{\citenamefont{Stump et~al.}(2003)\citenamefont{Stump, Huston,
  Pumplin, Tung, Lai, Kuhlmann, and Owens}}]{Stump:2003yu}
\bibinfo{author}{\bibfnamefont{D.}~\bibnamefont{Stump}},
  \bibinfo{author}{\bibfnamefont{J.}~\bibnamefont{Huston}},
  \bibinfo{author}{\bibfnamefont{J.}~\bibnamefont{Pumplin}},
  \bibinfo{author}{\bibfnamefont{W.-K.} \bibnamefont{Tung}},
  \bibinfo{author}{\bibfnamefont{H.~L.} \bibnamefont{Lai}},
  \bibinfo{author}{\bibfnamefont{S.}~\bibnamefont{Kuhlmann}}, \bibnamefont{and}
  \bibinfo{author}{\bibfnamefont{J.~F.} \bibnamefont{Owens}},
  \bibinfo{journal}{JHEP} \textbf{\bibinfo{volume}{10}}, \bibinfo{pages}{046}
  (\bibinfo{year}{2003}), \eprint{hep-ph/0303013}.

\bibitem[{\citenamefont{de~Florian et~al.}(2007)\citenamefont{de~Florian,
  Sassot, and Stratmann}}]{deFlorian:2007aj}
\bibinfo{author}{\bibfnamefont{D.}~\bibnamefont{de~Florian}},
  \bibinfo{author}{\bibfnamefont{R.}~\bibnamefont{Sassot}}, \bibnamefont{and}
  \bibinfo{author}{\bibfnamefont{M.}~\bibnamefont{Stratmann}},
  \bibinfo{journal}{Phys. Rev.} \textbf{\bibinfo{volume}{D75}},
  \bibinfo{pages}{114010} (\bibinfo{year}{2007}), \eprint{hep-ph/0703242}.

\bibitem[{\citenamefont{Gribov and Lipatov}(1972)}]{Gribov:1972ri}
\bibinfo{author}{\bibfnamefont{V.~N.} \bibnamefont{Gribov}} \bibnamefont{and}
  \bibinfo{author}{\bibfnamefont{L.~N.} \bibnamefont{Lipatov}},
  \bibinfo{journal}{Sov. J. Nucl. Phys.} \textbf{\bibinfo{volume}{15}},
  \bibinfo{pages}{438} (\bibinfo{year}{1972}), \bibinfo{note}{[Yad.
  Fiz.15,781(1972)]}.

\bibitem[{\citenamefont{Altarelli and Parisi}(1977)}]{Altarelli:1977zs}
\bibinfo{author}{\bibfnamefont{G.}~\bibnamefont{Altarelli}} \bibnamefont{and}
  \bibinfo{author}{\bibfnamefont{G.}~\bibnamefont{Parisi}},
  \bibinfo{journal}{Nucl. Phys.} \textbf{\bibinfo{volume}{B126}},
  \bibinfo{pages}{298} (\bibinfo{year}{1977}).

\bibitem[{\citenamefont{Dokshitzer}(1977)}]{Dokshitzer:1977sg}
\bibinfo{author}{\bibfnamefont{Y.~L.} \bibnamefont{Dokshitzer}},
  \bibinfo{journal}{Sov. Phys. JETP} \textbf{\bibinfo{volume}{46}},
  \bibinfo{pages}{641} (\bibinfo{year}{1977}), \bibinfo{note}{[Zh. Eksp. Teor.
  Fiz.73,1216(1977)]}.

\bibitem[{\citenamefont{Airapetian et~al.}(2009)}]{Airapetian:2009ae}
\bibinfo{author}{\bibfnamefont{A.}~\bibnamefont{Airapetian}}
  \bibnamefont{et~al.} (\bibinfo{collaboration}{HERMES Collaboration}),
  \bibinfo{journal}{Phys. Rev. Lett.} \textbf{\bibinfo{volume}{103}},
  \bibinfo{pages}{152002} (\bibinfo{year}{2009}), \eprint{0906.3918}.

\bibitem[{\citenamefont{Alekseev et~al.}(2009)}]{Alekseev:2008aa}
\bibinfo{author}{\bibfnamefont{M.}~\bibnamefont{Alekseev}} \bibnamefont{et~al.}
  (\bibinfo{collaboration}{COMPASS Collaboration}), \bibinfo{journal}{Phys.
  Lett.} \textbf{\bibinfo{volume}{B673}}, \bibinfo{pages}{127}
  (\bibinfo{year}{2009}), \eprint{0802.2160}.

\bibitem[{\citenamefont{Qian et~al.}(2011)}]{Qian:2011py}
\bibinfo{author}{\bibfnamefont{X.}~\bibnamefont{Qian}} \bibnamefont{et~al.}
  (\bibinfo{collaboration}{Jefferson Lab Hall A Collaboration}),
  \bibinfo{journal}{Phys. Rev. Lett.} \textbf{\bibinfo{volume}{107}},
  \bibinfo{pages}{072003} (\bibinfo{year}{2011}), \eprint{1106.0363}.

\bibitem[{\citenamefont{Aidala et~al.}(2014)\citenamefont{Aidala, Field,
  Gamberg, and Rogers}}]{Aidala:2014hva}
\bibinfo{author}{\bibfnamefont{C.~A.} \bibnamefont{Aidala}},
  \bibinfo{author}{\bibfnamefont{B.}~\bibnamefont{Field}},
  \bibinfo{author}{\bibfnamefont{L.~P.} \bibnamefont{Gamberg}},
  \bibnamefont{and} \bibinfo{author}{\bibfnamefont{T.~C.}
  \bibnamefont{Rogers}}, \bibinfo{journal}{Phys. Rev.}
  \textbf{\bibinfo{volume}{D89}}, \bibinfo{pages}{094002}
  (\bibinfo{year}{2014}), \eprint{1401.2654}.

\bibitem[{\citenamefont{Melis et~al.}(2015)\citenamefont{Melis, Boglione,
  Gonzalez~Hernandez, and Prokudin}}]{Melis:2015ycg}
\bibinfo{author}{\bibfnamefont{S.}~\bibnamefont{Melis}},
  \bibinfo{author}{\bibfnamefont{M.}~\bibnamefont{Boglione}},
  \bibinfo{author}{\bibfnamefont{J.}~\bibnamefont{Gonzalez~Hernandez}},
  \bibnamefont{and} \bibinfo{author}{\bibfnamefont{A.}~\bibnamefont{Prokudin}},
  \bibinfo{journal}{PoS} \textbf{\bibinfo{volume}{QCDEV2015}},
  \bibinfo{pages}{038} (\bibinfo{year}{2015}).

\bibitem[{\citenamefont{Collins et~al.}(2016)\citenamefont{Collins, Gamberg,
  Prokudin, Rogers, Sato, and Wang}}]{Collins:2016hqq}
\bibinfo{author}{\bibfnamefont{J.}~\bibnamefont{Collins}},
  \bibinfo{author}{\bibfnamefont{L.}~\bibnamefont{Gamberg}},
  \bibinfo{author}{\bibfnamefont{A.}~\bibnamefont{Prokudin}},
  \bibinfo{author}{\bibfnamefont{T.~C.} \bibnamefont{Rogers}},
  \bibinfo{author}{\bibfnamefont{N.}~\bibnamefont{Sato}}, \bibnamefont{and}
  \bibinfo{author}{\bibfnamefont{B.}~\bibnamefont{Wang}},
  \bibinfo{journal}{Phys. Rev.} \textbf{\bibinfo{volume}{D94}},
  \bibinfo{pages}{034014} (\bibinfo{year}{2016}), \eprint{1605.00671}.

\bibitem[{\citenamefont{Adolph et~al.}(2015)}]{Adolph:2014zba}
\bibinfo{author}{\bibfnamefont{C.}~\bibnamefont{Adolph}} \bibnamefont{et~al.}
  (\bibinfo{collaboration}{COMPASS Collaboration}), \bibinfo{journal}{Phys.
  Lett.} \textbf{\bibinfo{volume}{B744}}, \bibinfo{pages}{250}
  (\bibinfo{year}{2015}), \eprint{1408.4405}.

\bibitem[{\citenamefont{Accardi et~al.}(2016)}]{Accardi:2012qut}
\bibinfo{author}{\bibfnamefont{A.}~\bibnamefont{Accardi}} \bibnamefont{et~al.},
  \bibinfo{journal}{Eur. Phys. J.} \textbf{\bibinfo{volume}{A52}},
  \bibinfo{pages}{268} (\bibinfo{year}{2016}), \eprint{1212.1701}.

\bibitem[{\citenamefont{Dudek et~al.}(2012)}]{Dudek:2012vr}
\bibinfo{author}{\bibfnamefont{J.}~\bibnamefont{Dudek}} \bibnamefont{et~al.},
  \bibinfo{journal}{Eur. Phys. J.} \textbf{\bibinfo{volume}{A48}},
  \bibinfo{pages}{187} (\bibinfo{year}{2012}), \eprint{1208.1244}.

\bibitem[{\citenamefont{Kang and Qiu}(2009)}]{Kang:2008ey}
\bibinfo{author}{\bibfnamefont{Z.-B.} \bibnamefont{Kang}} \bibnamefont{and}
  \bibinfo{author}{\bibfnamefont{J.-W.} \bibnamefont{Qiu}},
  \bibinfo{journal}{Phys. Rev.} \textbf{\bibinfo{volume}{D79}},
  \bibinfo{pages}{016003} (\bibinfo{year}{2009}), \eprint{0811.3101}.

\bibitem[{\citenamefont{Vogelsang and Yuan}(2009)}]{Vogelsang:2009pj}
\bibinfo{author}{\bibfnamefont{W.}~\bibnamefont{Vogelsang}} \bibnamefont{and}
  \bibinfo{author}{\bibfnamefont{F.}~\bibnamefont{Yuan}},
  \bibinfo{journal}{Phys. Rev.} \textbf{\bibinfo{volume}{D79}},
  \bibinfo{pages}{094010} (\bibinfo{year}{2009}), \eprint{0904.0410}.

\bibitem[{\citenamefont{Braun et~al.}(2009)\citenamefont{Braun, Manashov, and
  Pirnay}}]{Braun:2009mi}
\bibinfo{author}{\bibfnamefont{V.~M.} \bibnamefont{Braun}},
  \bibinfo{author}{\bibfnamefont{A.~N.} \bibnamefont{Manashov}},
  \bibnamefont{and} \bibinfo{author}{\bibfnamefont{B.}~\bibnamefont{Pirnay}},
  \bibinfo{journal}{Phys. Rev.} \textbf{\bibinfo{volume}{D80}},
  \bibinfo{pages}{114002} (\bibinfo{year}{2009}), \bibinfo{note}{[Erratum:
  Phys. Rev.D86,119902(2012)]}, \eprint{0909.3410}.

\bibitem[{\citenamefont{Kang and Qiu}(2012)}]{Kang:2012em}
\bibinfo{author}{\bibfnamefont{Z.-B.} \bibnamefont{Kang}} \bibnamefont{and}
  \bibinfo{author}{\bibfnamefont{J.-W.} \bibnamefont{Qiu}},
  \bibinfo{journal}{Phys. Lett.} \textbf{\bibinfo{volume}{B713}},
  \bibinfo{pages}{273} (\bibinfo{year}{2012}), \eprint{1205.1019}.

\bibitem[{\citenamefont{Salam and Rojo}(2009)}]{Salam:2008qg}
\bibinfo{author}{\bibfnamefont{G.~P.} \bibnamefont{Salam}} \bibnamefont{and}
  \bibinfo{author}{\bibfnamefont{J.}~\bibnamefont{Rojo}},
  \bibinfo{journal}{Comput. Phys. Commun.} \textbf{\bibinfo{volume}{180}},
  \bibinfo{pages}{120} (\bibinfo{year}{2009}), \eprint{0804.3755}.

\bibitem[{\citenamefont{Prokudin}(2018)}]{Prokudin:hoppet}
\bibinfo{author}{\bibfnamefont{A.}~\bibnamefont{Prokudin}},
  \bibinfo{journal}{private communication}  (\bibinfo{year}{2018}).

\bibitem[{\citenamefont{Aybat et~al.}(2012{\natexlab{b}})\citenamefont{Aybat,
  Prokudin, and Rogers}}]{Aybat:2011ta}
\bibinfo{author}{\bibfnamefont{S.~M.} \bibnamefont{Aybat}},
  \bibinfo{author}{\bibfnamefont{A.}~\bibnamefont{Prokudin}}, \bibnamefont{and}
  \bibinfo{author}{\bibfnamefont{T.~C.} \bibnamefont{Rogers}},
  \bibinfo{journal}{Phys. Rev. Lett.} \textbf{\bibinfo{volume}{108}},
  \bibinfo{pages}{242003} (\bibinfo{year}{2012}{\natexlab{b}}),
  \eprint{1112.4423}.

\bibitem[{\citenamefont{Moffat et~al.}(2017)\citenamefont{Moffat, Melnitchouk,
  Rogers, and Sato}}]{Moffat:2017sha}
\bibinfo{author}{\bibfnamefont{E.}~\bibnamefont{Moffat}},
  \bibinfo{author}{\bibfnamefont{W.}~\bibnamefont{Melnitchouk}},
  \bibinfo{author}{\bibfnamefont{T.~C.} \bibnamefont{Rogers}},
  \bibnamefont{and} \bibinfo{author}{\bibfnamefont{N.}~\bibnamefont{Sato}},
  \bibinfo{journal}{Phys. Rev.} \textbf{\bibinfo{volume}{D95}},
  \bibinfo{pages}{096008} (\bibinfo{year}{2017}), \eprint{1702.03955}.

\bibitem[{\citenamefont{Boglione
  et~al.}(2017{\natexlab{b}})\citenamefont{Boglione, Gonzalez-Hernandez, and
  Taghavi}}]{Boglione:2017jlh}
\bibinfo{author}{\bibfnamefont{M.}~\bibnamefont{Boglione}},
  \bibinfo{author}{\bibfnamefont{J.~O.} \bibnamefont{Gonzalez-Hernandez}},
  \bibnamefont{and} \bibinfo{author}{\bibfnamefont{R.}~\bibnamefont{Taghavi}},
  \bibinfo{journal}{Phys. Lett.} \textbf{\bibinfo{volume}{B772}},
  \bibinfo{pages}{78} (\bibinfo{year}{2017}{\natexlab{b}}),
  \eprint{1704.08882}.

\bibitem[{\citenamefont{Collins and Rogers}(2018)}]{Collins:2018teg}
\bibinfo{author}{\bibfnamefont{J.}~\bibnamefont{Collins}} \bibnamefont{and}
  \bibinfo{author}{\bibfnamefont{T.~C.} \bibnamefont{Rogers}}
  (\bibinfo{year}{2018}), \eprint{arXiv:1801.02704}.

\end{thebibliography}


\end{document}